\documentclass[
  prx,aps,floatfix,
  showpacs,twocolumn,
  preprintnumbers,
  amsmath,amssymb,
  superscriptaddress,
  longbibliography
]{revtex4-2}
\usepackage[utf8]{inputenc}
\usepackage{amsmath}
\usepackage{amsthm}
\usepackage{amsfonts}
\usepackage{amssymb}
\usepackage{amssymb}
\usepackage{bm}
\usepackage[colorlinks,citecolor=blue,urlcolor=blue,linkcolor=blue]{hyperref}
\usepackage{algorithm}
\usepackage{algpseudocode}
\usepackage[capitalise]{cleveref}
\usepackage{physics}
\usepackage{graphicx}
\usepackage{color}
\usepackage[dvipsnames]{xcolor}
\usepackage{url}
\usepackage{bbm}
\usepackage{rotating}
\usepackage[normalem]{ulem}

\crefname{supp}{supplementary material}{supplementary materials}
\Crefname{supp}{Supplementary Material}{Supplementary Materials}

\theoremstyle{definition}
\newtheorem{definition}{Definition}
\newtheorem{example}{Example}
\newtheorem{lemma}{Lemma}
\newtheorem{corollary}{Corollary}
\newtheorem{proposition}{Proposition}

\theoremstyle{plain}
\newtheorem{theorem}{Theorem}

\theoremstyle{remark}
\newtheorem{notation}{Notation}

\newcommand{\expect}[1]{\left\langle #1 \right\rangle}

\newcommand{\lie}{ad}
\newcommand{\hilbert}[1]{\mathcal{H}(\mathbb{C}^d)^{\otimes #1}}
\newcommand{\step}{l}
\newcommand{\hermitian}[1]{\mathcal{A}_{#1}}
\newcommand{\dom}[1]{\mathrm{dom}(#1)}
\newcommand{\locality}{w}
\newcommand{\opmap}[2]{#1[#2]}

\newcommand{\conn}[1]{\opmap{R}{#1}}
\newcommand{\connsub}[2]{\opmap{R_{#1}}{#2}}
\newcommand{\sbset}{(\partial H_n)^{G_{\step}}}
\newcommand{\batch}{b}
\newcommand{\boundary}{B}
\newcommand{\isometric}[2]{#1^{\dagger} #2 #1}
\newcommand{\opte}[2]{#1^{\dagger} #2 #1}
\newcommand{\ansatz}[1]{f_{#1}^{\theta}}
\newcommand{\omm}[1]{\text{OMM}_{#1}^{\theta}}
\newcommand{\hem}[1]{\text{HEM}_{#1}}
\newcommand{\device}[1]{H_{#1}^{\text{dev}}(\theta_{#1}, t_{#1})}
\newcommand{\rydberg}[1]{H_{#1}^{\text{ryd}}(\theta_{#1}, t_{#1})}

\makeatletter
\newcommand*\bigcdot{\mathpalette\bigcdot@{.5}}
\newcommand*\bigcdot@[2]{\mathbin{\vcenter{\hbox{\scalebox{#2}{$\m@th#1\bullet$}}}}}
\makeatother

\begin{document}

\preprint{APS/123-QED}

\title{Operator Learning Renormalization Group}% Force line breaks with \\

\author{Xiu-Zhe Luo}
\email{paper@rogerluo.dev}
\affiliation{%
	Department of Physics and Astronomy, University of Waterloo, Waterloo N2L 3G1, Canada
}%
\affiliation{
	Perimeter Institute for Theoretical Physics, Waterloo, Ontario N2L 2Y5, Canada
}%
\author{Di Luo}%
\affiliation{
	Center for Theoretical Physics, Massachusetts Institute of Technology, Cambridge, MA 02139, USA
}
\affiliation{
	The NSF AI Institute for Artificial Intelligence and Fundamental Interactions
}
\affiliation{
	Department of Physics, Harvard University, Cambridge, MA 02138, USA
}
\author{Roger G. Melko}
\affiliation{%
	Department of Physics and Astronomy, University of Waterloo, Waterloo N2L 3G1, Canada
}%
\affiliation{
	Perimeter Institute for Theoretical Physics, Waterloo, Ontario N2L 2Y5, Canada
}%

\date{\today}% It is always \today, today,
\begin{abstract}
    In this paper, we present a general framework for quantum many-body simulations called the operator learning renormalization group (OLRG). Inspired by machine learning perspectives, OLRG is a generalization of Wilson's numerical renormalization group and White's density matrix renormalization group, which recursively builds a simulatable system to approximate a target system of the same number of sites via operator maps. OLRG uses a loss function to minimize the error of a target property directly by learning the operator map in lieu of a state ansatz. This loss function is designed by a scaling consistency condition that also provides a provable bound for real-time evolution. We implement two versions of the operator maps for classical and quantum simulations. The former, which we call the Operator Matrix Map, can be implemented via neural networks on classical computers. The latter, which we call the Hamiltonian Expression Map, generates device pulse sequences to leverage the capabilities of quantum computing hardware. We illustrate the performance of both maps for calculating time-dependent quantities in the quantum Ising model Hamiltonian.
\end{abstract}

\maketitle

\section{Introduction}
Simulating quantum many-body systems is a fundamental problem in physics with many applications, including the understanding and design of quantum materials, molecules and matter~\cite{altland2010condensed,heisenberg1985theorie,chan2011density}. However, the general simulation problem has been proven to be hard~\cite{kitaev2002classical,gharibian2015quantum}. This has motivated the development of various classical frameworks to tackle the problem heuristically, including Wilson's Numerical Renormalization Group (NRG)~\cite{wilson1975renormalization}, White's Density Matrix Renormalization Group (DMRG)~\cite{white1992density,White1993DensitymatrixAF} and Variational Monte Carlo (VMC)~\cite{mcmillan1965ground,Becca_Sorella_2017}. 
It has also motivated strategies for leveraging quantum hardware for simulation, where frameworks
such as quantum phase estimation~\cite{kitaev1995quantum,nielsen2010quantum}, Hamiltonian simulation~\cite{childs2009universal,gilyen2019quantum,doi:10.1137/18M1231511} and variational quantum algorithms (VQA)~\cite{peruzzo2014variational,wecker2015progress,mcclean2016theory} have been proposed to take advantage of devices with potential quantum advantage.

Despite the hardness of the problem, a useful observation is that, upon scaling the system size, many properties of interest (i.e., observables, entanglement entropy, spectrum, etc.) can exhibit minimal fluctuations and demonstrate consistent behavior across adjacent system sizes.~\cite{lieb1972finite,hastings06spectral,wang2021bounding}. This hints that, given an oracle to query observable properties from the $n-1,n-2,\cdots,1$-site system with tractable cost, predicting a property in $n$-site system might be possible. Technically, predicting larger system properties by solving smaller system properties is an appealing direction, allowing the observations and theory relevant to small-system solvers to be transferred into large-scale many-body system solvers.  Historically, numerical renormalization formulations 
such as NRG and DMRG 
have been motivated by this observation. 

\begin{figure*}[t]
	\includegraphics[scale=0.9]{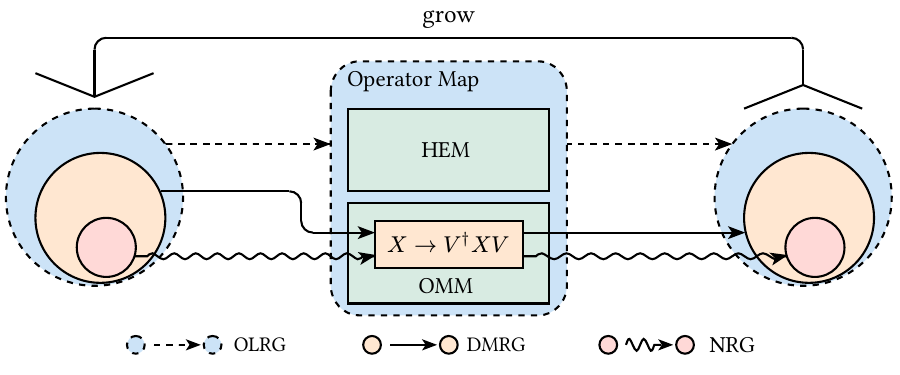}
	\caption{
		Workflow of NRG, DMRG, and OLRG. The colored circles represent the relevant set of operators for each framework. Left is real and right is virtual. (red) NRG keep track of the Hamiltonian operator; (orange) DMRG keep track of operators relevant to construct a superblock; $X \rightarrow \isometric{V}{X}$ is the basis transform into low-energy spectrum subspace or chosen-state subspace used in NRG and DMRG; (blue) OLRG generalizes to set of relevant operators and arbitrary operator maps.
	}\label{fig:nrg_and_dmrg}
\end{figure*}

The success of NRG and especially DMRG relies largely on the linear ansatz for the operator map and the loss function of the spectrum error or average expectation error of a chosen state~\cite{wilson1975renormalization,white1992density,White1993DensitymatrixAF,schollwock2005density}, which can be solved directly using an eigensolver. Linearity is one of the key reasons behind the fast convergence of DMRG, where the local optimal point of the loss function can be identified directly using an eigensolver. However, this linearity also limits the expressiveness, leading to the limitations encountered by matrix product states (MPS) in simulating high-dimensional systems or long-time dynamics. While the tensor network formalism~\cite{penrose1971applications,deutsch1989quantum,schollwock2011density,white2004real,paeckel2019time,evenbly2011tensor,zaletel2020isometric,orus2014practical,vidal2009entanglement,PhysRevLett.101.110501,verstraete2008matrix,PhysRevLett.93.040502,PhysRevLett.93.207204,PhysRevB.91.165112} has been developed to address this issue, it is possible that the limitation of expressiveness in linear functions may be fundamental~\cite{gao2017efficient,pascanu2013number,montufar2014number,eldan2016power,khrulkov2017expressive,raghu2017expressive,PhysRevB.97.085104,li2021boltzmann}. As for the loss function, the spectrum error or average expectation error of a chosen state is a natural choice. However, compared to optimizing the error of the target property directly, the spectrum error or average expectation error of a chosen state may introduce a bias when these are not the target property themselves~\cite{Becca_Sorella_2017}. 
Given these arguments, it is natural to ask whether one could generalize the NRG and DMRG algorithms, 
%Naturally, we ask if one can generalize NRG and DMRG 
such that instead of using a linear operator map for an intermediate target, % (i.e.~the spectrum or arbitrary observable error), 
an arbitrary operator map can be used as an ansatz for the final target.  %, such as observable expectation, entanglement entropy, and phase transition point.

%This proposition is affirmative by reviewing these algorithms from a machine learning perspective. 
We view this question through a modern machine learning perspective.
For both ground state and dynamics, predicting properties of a $n$-site system from smaller system properties can be framed as a machine-learning task. An algorithm can learn from a data set of $n-1,\cdots,1$-site properties and predict the $n$-site property. Looking closely at NRG and DMRG, the algorithms take a set of $n$-site operators as input and generate a compact representation of these operators as output. Then, the compact operators are grown by one site and used as the operators for an $n+1$-site system as shown in \cref{fig:nrg_and_dmrg}. From a learning perspective, aside from the details of the loss function for a specific problem, this can be seen as a generative learning procedure, where the model tries to generate a set of virtual compact operators relevant to calculating the target property. Thus, we take as the key idea of our algorithm the perspective of learning operator maps, i.e., generalizations of the linear operator maps in NRG and DMRG, as opposed to learning parameterized states as in the tensor network formalism.
%instead of viewing as a state in the tensor network formalism, the linear ansatzes in NRG and DMRG, as defined traditionally, are ansatzes for operator maps. This is the key idea of our generalization.

We call the algorithm thus introduced in this paper 
%In this paper, we introduce an alternative variational principle, 
the Operator Learning Renormalization Group (OLRG). 
%This framework refreshes the understanding of NRG and DMRG from a machine-learning perspective (compared in~\cref{fig:nrg_and_dmrg}). 
In a similar spirit to the renormalization group and embedding theory~\cite{doi:10.1021/ct301044e,doi:10.1021/acs.jctc.6b00316,doi:10.1021/acs.accounts.6b00356}, the OLRG framework provides a route to leverage the techniques developed for small-system solvers for large-scale many-body systems. While our framework is general enough to address arbitrary simulation problems, we prove in particular rigorous bounds for the loss function of real-time dynamics. The loss function is designed to minimize the error of a target property directly instead of an intermediate target. The philosophy of removing intermediate steps is commensurate with end-to-end (e2e) learning~\cite{muller2005off,end2end}, which has been a core concept in the success of modern deep learning, leading to many state-of-the-art results~\cite{collobert2011natural,krizhevsky2012imagenet,mnih2015human,silver2017mastering}. As a result, instead of modeling a quantum state, arbitrary operator maps are allowed as ansatzes by this variational principle. For classical simulation, the operator map is called an Operator Matrix Map (OMM). In this paper, we will focus on demonstrating OMM implemented by a neural network.
Furthermore, this variational principle has a broader application when considering operators not represented by matrices, such as a pulse sequence in a real quantum device. Considering an operator map of a problem Hamiltonian to a device Hamiltonian expression, this leads to an alternative quantum simulation algorithm %that bridges the product formula and variational quantum algorithm (VQA) 
for near-term devices that are not fully fault-tolerant~\cite{bluvstein2023logical,doi:10.1126/science.abo6587,daley2022practical,arute2019quantum}, which we call the Hamiltonian Expression Map (HEM).

This paper is organized as follows. We first review the NRG and DMRG algorithms in their traditional setting in \cref{sec:review}. In \cref{sec:framework}, we introduce the general framework of OLRG, including the {\it scaling consistency} condition, a general principle guiding the design of the loss function. Then, we explore the concrete scaling consistency condition for the real-time evolution of a geometrically local Hamiltonian. In \cref{sec:algorithm}, we introduce two variant algorithms of OLRG for classical and quantum simulation of real-time dynamics. By viewing the operator map as OMM, we discuss using OLRG as a variational algorithm on conventional computers. 
%More specifically, we focus on introducing OMM implemented by a neural network, which can be more expressive than matrix product state (MPS).
By viewing the operator map as HEM, we discuss using OLRG as a variational quantum algorithm for near-term digital-analog quantum devices. In \cref{sec:results}, we study the two-point correlation function of a one-dimensional ($1D$) Transverse Field Ising Model (TFIM) to demonstrate our theory and the effects of different hyperparameters for OMM and HEM. Finally, we discuss open questions and potential improvements in \cref{sec:discussion}.

\section{NRG and DMRG in the Traditional Formulation}\label{sec:review}

To further understand the motivation and thought process of the NRG and DMRG algorithms, we will review them in their traditional formulations from an operator map perspective. Wilson's NRG starts with a simple idea: to obtain the low-energy properties of a $N$-site system, where $N$ is a large number or infinity. We can start by dividing the $N$-site system into identical $n$-site small systems named a block, assuming $N = 2^q n$. Then, the block Hamiltonian $H_{S_n}$ on a small system $S_n$ of $n$ sites can be compressed from $2^n\times 2^n$ to some size $M\times M$ by finding an approximation of the matrix $H_{S_n}$. Wilson proposed to use a low-rank approximation of the Hamiltonian $\isometric{V_n}{H_{S_n}}$ such that $\isometric{V_n}{H_{S_n}}$ preserves the low-energy eigenstates of $H_{S_n}$. Naturally, $V_n$ are the $M$ lowest eigenstates of $H_{S_n}$, which preserves the low-energy spectrum. Then, we can grow the system by copying the $n$-site system to form a $2n$-site system and repeat the process. For single particle models such as $H_{S_n} = \sum_i X_i$, this is relatively straightforward. Since each small system of $n$ sites does not interact with each other, the $2n$-site Hamiltonian $H_{S_n} H_{S_n}$ can be written as $H_{S_n}\otimes I + I\otimes H_{S_n}$, and with the compressed Hamiltonian $\isometric{V_n}{H_{S_n}} \otimes I + I\otimes \isometric{V_n}{H_{S_n}}$. With $q$ steps, this process should eventually lead to a $n 2^q$-site system, approximating an infinite system. %This is the basic idea of NRG. 
In summary, NRG uses the error of the low-energy spectrum as the optimization target and a basis transform $V_n$ as the ansatz. Thus, at each step, we produce a virtual Hamiltonian $\isometric{V_n}{H_{S_n}}$ to replace the original one. However, such approximation is sub-optimal for two reasons: (i) the choice of low-energy eigenstates is suboptimal when the only properties of interest are the ground-state properties. (ii) copying the small system does not reflect the effect of boundary conditions. As a result, NRG works well for low-energy spectrum problems without a strong effect on boundary condition~\cite{wilson1975renormalization} but fails for more general quantum lattice ground-state problems in real-space form~\cite{white1992real}.

Historically, White's DMRG was presented as a generalization of NRG. We will explain the process using $1D$ TFIM Hamiltonian $H_n = \sum_{i} Z_i Z_{i+1} + h\sum_i X_i$. Assuming a chain of ``good'' compression $V_1,V_2,\cdots,V_{N-1}$ in a similar RG process has been found but for arbitrary ground state observables in the infinite system. Then, given a $n$-site system $S_n$ and environment $E_n$, $V_{n+1}$ should produce a good approximation of $S_n \bigcdot \bigcdot\, E_n$ named a superblock, where $\bigcdot$ means a new physical site, and its Hamiltonian is written as $H_{S_n}\otimes I^{n+2} + H_{S_n\bigcdot} + H_{\bigcdot\bigcdot} + H_{\bigcdot E_n} + I^{n+2}\otimes H_{E_n}$, and without compression
\begin{equation}
\begin{aligned}
    H_{S_n\bigcdot} &= I^{\otimes n-1}\otimes Z \otimes Z \otimes I^{\otimes n+1}\\
    H_{\bigcdot\bigcdot} &= I^{\otimes n}\otimes Z\otimes Z\otimes I^{\otimes n}\\
    H_{\bigcdot E_n} &= I^{\otimes n+1}\otimes Z\otimes Z\otimes I^{\otimes n-1}\\
\end{aligned}
\end{equation}
The construction of $S_n \bigcdot \bigcdot\, E_n$ requires addressing the effect of a neighboring site at the boundary $S_n \bigcdot$ then addressing the effect of environment bath $\bigcdot\, E_n$. Thus, a superblock can be a good test of the boundary and environment effect. Then, applying $V_{n+1}$ on $S_n \bigcdot$ and $\bigcdot\, E_n$ will result in a virtual $S_n \bigcdot \bigcdot\, E_n$ system. Comparing an arbitrary ground state observable $A$ on $S_n\bigcdot$ results in the following error estimation,
\begin{equation}
	\norm{\tr(\rho_{S_n \bigcdot} A) - \tr(\isometric{V_{n+1}}{\rho_{S_n \bigcdot} A})}
\end{equation}
where $\rho = \tr_{\bigcdot\, E_n}(\ket{\psi_0})$ is the ground state on $S_n\bigcdot$. Since $A$ is an arbitrary observable, the optimal $V_{n+1}$ should be the isometric map to the low-rank approximation of $\rho$~\cite{white1992density,schollwock2005density}, namely a basis transform into the ground state subspace. To build the Hamiltonian of the next $2n+4$-site superblock, except the virtual Hamiltonian from the $n+1$-site system and environment, we also need the virtual operator $H_{S_{n+1}\bigcdot}$, $H_{\bigcdot\,\bigcdot}$ and $H_{\bigcdot\, E_{n+1}}$, which are $I^{\otimes n} \otimes Z$, $Z\otimes I^{\otimes n}$ and $I^{\otimes n+1}$ in the $n+1$-site space. Then, one can repeat this process until the target system size is $N$. In summary, in DMRG, the superblock is used instead of a block to test the effect of boundary and bath. Besides the Hamiltonian itself, we keep track of some extra virtual operators to build the superblock. The transform $V_{n+1}$ is then optimized based on the loss function defined on the superblock. A comparison of loss functions is shown in \cref{tab:loss_functions}. This thought process of generating virtual operators describing the same $n$-site system is the key idea of our generalization.
\begin{table}[t]
	\begin{ruledtabular}
		\begin{tabular}{ccc}
			Method                                                & Loss function                              & RG transformation   \\
			\hline
			NRG~\cite{wilson1975renormalization}                  & low-energy spectrum error      & isometry    \\
			DMRG~\cite{white1992density,White1993DensitymatrixAF} & $\norm{\rho - \hat{\rho}}_{F}\quad \rank{\hat{\rho}}\leq M$ & isometry    \\
			OLRG                                                  & scaling consistency                          & arbitrary \\
		\end{tabular}
	\end{ruledtabular}
        \caption{\label{tab:loss_functions}%
		A review of previous RG-like variational methods by loss function at each scale and RG transformation. $H$ denotes the Hamiltonian. $\rho$ denotes the density matrix. $\norm{\cdot}_{F}$ is the trace norm (Frobenius norm). $M$ denotes the maximum rank of the low-rank approximation.}
\end{table}

\section{Operator Learning RG Framework}\label{sec:framework}

The procedure in \cref{sec:review} can be summarised as follows:
\begin{center}
	\textit{Instead of considering all $n$-site operators and having a static definition of operators in the block object, we only focus on the subset of operators required to calculate the target output.}
\end{center}
We call these operators as the set of ``relevant'' operators and denote them as a set $S_n$ for a $n$-site system. And denote $S_n^{(0)}$ as the ground truth without altering any relevant operators. In NRG, this is only the Hamiltonian $S_n = \{H_n\}$, and in our $1D$ TFIM DMRG example, this is $S_n = \{H_n, I^{\otimes n-1}\otimes Z, Z\otimes I^{\otimes n-1}\}$. We then look at the target output and rough RG procedure to trace back the minimum required operations by removing the details from \cref{sec:review}.

\begin{figure*}
	\includegraphics[scale=0.65]{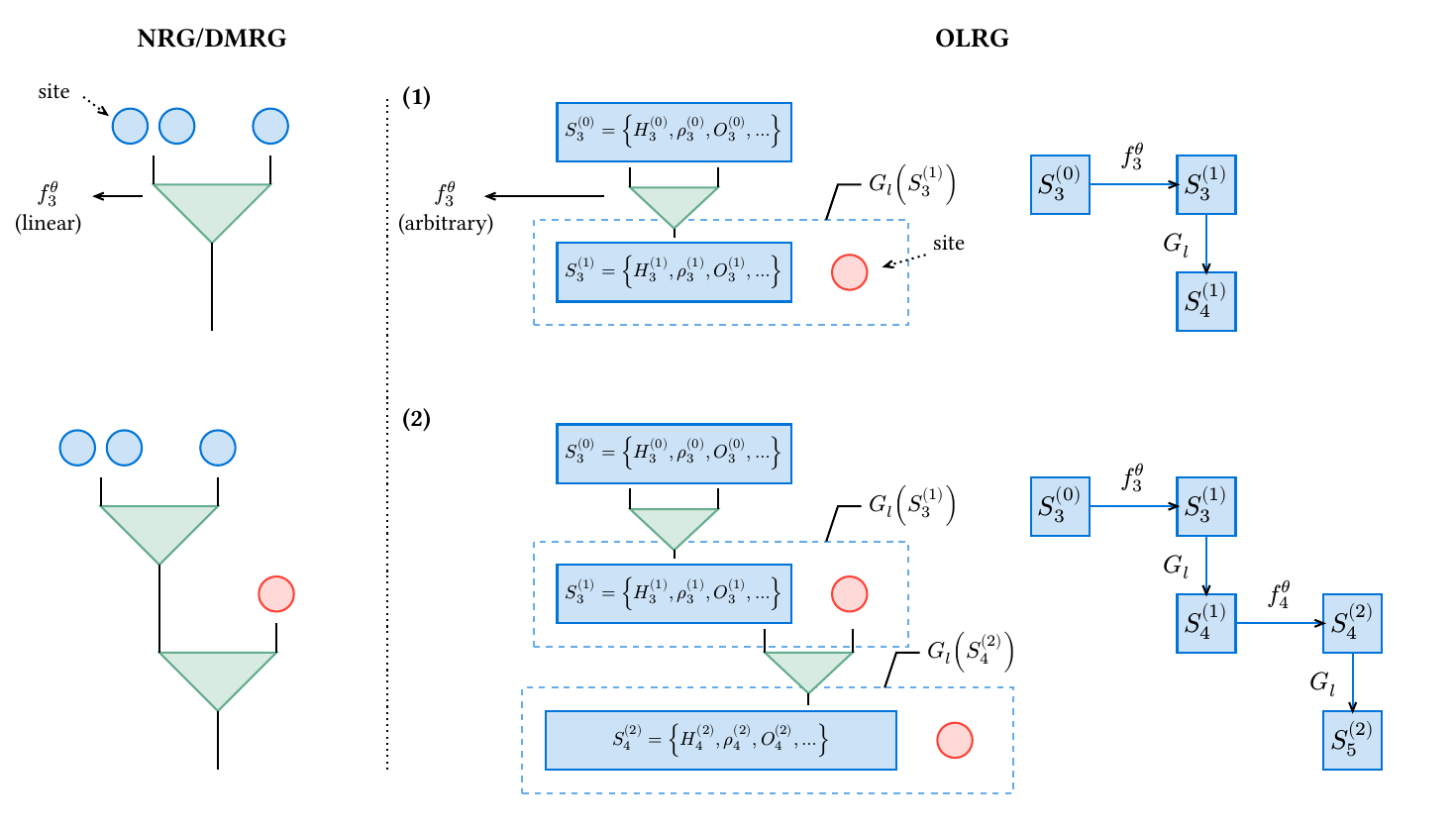}
	\caption{Illustration of three OLRG growing steps starts from a 3-site system. $G_{\step}$ denotes the operation of adding $k$ sites into the system. $\ansatz{n}$ denotes the operator map of $n$-site system with parameters $\theta$.
	(\textit{Left}) When $\ansatz{n}$ is an isometric matrix, this process is equivalent to a canonical MPS. The red circles mark the physical legs, and the blue triangle denotes the isometric matrix.
	(\textit{Middle}) The blue box depicts the set of operators that are used to calculate the target property. The dashed box denotes the grown box. The arrow represents the operator map $\ansatz{n}$. (\textit{Right}) The flow chart of this process. $S_5^{(2)} = \opmap{G_1}{S_4^{(2)}} = \opmap{(G_1\circ \ansatz{4})}{S_4^{(1)}} = \opmap{(G_1\circ \ansatz{4} \circ G_1 \circ \ansatz{3})}{S_3^{(0)}}$.
	}\label{fig:generic-workflow}
\end{figure*}

First, we denote the target output at $n$-site system calculated by these operators as $\opmap{p_n}{S_n}$, where $p_n$ is called a property function. Here by ``property function'', we mean a function that maps the set $S_n$ to a scalar value, such as the ground state energy, the two-point correlation function, the entanglement entropy, etc. A formal definition of $p_n$ is introduced in \cref{appx:scale-consistency}. Denote the operator map as $\ansatz{n}: \hermitian{n}\rightarrow\hermitian{n}$, where $\hermitian{n}$ is the space of Hermitian operators and $\theta$ are the parameters. The operator map $\ansatz{n}$ maps a Hermitian operator to another Hermitian operator of the same number of sites. For example, when $p_n$ is the expectation of a two-point correlator at time $T$ evolved by the TFIM Hamiltonian,
\begin{equation}
	S_n^{(0)} = \{\rho_0, H_n, \boundary_n, O_n^{ab}\}
\end{equation}
where $\boundary_n = I^{\otimes n-1} Z$, applying $\ansatz{n}$ onto $S_n^{(0)}$ result in
\begin{equation}
	S_n^{(1)} = \opmap{\ansatz{n}}{S_n^{(0)}} = \{\opmap{\ansatz{n}}{\rho_0}, \opmap{\ansatz{n}}{H_n}, \opmap{\ansatz{n}}{\boundary_n}, \opmap{\ansatz{n}}{O_n^{ab}}\}
\end{equation}

In NRG and DMRG, this is the basis transform $\opmap{\ansatz{n}}{X} = \isometric{V_n}{X}$. Constructing a $n+1$-site operator from a $n$-site operator is denoted as $G_1$. In DMRG, this corresponds to the step that adds one site $\bigcdot$. Then, starting from an operator $X$ in $n$-site system, we can write down the output operator in the $N$-site system after performing the entire RG process as $\opmap{\underbrace{G_1\circ \ansatz{N-1} \circ \cdots \circ G_1 \circ \ansatz{n}}_{q\text{ times }}}{X}$. This is the corresponding virtual operator in the $N$-site system, where $N=n + q$. We can then obtain the entire set of virtual relevant operators in the $N$-site system, denoted as $S_N^{(q)} = \opmap{\underbrace{G_1\circ \ansatz{N-1} \circ \cdots \circ G_1 \circ \ansatz{n}}_{q\text{ times}}}{S_n^{(0)}}$, where $(q)$ means we transformed the system for $q$ times by applying $\ansatz{n},\cdots,\ansatz{N-1}$. Thus the error of output is $\norm{p_N(S_N^{(0)}) - p_N(S_N^{(q)})}$, where $S_N^{(0)} = \opmap{G_1^q}{S_n^{(0)}}$ by definition. If we can minimize this error, we will obtain a set of virtual operators $S_N^{(q)}$ resulting a similar property value. In summary, it doesn't matter if our set of virtual operators $S_N^{(q)}$ is a complete set of real operators describing the $N$-site system properties. As long as it can compute the property $p_N$ with a good error, it is probably a set of real operators. 

Thus, instead of approximating states, the NRG and DMRG algorithms can be viewed as generative learning algorithms~\cite{goodfellow2020generative} that generate a set of virtual relevant operators at each scale. As shown in \cref{fig:generic-workflow} (right), \textbf{(1)} starting from $S_3^{(0)} = \{H_3^{(0)}, \rho_3^{(0)}, O_3^{(0)}, \cdots\}$, we generate $S_3^{(1)} = \opmap{\ansatz{3}}{S_3^{(0)}} = \{H_3^{(1)}, \rho_3^{(1)}, O_3^{(1)}, \cdots\}$. Assuming this new set of operators is sufficient to approximate the properties we would like to calculate, we use this set of operators as if they were the ground truth. If they are ``good'' approximations, we should be allowed to grow the virtual system by 1 site (marked by a dashed box). We can obtain the next 4-site system as $S_4^{(1)} = \opmap{G_1}{S_3^{(1)}}$. \textbf{(2)} We then use $S_4^{(1)}$ as the input to generate another set of operators $S_4^{(2)} = \opmap{\ansatz{4}}{S_4^{(1)}}$ and grow it into $S_5^{(2)}$ to calculate $\opmap{p_5}{S_5^{(2)}}$ as an approximation of $\opmap{p_5}{S_5^{(0)}}$. 

In the case of classical simulation, the virtual relevant operators should use less storage than the original to keep the algorithm running within a constant memory. Thus in NRG and DMRG, they are generated by linear isometric functions $\ansatz{n_q}(X) = \isometric{V_{n_q}}{X}, n_q = n,n+1,\cdots,N$. These functions take the matrix of the original operator and return a compressed matrix at each scale. As shown in~\cref{fig:generic-workflow} (left), the chain of $f_{n_q}$ forms the canonical MPS. Next, the functions $\ansatz{n_q}$ are optimized by a loss function that is defined on the data of block (NRG) or superblock (DMRG) generated by a small-system solver, e.g., the low-energy spectrum error or average expectation error of a chosen state. The loss function heuristically controls the final error. Thanks to the linear nature of $\ansatz{n_q}$, the optimal point of such loss functions can be identified directly using an eigensolver without actually generating the whole data set of operators. This loss function heuristically allows the set $S_n^{(1)} = \opmap{\ansatz{n}}{S_n^{(0)}}$ to grow into the set $S_{n+1}^{(2)} = \opmap{G_1}{S_n^{(1)}}$ with the final error controlled. Thus, the learning process can be repeated until the target system size is reached.

This perspective further guides us to investigate the requirement of implementing proper loss functions, such that the requirement of a linear $\ansatz{n}$ can be extended. As a result, we suggest a fundamental principle for creating such loss functions, which we call the \textit{scaling consistency condition}. This principle is outlined and compared with other heuristic approaches in \cref{tab:loss_functions}. In the following, we define this process and its underlying concepts through formal definitions and corresponding examples. Then, we introduce the error upper bound due to satisfying the scaling consistency condition. Next, we look into the real-time evolution of a geometrically local Hamiltonian and further reduce the loss function to local-observable errors. Last, we discuss these local observables and the corresponding evaluation of the loss function. In fact, this paradigm above shares the same philosophy as so-called \texttt{duck typing} in programming languages~\cite{roger2024duck,wiki:Duck_typing} (e.g as used in a DMRG tutorial~\cite{simple-dmrg}). By way of definition,

\begin{center}
\textit{If it looks like a duck, swims like a duck, and quacks like a duck, then it probably is a duck.}
\end{center}

\subsection{The Scaling Consistency Condition}\label{sec:scale-consistency}

Next, we explain how to define a tractable loss function such that for arbitrary $\ansatz{n}$ we can (a) preserve the properties we want to calculate in the target system and (b) allow the system to grow to the target size with final error controlled. This is the main goal of the following definitions and theorems. We first need to formally define $G_{\step}$ to understand what it means to grow the system by $\step$ sites. Before diving into the formal definition, we can look at how one rewrites the $n$-site Hamiltonian $H_{n}$ of the $1D$ TFIM as the $n-1$-site Hamiltonian

\begin{example}[Growing the TFIM Hamiltonian]
For example, for the $1D$ TFIM model, the Hamiltonian of an $n$-site $1D$ system is constructed by extending the Hamiltonian of an $n-1$-site $1D$ system and adding terms that incorporate the $n$-th site:
\begin{equation}\label{ex:tfim}
	\begin{aligned}
		H_n
		 & = \opmap{G_1}{H_{n-1}}                                                                      \\
		 & = H_{n-1}\otimes I + \underbrace{I^{\otimes n-1}\otimes Z \otimes Z}_{\text{new site interaction}} + \underbrace{h I^{\otimes n} \otimes X}_{\text{new site field}},
	\end{aligned}
\end{equation}
and more generally we can rewrite the $n+\step$-site $1D$ TFIM Hamiltonian as $n$-site $1D$ TFIM Hamiltonian as follows
\begin{equation}
    \begin{aligned}
        H_{n+\step} = & H_{n}\otimes I^{\otimes \step} +
        I^{\otimes n-1}\otimes Z\otimes Z \otimes I^{\otimes \step-1} \underbrace{+\cdots +}_{\text{$\step-2$ terms}}\\
        &I^{\otimes n+\step-2}\otimes Z\otimes Z + I^{\otimes n}\otimes h\cdot X \otimes I^{\otimes \step-1} +\\
        &I^{\otimes n}\otimes h\cdot X \otimes I^{\otimes \step-1} \underbrace{+\cdots +}_{\text{$\step-2$ terms}} I^{\otimes n+\step-1}\otimes h\cdot X
    \end{aligned}
\end{equation}

\end{example}

We can see the \cref{ex:tfim} as breaking the entire system into 1-site fragment, then each time  $G_{\step}$ is applied, it puts $\step$ fragments back. Naturally, one can define the growing operator as a building operation that puts $\step$ fragments back after dividing the total $N$-site operator. This is the main idea of the following definition.

\begin{definition}[Growing operator, informal]
	A growing operator $G_{\step}$ is a superoperator that increases the size of the system by $\step$-sites. This superoperator formalizes how one grows a given operator $A_n$ of $n$ sites by $\step$ sites. In general, $G_{\step}$ can be represented as follows,
	\begin{equation}
		\opmap{G_{\step}}{A_n} = A_n\otimes \conn{A_n} + \sum_i \boundary_n^i \otimes \conn{\boundary_n^i},
	\end{equation}
	where $\boundary_n^i$ and $\conn{\boundary_n^i}$ are pairs of operators that connect the $n$-site system and the $\step$-site environment. We call the operators $\boundary_n^i$ the boundary operators. The index $i$ goes over all possible decomposition and thus can be exponentially large in the most general case. A more detailed definition will be given in \cref{appx:scale-consistency}.
\end{definition}

In summary, the concept of a growing operator is pivotal in understanding how an operator of a $n+\step$-site system can be expressed in terms of an operator of a $n$-site system by first dividing the total system of $N$ sites into $N/\step$ fragments. This will be particularly clear to those familiar with tensor networks: the growing operator can be analogously represented as a tensor within the Tensor Network Operator (TNO) formalism~\cite{Pirvu_2010,PhysRevB.95.035129}. Each time applying the tensor creates a few new physical legs. However, in our subsequent theorem, we opt not to use the TNO formalism. Our rationale is to present our proof from an algebraic standpoint, which we find more suitable for our generalization purposes. To further elucidate this concept, the growing operator can also be applied to other operators, such as the density matrix operator of the zero-state.

\begin{example}[Growing the zero state]
$\rho_n = (\ket{0}\bra{0})^{\otimes n}$ can be written as,
\begin{equation}
	\rho_n = \opmap{G_1}{\rho_{n-1}} = \rho_{n-1}\otimes \ket{0}\bra{0} .
\end{equation}

\end{example}

As mentioned, the growing operator is defined by dividing the total system into fragments and combining them. The two-point correlation function is a typical example of an operator that requires dividing the total system into fragments otherwise the definition of the two-point correlation function could be ambiguous.

\begin{example}[Growing the two-point correlator]
In a similar vein, and without loss of generality, consider a two-point correlator expressed as
\begin{equation}
	O_n^{xy} = I^{\otimes x} \otimes X \otimes I^{\otimes y} \otimes Y \otimes I^{\otimes n - x - y - 2}.
\end{equation}
The growing operator from a smaller size $n-1$ to a larger size $n$ can be written as
\begin{equation}
	O_{n}^{xy} = \opmap{G_1}{O_{n-1}^{xy}} = \begin{cases}
		O_{n-1}^{xy}\otimes I & \begin{aligned}[t]
			                         & \text{if } n < x     \\
			                         & \text{or } x < n < y \\
			                         & \text{or } n > y
		                        \end{aligned} \\
		O_{n-1}^{xy}\otimes X & \text{if } n = x        \\
		O_{n-1}^{xy}\otimes Y & \text{if } n = y        \\
	\end{cases}
\end{equation}

\end{example}

The e2e-style loss function can be written as the error $\norm{\opmap{p_N}{S_{N}^{(0)}} - \opmap{p_N}{S_N^{(q)}}}$, as demonstrated in \cref{fig:generic-workflow} (blue nodes). Then, our goal will be minimizing this loss function by optimizing the parameters $\theta$ in the OLRG steps. The parameters $\theta$ can appear in two places: (a) the operator map $\ansatz{n_q}$ itself, similar to NRG and DMRG; (b) the output operator $\opmap{\ansatz{n_q}}{X}$. We will discuss them in \cref{sec:algorithm}. However, this quantity, as the loss function, is infeasible to calculate. We wish to simplify it into a more tractable form within each growing step. Naively, as shown in \cref{fig:generic-workflow} (blue nodes), for calculating 5-site system starting from 3-site system, one may use $\norm{\opmap{p_{3}}{S_{3}^{(0)}} - \opmap{p_{3}}{S_{3}^{(1)}}} + \norm{\opmap{p_{4}}{S_{4}^{(1)}} - \opmap{p_{4}}{S_{4}^{(2)}}}$ as the loss function instead. However, this does not necessarily bound the final error. In fact, we show that such a loss function fails to bound the error in \cref{sec:results} as the 0$^{\rm th}$-order loss function in our theory. This motivates us to introduce the following definition and theorem.

\begin{definition}[$\epsilon$-scaling consistency]\label{def:epsilon-scaling-consistency}
	An operator map $\ansatz{n}: \mathcal{A}_n \rightarrow \mathcal{A}_n$ is said to satisfy $\epsilon$-scaling consistency for a set of relevant operators $S_n$ and property $p_N$ where $N\geq n$, if $\exists \epsilon > 0, \forall q = 1,2,\cdots,(N - n)/\step$ we always have
	\begin{equation}
		\norm{p_N[\opmap{G_{\step}^q}{S_n}] - p_N[\opmap{(G_{\step}^{q}\circ \ansatz{n})}{S_n}]} \leq \epsilon.
	\end{equation}
\end{definition}

The $\epsilon$-scaling consistency condition measures the error caused by applying the operator map $\ansatz{n_q}$ at each OLRG step. This error appears because the $n$-site part within a $N$-site system is transformed by $\ansatz{n}$, thus resulting in an error between two $N$-site systems ($\opmap{G_l^q}{S_n}$ and $\opmap{(G_l^{q}\circ \ansatz{n})}{S_n}$). It is worth noting that the conceptualization of growing operators shares numerous commonalities with Density Matrix Embedding Theory (DMET)~\cite{doi:10.1021/ct301044e,doi:10.1021/acs.jctc.6b00316}. For readers versed in DMET, the terminology of ``scaling consistency'' also takes inspiration from the self-consistency principle in DMET. In OLRG, the optimization of the operator map is directed not toward aligning the properties of individual fragments with the original system but rather toward achieving consistency in properties across varying scales. Denote one step of growing and transforming as $D_{\step} = G_{\step}\circ \ansatz{n_q}$ called an OLRG step. For convenience, we let $n_q$ being adaptive to the number of sites in $D_{\step}$. The target property can be written as $p_N(\opmap{D_{\step}^q}{S_{n}})$. Then we have the following theorem:

\begin{theorem}[System scaling error]\label{thm:sc-error}
	For target system size $N$ and starting system size $n$, where $N\geq n$, if for $n_q=n,n+l,\cdots,N$, the operator map $\ansatz{n_q}$ satisfy the $\epsilon$-scaling consistency condition for $S_n,S_{n+1},\cdots,S_{N-1}$, and $q = (N - n)/\step$ then $\exists\epsilon > 0$ such that
	\begin{equation}
		\norm{p_N[\opmap{G_{\step}^q}{S_n}] - p_N[\opmap{D_{\step}^q}{S_n}]} \leq q\epsilon.
	\end{equation}
\end{theorem}

\begin{figure}[t]
	\includegraphics[scale=0.7]{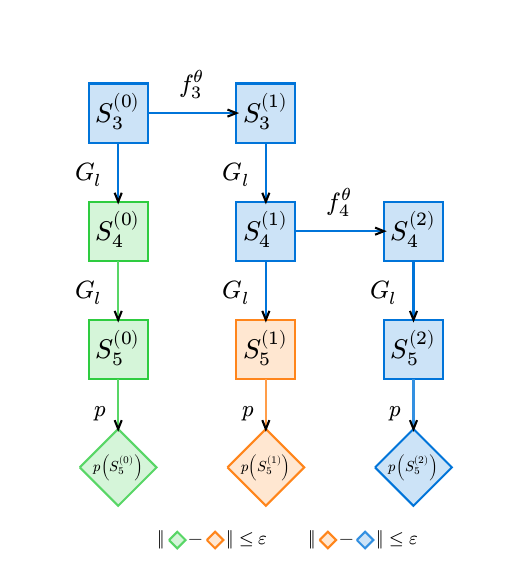}
	\caption{Comparing 2 OLRG growing steps and the ground truth starts from a 2-site system.
	Denote $S_n^{(q)}$ as the system of size $n$ by applying $D_l$ for $q$ times.
	Green nodes depict the ground truth. Blue nodes represent algorithm growing steps. Orange nodes represent the 5-site system applying only $\ansatz{3}$. The red nodes denote the computed property at each 5-site system $p(\opmap{G_{\step}}{S_n^{(q)}})$.}\label{fig:rg-workflow}
\end{figure}

\begin{proof}
	This is obvious by inserting zeros of neighboring values $p(\opmap{G_{\step}^{q-j} D_{\step}^j}{S_n}) - p(\opmap{G_{\step}^{q-j} D_{\step}^j}{S_n})$ into the left-hand-side then use triangular inequality. A visual proof is shown in \cref{fig:rg-workflow}. \cref{appx:scale-consistency} provides a more detailed proof.
\end{proof}

\cref{thm:sc-error} suggests that to implement an e2e-style loss function rather than minimizing the differences in properties at the current system size, one should optimize the discrepancy between properties at the target system size $N$ at each OLRG step. This concept is exemplified by the error observed between the last blocks of each column in \cref{fig:rg-workflow}. Instead of optimizing properties from the blue blocks ($S_3^{(0)}$ and $S_3^{(1)}$, $S_4^{(1)}$ and $S_4^{(2)}$), one should optimize the properties from the blocks at the bottom ($S_5^{(0)}$ and $S_5^{(1)}$, $S_5^{(1)}$ and $S_5^{(2)}$). With \cref{thm:sc-error}, we convert the problem of reducing the error $\norm{p(\opmap{G_{\step}^q}{S_{n}}) - p(\opmap{D_{\step}^q}{S_{n}})}$ into reducing the error defined by $\epsilon$-scaling consistency (\cref{def:epsilon-scaling-consistency}). While the quantity in $\epsilon$-scaling consistency is still infeasible to evaluate, intuitively, such error is caused by applying $\ansatz{n}$ to the $n$-site system. Thus, the error must come from the change of some operators in the $n$-site system. To control the error, we only need to expand our set of relevant operators to include these operators. In the next subsection, while the rigorous $\epsilon$-scaling consistency condition for the ground state and imaginary time dynamics remains an open question, we show what kind of operators in the $n$-site system will contribute to this error for the real-time evolution of a geometrically local Hamiltonian.

\subsection{Loss Function for Real-Time Evolution}\label{sec:sc-rt-loss}

\begin{figure}[t]
	\includegraphics[scale=0.30]{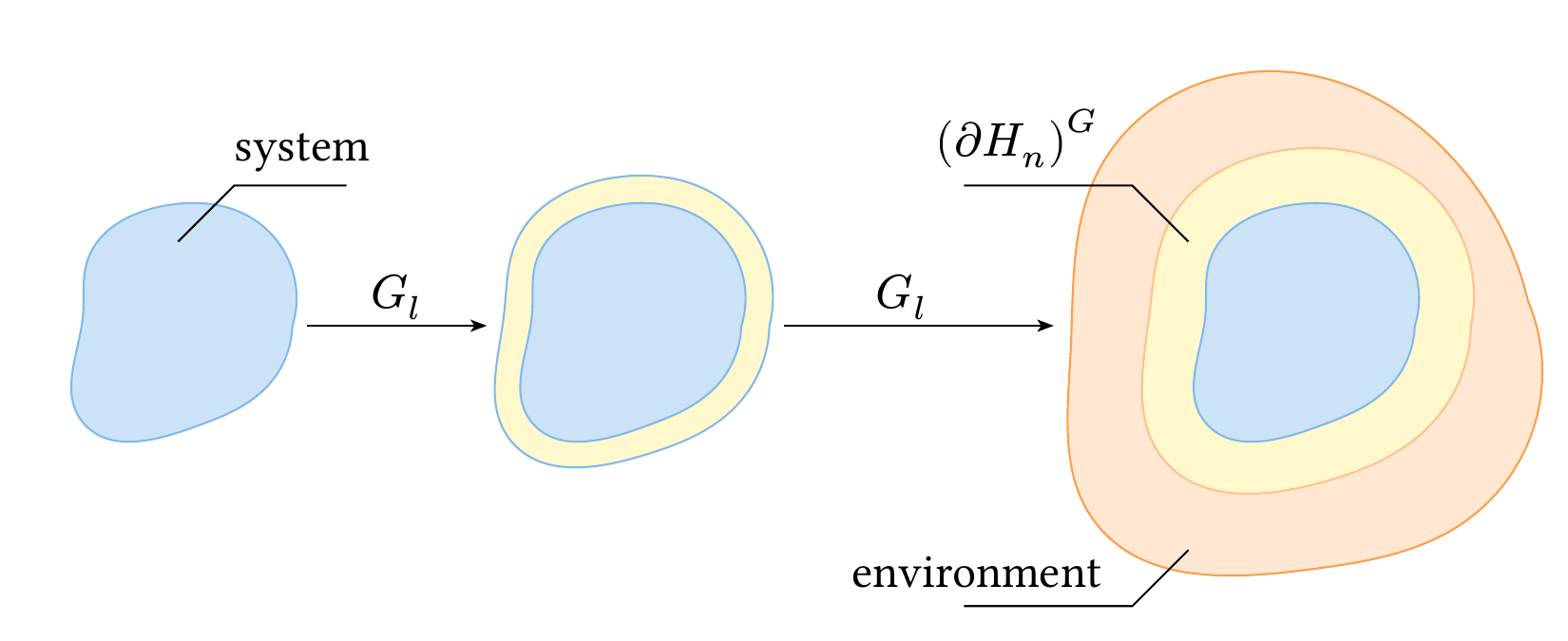}
	\caption{
		For a geometrically $\locality$-local Hamiltonian, the growing operator stops changing the system after it grows outside the boundary band of stretch $\locality$ after applying $G_{\step}^2 = G_{2\step}$. This results in a saturated yellow band where only terms within this yellow band interact with the system Hamiltonian.
	}
	\label{fig:grow-demo}
\end{figure}

Nevertheless, when we look closer to a more realistic system, it is usually geometrically local. More specifically, geometrically $\locality$-local means given a Hamiltonian of the form $H_n = \sum_a H_a$, each term $H_a$ can only act on neighboring $\locality$ sites geometrically. In this case, applying $G_{\step}$ for $q$ times will result in the following equation, where by definition, $G_{\step}^q = G_{q\step}$ and,
\begin{equation}\label{eq:G-kq}
	\begin{aligned}
		\opmap{G_{q\step}}{H_n}  &=\\
  &H_n\otimes I^{\otimes kq} + \sum_{i\in \sbset} \boundary_n^i \otimes R(\boundary_n^i) + I^{\otimes n}\otimes K,
	\end{aligned}
\end{equation}
where $\sbset$ denotes a set of operators acting on the boundary of $H_n$. The set $\sbset$ will saturate once the growing operator applies outside the system boundary as demonstrated in \cref{fig:grow-demo}. The size of $\sbset$ is proportional to the boundary size of the system $S_n$ and the number of operators $\boundary_n^i \otimes \conn{\boundary_n^i}$, as previously defined in the context of a growing operator $G_{\step}$ (depicted in the yellow band in \cref{fig:grow-demo}). Furthermore, $K$ represents the Hamiltonian of the environment. A more detailed and formal discussion about the set $\sbset$ is included in \cref{appx:grow-op}, where the geometrically local Hamiltonian is generalized into the geometrically local Hamiltonian with constant non-geometrically local terms, such as systems with periodic boundary condition. This approach simplifies the criterion for scaling consistency, necessitating consistency only within $\sbset$, since geometrically, the operator transformation $\ansatz{n}$ affects only operators within this range. This leads to the following proposition.

\begin{proposition}\label{prop:series-expansion}
If we can effectively break down the property function $p_N(\opmap{G_{l}^{q}}{S_n})$ and $p_N(\opmap{G_{l}^{q}}{\ansatz{n}(S_n)})$ into expectation values separately on system and environment, this might offer a more practical way to develop the e2e-style loss function:
\begin{equation}
	\begin{aligned}
		p_N(\opmap{G_{\step}^{q}}{S_n}) &= \sum_{i=0}^{\infty} \alpha_i \expect{A_i}\expect{B_i}\\
		p_N(\opmap{G_{\step}^{q}}{\ansatz{n}(S_n)}) &= \sum_{i=0}^{\infty} \alpha_i \expect{A_i^{\prime}}\expect{B_i}\\
	\end{aligned}
\end{equation}
where $i$ is the index of the series expansion, $\alpha_i$ represents the scalar factor at each order, $ \expect{A_i} $ represents observables on the $n$-site system and $ \expect{B_i} $ corresponds to observables on the $kq$-site environment. $\expect{A_i^{\prime}}$ is the transformed observable on the $n$-site system. For example, if $\expect{A_i} = tr(\rho_0 \exp{it H_n} A_i \exp{-it H_n})$ is the expectation of a time evolved observable, then $\expect{A_i^{\prime}} = tr(\opmap{\ansatz{n}}{\rho_0} \opte{U^{\prime}(t)}{\opmap{\ansatz{n}}{A_i}} )$ is the expectation value re-calculated using the virtual operators, where $U^{\prime}(t) = \exp{-it \opmap{\ansatz{n}}{H_n}}$. If we optimize our operator mapping function such that
$ \norm{\expect{A_i} - \expect{A_i^{\prime}}} \leq \epsilon $, it follows that
$ \norm{p(\opmap{G_{\step}^{q}}{S_n^{(0)}}) - p(\opmap{G_{\step}^{q}}{\ansatz{n}(S_n^{(0)})})} \leq \epsilon \sum_{i=0}^{\infty} \alpha_i \expect{B_i} $. Provided that $ \sum_{i=0}^{\infty} \alpha_i \expect{B_i} $ converges to a finite value, the convergence
of the e2e-style loss function can be ensured.
\end{proposition}

We further explore the expectation value of an observable $O_n(T)$ in the Heisenberg picture $O_n(T) = e^{i T H_n}O_n e^{-i T H_n}$, where $T$ is the total evolution time, $H_n$ is a geometrically local Hamiltonian, with a product state $\rho_n$ as initial state,
\begin{equation}
	p(S_{n}) = tr(\rho_0 e^{iH_n T}O_n e^{-iH_n T}).
\end{equation}
Without loss of generality, we assume $O_n$ is local and $G_{\step}(O_n) = O_n\otimes \conn{O_n}$. Because $\rho_n$ is a product state such that $G_k(\rho_n) = \rho_n\otimes\conn{\rho_n}$. This expectation $p(S_{n})$ can expand into a series of expectation values in the $n$-site system and the environment (detailed in \cref{appx:sc-rt-proof}). Thus, we find the desired series expansion proposed in \cref{prop:series-expansion}. This leads us to the subsequent theorem:

\begin{theorem}[Real-time $\epsilon$-scaling consistency]\label{thm:rt-scaling-consistency}
	Given that a $\locality$-local Hamiltonian $H_n$ and its growing operator $G_{\step}$ will saturate, denote the set as $\sbset$. $\exists \epsilon > 0$ and expectation values $\chi(S_n)$, such that if $\forall \chi$ we have
	\begin{equation}
		\norm{\chi(S_n) - \chi(\opmap{\ansatz{n}}{S_n})} \leq \epsilon.
	\end{equation}
        For $S_n = \{H_n,B_n^i,\rho = \rho_n\otimes\conn{\rho_n}, O = O_n\otimes \conn{O_n}\}$ and $N=n+kq$ then the error of expectation $\opmap{p_N}{\opmap{G_k^q}{S_n}} = \expect{\rho e^{iTH_{N}} O e^{-iTH_{N}}}$ is bounded by
	\begin{equation}
		\begin{aligned}
			&\norm{\opmap{p_N}{\opmap{G_{\step}^q}{S_n}} - \opmap{p_N}{\opmap{(G_{\step}^{q}\circ \ansatz{n})}{S_n}}}\\
			&\leq \epsilon C \exp{T \norm{(\partial H_n)^{G_{\step}}} C / 2},
		\end{aligned}
	\end{equation}
 where $C$ is a constant, $T$ is the total evolution time. A detailed theorem and its proof can be found in \cref{appx:sc-rt-proof}
\end{theorem}

\cref{thm:rt-scaling-consistency} gives a single step error, thus combined with \cref{thm:sc-error}, we have the total error of $q$ steps upper bounded by
\begin{equation}
	q\epsilon C \exp{T \norm{(\partial H_n)^{G_{\step}}} C / 2}.
\end{equation}
This indicates that if we can optimize the error of these expectations $\chi$ at $n_q = n,n+l,\cdots,N$-site system due to applying $\ansatz{n_q}$, we should be able to optimize the error of the target property at the target system size $N$. Since the norm $\norm{\sbset}$ is a constant, this error is independent of system size $N$ and only accumulates linearly with the number of OLRG steps.

Thus, we can tailor the loss function's design for real-time evolution by considering it as the cumulative error of all observables, as detailed in \cref{thm:rt-scaling-consistency} with an order cutoff in the series. Then \cref{thm:rt-scaling-consistency} can guarantee as we increase the order the output will directly move towards the ground truth. This aligns with the e2e learning. \cref{thm:rt-scaling-consistency} has a very similar bound as Lieb-Robinson bound~\cite{lieb1972finite} and other results derived from it~\cite{childs2019theory}. Intuitively, the reason why real-time dynamics can have this bound is also due to the limitation of propagating correlations. However, we do not use Lieb-Robinson bound in our proof in~\cref{appx:sc-rt-proof}. It is interesting to see if we can derive a similar bound using the Lieb-Robinson bound. This will provide a more general understanding of the error bound in our framework.

Next, based on the proof in~\cref{appx:sc-rt-proof}, we introduce the definition of $\chi$. Denote the operator $\boundary_n^i$ from \cref{eq:G-kq} in the Heisenberg picture as $\boundary_n^i(t) = e^{iH_n t} \boundary_n^i e^{-iH_n t}$ where $0\leq t \leq T$. The proof of this theorem reveals that the observables are essentially time correlation functions defined on the operator $\boundary_n^i(t)$ and the part of our target observable on the system $O_n(T)$. We refer to these as the Time-Ordered Boundary Correlation (TOBC) denoted as $\chi$:
\begin{equation}
	\expect{\chi_{\bm{i,t,\sigma}}(S_n, T)} = tr(\rho_n \prod_{\bm{i,t,\sigma}} \lie_{\boundary_n^i(t),\sigma}[O_n(T)]),
\end{equation}
where $\rho_n$ is the initial state of the system, the multi-index $\bm{i,t,\sigma} = i_1,i_2,\cdots,i_k, t_1,t_2,\cdots, t_k,\sigma_1,\sigma_2,\cdots,\sigma_k$, each index in $\mathbf{i}$ iterates over $(\partial H)^G$, $t$ are the checkpoints in the time evolution, and $\sigma = \pm 1$. As mentioned, the input $S_n$ denotes the set of relevant operators at $n$-site system. For TOBC specifically, $S_n$ are the primitive operators required to calculate TOBCs defined as $S_n = \{\rho, O_n, H_n, \boundary_n^i\}$ where $\boundary_n^i\in\sbset$. The notation $\lie_{A,\sigma}(B) = AB + \sigma BA$ and $\lie_{A,+1}(B) = \{A, B\} = AB + BA, \lie_{A,-1}(B)=[A, B]=AB - BA$, their composition denotes the recursive commutators and anti-commutators $\lie_{A,+1}\lie_{B,+1}(C) = \{A, \{B, C\}\}$, $\lie_{A,-1}\lie_{B,+1}(C) = [A, \{B, C\}]$. For the $k$th-order TOBC, the notion of $\prod_{\bm{i,m,\sigma}}$ denotes the following product
\begin{equation}
	\lie_{\boundary_n^{i_1}(t_1),\sigma_1}\lie_{\boundary_n^{i_2}(t_2),\sigma_2}\cdots \lie_{\boundary_n^{i_k}(t_k),\sigma_k}[O_n(T)].
\end{equation}
For example, we can write down the TOBC at different orders. For the 0-th order, this refers to the observable $O_n(T)$. For the 1st order, for $0\leq t\leq T$, we have,
\begin{equation}
    \begin{aligned}
        \chi_{i,t,-1}(S_n, T) &= [\boundary_n^i(t), O_n(T)]\\
        \chi_{i,t,+1}(S_n, T) &= \{\boundary_n^i(t), O_n(T)\}.
    \end{aligned}
\end{equation}
For the 2nd order, for $0\leq t_1 \leq t_2\leq T$, we have,
\begin{equation}
    \begin{aligned}
        \chi_{\mathbf{i},\mathbf{t},\{-1,-1\}}(S_n, T) &= [\boundary_n^{i_1}(t_1), [\boundary_n^{i_2}(t_2), O_n(T)]]\\
        \chi_{\mathbf{i},\mathbf{t},\{-1,+1\}}(S_n, T) &= [\boundary_n^{i_1}(t_1), \{\boundary_n^{i_2}(t_2), O_n(T)\}]\\
        \chi_{\mathbf{i},\mathbf{t},\{+1,+1\}}(S_n, T) &= \{\boundary_n^{i_1}(t_1), \{\boundary_n^{i_2}(t_2), O_n(T)\}\}\\
        \chi_{\mathbf{i},\mathbf{t},\{+1,-1\}}(S_n, T) &= \{\boundary_n^{i_1}(t_1), [\boundary_n^{i_2}(t_2), O_n(T)]\} . \\
    \end{aligned}
\end{equation}

\begin{figure}[t]
	\includegraphics[scale=0.4]{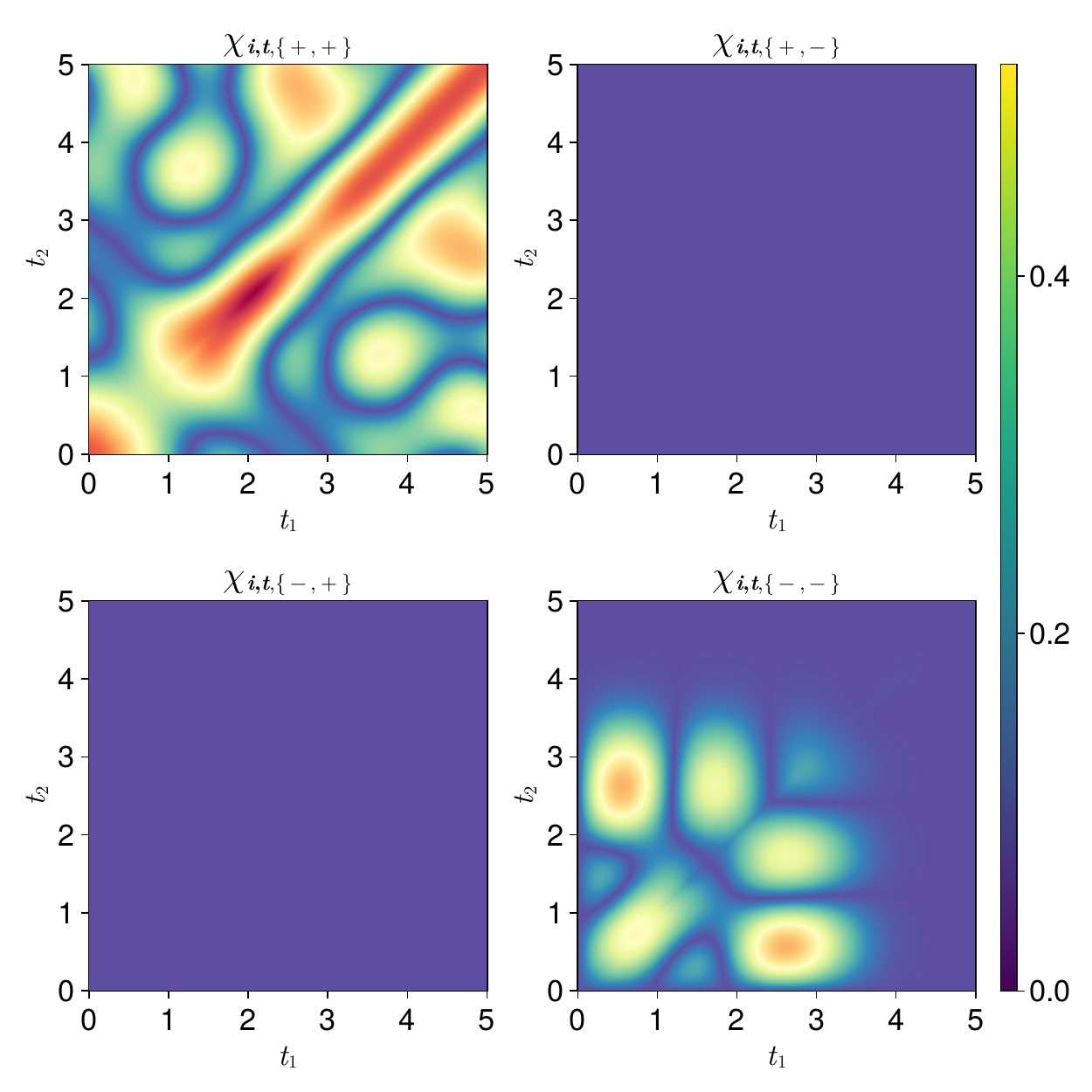}
	\caption{
		2nd-order TOBC for 5-site $1D$ TFIM at $T=5.0$ for the two-point correlation function $\expect{Z_1 Z_2}_{T=5.0}$ with $\ket{00000}$ as the initial state and $h = 1.0$.
	}
	\label{fig:tobc-2}
\end{figure}

In practice, the time points $t_1,t_2,\cdots$ are checkpoints from the small-system solver. Many correlators in this setup are nearly zero. These can be further pinpointed by introducing a specific Hamiltonian and observables into the correlator expression. For example, for a 5-site $1D$ TFIM model, where $\boundary_5^1 = Z_5 = I^{\otimes 4} \otimes Z$, two of its 2nd order TOBC,
\begin{equation}
    \begin{aligned}
        \expect{[Z_5(t_1), \{Z_5(t_2), (Z_1 Z_2)(5.0)\}]} &\approx 0\\
        \expect{\{Z_5(t_1), [Z_5(t_2), (Z_1 Z_2)(5.0)]\}} &\approx 0,
    \end{aligned}
\end{equation}
are nearly zero, as shown in \cref{fig:tobc-2}. However, in a worst-case scenario, the number of potentially non-zero TOBC increases exponentially with the order $l$. Assuming there are $M$ checkpoints, there are $O(2 M \Vert (\partial H_n)^{G_{\step}} \Vert )^l$ TOBCs. This exponential rise renders a comprehensive evaluation of the entire loss function at higher orders impractical. A uniform sampling at each order without time ordering is proposed as an effective solution. This is because we optimize a summation of the correlators' error, thus resulting in a uniform distribution. When the original dynamics are faithfully approximated, the extra correlators not in time order should also have a small error. In practical terms, this approach involves selecting a batch of operators for sampling, akin to using data batches in Stochastic Gradient Descent (SGD) in conventional deep learning algorithms. Significantly, this sampling technique is highly compatible with advanced accelerated computing frameworks, such as CUDA~\cite{cuda}, which are designed for batch operations.

There might be concerns regarding the typically large value of $M$ and the consequent size of each operator batch, potentially leading to an extensive sampling requirement. As shown in \cref{fig:tobc-2}, for nonzero TOBC, in practice, many points are nearly zero and thus contribute little to the loss function. Thus, since $M$ is only a factor in the loss function rather than the small-system solver, if the dynamics of interest are smooth, a fine step size is not necessary in practice to obtain satisfactory results. In \cref{appx:hyper-step-size}, we include some additional results on the step size of the TOBC sampling, which does not show a significant difference in the final relative error.

\section{OLRG Algorithms}\label{sec:algorithm}

We have successfully relaxed the linearity constraint on the functions $\ansatz{n_q}, n_q=n,n+l\cdots,N-l$ for a specific observable $O_N$ at $N$-site system. Since $\ansatz{n_q}$ can now be an arbitrary operator map, it can act as the map between operator matrices, as well as the map between operator expressions. This allows parameterized $\ansatz{n_q}$ or parameterized output $\opmap{\ansatz{n_q}}{X}$ when the output is an expression. This leads to the classical and quantum algorithms we introduce in this section. 
% A more technical discussion about parameterizing the operator map $\ansatz{n_q}$ and the output $\opmap{\ansatz{n_q}}{X}$ is included in~\cref{appx:transform-hamiltonian}.

One can then search for the optimal parameters for $\ansatz{n_q}$ or the output $\opmap{\ansatz{n_q}}{X}$. For special $\ansatz{n_q}$ or $\opmap{\ansatz{n_q}}{X}$, like the linear map in NRG and DMRG, the optimal point can be identified directly. In the general case, we employ gradient-based optimization~\cite{DBLP:journals/corr/KingmaB14} to search for optimal parameters $\theta$. The gradient can be obtained using modern differentiable programming frameworks~\cite{jax2018github,NEURIPS2020_9332c513,10.1145/3458817.3476165,innes2018don,paszke2019pytorch,abadi2016tensorflow,innes2018flux,Luo2020yaojlextensible,bergholm2018pennylane} and their automatic differentiation algorithms~\cite{griewank2008evaluating,chen2018neural,giles2008extended,seeger2017auto,xie2020automatic} that work not only on classical computers but also on quantum devices. This leads to the following general variational algorithm (\cref{alg:general}) that starts with a small system and iteratively enlarges the system until it reaches the target system size (the blue blocks in \cref{fig:rg-workflow}).

\begin{algorithm}[H]
	\caption{OLRG Algorithm}\label{alg:general}
	\begin{algorithmic}
		\Require target size $N$, initial system $S_n$, small classical or quantum solver, batch size $b$, growing size $\step$, MAX ITERATION
		\State $i,q \gets 0$
		\State $\ansatz{n} \gets \theta$
		\While{$i < \text{MAX ITERATION}$}
			\State $\mathcal{L} \gets 0$
			\While{$n+q \step \leq N$}
				\State $S_{n+q \step}^{(q+1)} \gets \ansatz{n}(S_{n+q \step}^{(q)})$ \Comment{obtain virtual operators}
				\State Sample $\bm{i,t,\sigma}$ for TOBCs $\chi_{\bm{i,t,\sigma}}$ on $S_{n+q \step}^{(q)}$ and $S_{n+q \step}^{(q+1)}$
				\State Evaluate $\chi_{\bm{i,t,\sigma}}(S_{n+q \step}^{(q)}, T)$ and $\chi_{\bm{i,t,\sigma}}(S_{n+q \step}^{(q+1)}, T)$
				\State $\mathcal{L}_q \gets \frac{1}{b} \sum_{\bm{i,t,\sigma}} \norm{\chi_{\bm{i,t,\sigma}}(S_{n+q \step}^{(q)}, T) - \chi_{\bm{i,t,\sigma}}(S_{n+q \step}^{(q+1)}, T)}$
				\State $\mathcal{L} \gets \mathcal{L} + \mathcal{L}_q$
				\State $S_{n+(q+1) \step}^{(q+1)} \gets G_{\step}(S_{n+q \step}^{(q+1)})$ \Comment{grow}
				\State $q \gets q + 1$
			\EndWhile
			\State Optimize $\theta$ to minimize $\mathcal{L}$ by calling an optimizer
		\EndWhile
	\end{algorithmic}
	% \begin{enumerate}
	% 	\item \textbf{Input}: a small system of $n$ sites with the set of relevant operators $S_n$, and a classical or quantum solver.
	% 	\item Applying the operator map $\ansatz{n}$ to the set of relevant operators $S_n$ to obtain the set of virtual operators $\ansatz{n}(S_n)$.
	% 	\item Sampling a batch of the index $\bm{i,t,\sigma}$ for the TOBCs $\chi_{\bm{i,t,\sigma}}$.
	% 	\item Evaluate the value $\chi_{\bm{i,t,\sigma}}(S_n, T)$ by calling the solver.
	% 	\item Evaluate the value $\chi_{\bm{i,t,\sigma}}(\ansatz{n}(S_n), T)$ by calling the solver.
	% 	\item Evaluate the average error of the sampled TOBCs $\mathcal{L}_n = \sum_{\bm{i,t,\sigma}} \norm{\chi_{\bm{i,t,\sigma}}(S_n, T) - \chi_{\bm{i,t,\sigma}}(\ansatz{n}(S_n), T)}$.
	% 	\item Enlarge the set of virtual operators and use it as the $n+\step$-site relevant operators: $S_{n+\step} = G_{\step}(\ansatz{n}(S_n))$.
	% 	\item Reiterate from step 1, accumulating the loss function $\mathcal{L} \leftarrow \mathcal{L} + \mathcal{L}_n$ until reach target system size $N$.
	% 	\item Optimize the loss function concerning the parameters of $\ansatz{n}$. Compute the update $\Delta$ of the parameters $\theta$ by calling an optimizer, e.g ADAM~\cite{DBLP:journals/corr/KingmaB14}. Applying the update to parameters $\theta\leftarrow \theta + \Delta$.
	% 	\item Repeat the above steps until the loss function converges.
	% \end{enumerate}
\end{algorithm}

The OLRG algorithm is a general variational algorithm that can be applied to both classical and quantum systems. Before introducing more details about the operator map $\ansatz{n}$ for the classical and quantum cases, to illustrate the algorithm further, we will go through a concrete example of the algorithm in the context of calculating the real-time evolution of the two-point correlation function $\expect{Z_1 Z_2}_T$ in a $1D$ TFIM model. Starting from a 2-site system, the relevant operators are $S_2 = \{H_2, \boundary_2 = IZ, \rho_2 = \ket{00}\bra{00}, O_2 = ZZ\}$. Applying our operator map $\ansatz{2}$ we have $\opmap{\ansatz{2}}{S_2} = \{\opmap{\ansatz{2}}{H_2}, \opmap{\ansatz{2}}{IZ}, \opmap{\ansatz{2}}{\ket{00}\bra{00}}, \opmap{\ansatz{2}}{ZZ}\}$, then we can sample a batch of indices $\bm{i,t,\sigma}$ with batch size $\batch$, evaluate the TOBCs by solving the Heisenberg equation of $IZ,ZZ$ for the 2-site system. One must save the checkpoints at $\bm{t}$ for the boundary operator $IZ$. This allows us to calculate the loss function
\begin{equation}\label{eq:general-loss-2}
	\mathcal{L}_2 = \frac{1}{\batch} \sum_{\bm{i,t,\sigma}} \norm{\chi_{\bm{i,t,\sigma}}(S_2, T) - \chi_{\bm{i,t,\sigma}}(\ansatz{2}(S_2), T)}.
\end{equation}
This batch of sampled TOBCs is usually referred as the mini-batch in deep learning. The size of this batch is a crucial hyperparameter controlling the variance of the gradient and thus impacts the optimizer's behavior. When $b = 1$, the algorithm is called Stochastic Gradient Descent (SGD), and when $b$ is all the TOBCs, the algorithm is plain gradient descent. Then, assuming we are taking the simplest growing strategy that grows the system by 1 site at each step, we can grow the relevant operators using $G_1$, resulting in a new set of relevant operators $S_3 = \{H_3, \boundary_3, \rho_3, O_3\}$ defined as,
\begin{equation}
	\begin{aligned}
		H_3 &= \opmap{G_1}{\opmap{\ansatz{2}}{H_2}}\\
				&= \opmap{\ansatz{2}}{H_2}\otimes I + \opmap{\ansatz{2}}{\boundary_2}\otimes Z + I\otimes (h\cdot X)\\
		\boundary_3 &= I\otimes Z\\
		\rho_3 &= \opmap{\ansatz{2}}{\ket{00}\bra{00}}\otimes \ket{0}\bra{0}\\
		O_3 &= \opmap{\ansatz{2}}{O_2}\otimes I .
	\end{aligned}
\end{equation}
Here, we assume $\opmap{\ansatz{2}}{I} = I$, thus applying $I$ without calculating $\opmap{\ansatz{2}}{I}$ in $\boundary_3$. $I$ automatically adjusts to the size of the corresponding system, e.g. $\boundary_3 = I\otimes Z$; $I$ should share the same size as $\opmap{\ansatz{2}}{ZZ}$, and in $O_3$, $I$ should share the same size as $\ket{0}\bra{0}$. Then we can repeat the previous steps to obtain $\mathcal{L}_3$, and $S_4$ until we get $\mathcal{L}_{10}$ and $S_{10}$. We then calculate the total loss as $\mathcal{L} = \mathcal{L}_2 + \cdots + \mathcal{L}_{10}$. Finally, we can differentiate the loss function $\mathcal{L}$ with respect to the parameters $\theta$ and update the parameters $\theta$ using a gradient-based optimizer. This is called \textit{one epoch} of the algorithm. We can repeat the above steps until the loss function converges.

However, this training process directly optimizes towards a target observable at time $T$. If one is interested in the time points $0 \leq t_1\leq t_2\leq \cdots \leq T$, a transfer learning~\cite{pan2009survey} strategy can be employed. We can optimize the parameters $\theta$ at the first time point $t_1$ resulting in the optimized parameters $\theta_{t_1}$. Then, we use $\theta_{t_1}$ as the initial point for optimizing $t_2$, and so on. This method is similar to the strategy employed in other variational algorithms, such as MPS time-dependent variational principle (TDVP) and time-dependent VMC~\cite{kramer2008review,broeckhove1988equivalence,Yuan2019theoryofvariational,carleo2017solving}. However, in our setup, the parameters $\theta$ do not always represent an explicit state. In other variational algorithms, the states are passed through to the next time point explicitly by evolving in the parameter space. In our setup, the states are implicitly passed through as the parameters $\theta$. In special cases, an explicit state can be constructed from $\ansatz{n_q}$, which results in a similar algorithm as MPS TDVP. This therefore leads to potential improvements of the MPS TDVP algorithm to address long-time correlations. The detailed relation between this transfer learning strategy and the MPS TDVP algorithm is discussed in \cref{appx:mps-tdvp}.

Based on how one defines $\ansatz{n_q}$ and the representation of operators, the general algorithm can find different applications. In the following, we will introduce two specific algorithms for the classical and quantum case. For the classical case, we will use $\ansatz{n_q}$ as a parameterized Operator Matrix Map (OMM). For the quantum case, we will use $\ansatz{n_q}$ as a map from the problem Hamiltonian expression to the device Hamiltonian expression with parameters, namely Hamiltonian Expression Map (HEM).

\begin{figure}[b]
	\includegraphics[scale=0.85]{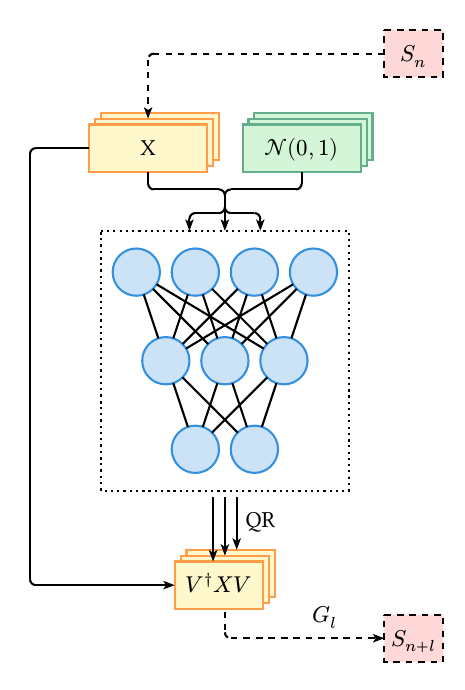}
	\caption{Illustration of neural OMM. $X$ is a batch of input relevant operators, \textit{QR} is the QR decomposition, $\isometric{V}{X}$ is a batch of output relevant operators by applying the batch of isometric matrices onto $X$. $G_l$ is the growing operator, $S_n$ is the input set of relevant operators and $S_{n+l}$ is the output set of relevant operators}\label{fig:nn}
\end{figure}

\subsection{Classical Algorithm: Operator Matrix Map}\label{sec:classical-algo}

The simulation challenge is more pronounced in real-time dynamics on conventional computers than in ground state, where no efficient classical algorithm is known for simulating the general real-time evolution of a quantum system. This further motivates the development of previous variational frameworks for real-time dynamics by employing a variational ansatz to model the system's state and subsequently evolving this ansatz over time by optimizing the variational parameters through the time-dependent variational principle (TDVP)~\cite{kramer2008review,broeckhove1988equivalence,Yuan2019theoryofvariational}. Born from the development of DMRG, the matrix product state (MPS) TDVP is the most successful strategy for solving $1D$ many-body physics. A holy grail of the field is to find an algorithm that performs as well in $D>1$, with contenders like tensor network states (TNS)~\cite{penrose1971applications,deutsch1989quantum,schollwock2011density,white2004real,paeckel2019time,evenbly2011tensor,zaletel2020isometric,orus2014practical,vidal2009entanglement,PhysRevLett.101.110501,verstraete2008matrix,PhysRevLett.93.040502,PhysRevLett.93.207204,PhysRevB.91.165112} and neural network states (NNS)~\cite{carleo2012localization,carleo2014light,carleo2017solving,hermann2020deep,torlai2018neural,PhysRevB.97.035116,carrasquilla2021use,hibat2020recurrent,nys2023realtime,PhysRevLett.128.090501,hartmann2019neural,choo2019two,carleo2018constructing,choo2020fermionic,sinibaldi2023unbiasing} continually making progress.

Our approach mirrors the workflows of NRG and DMRG but relaxes the linearity constraint on the operator map $\ansatz{n_q}$, allowing for a more expressive operator map. We also propose a loss function that directly minimizes the error of the target property, thus allowing us to use the same workflow for real-time dynamics. This is achieved by optimizing the error of the TOBCs, as detailed in \cref{sec:sc-rt-loss}.

In the same spirit as DMRG, we design $\ansatz{n_q}$ as a parameterized compression function $\opmap{\omm{n_q}}{X}$ of the operator matrices $X$ and parameters $\theta$. However, if $\opmap{\omm{n_q}}{X}$ is a dense linear function with a fixed size, the operator map is equivalent to an MPS, thus providing no advantage in expressiveness over MPS. From a physics perspective, approximating a larger pure system using a smaller pure system is not always possible. On the other hand, from the expressiveness perspective, the sum of matrix product states requires exponentially more parameters to represent the same state, despite that in certain cases when the tensors contain more structure, one can further compress the MPS via singular value decomposition~\cite{paeckel2019time}. Both suggest that letting $\opmap{\omm{n_q}}{X}$ be an ensemble of linear maps, which creates an ensemble of small pure systems, will be more expressive. This shares the idea of using an ensemble of MPS generated by a recurrent neural network to represent the wave function in VMC~\cite{chen2023autoregressive}. In summary, instead of generating a single set of relevant operators from input $S_n$, we will generate an ensemble of relevant operators sampled by a probability based on the input $S_n$. Starting from $z$ copies of the pure system, we can sample a single set of relevant operators from each copy and then forward them to the next step. This allows us to sample a chain of ensemble systems while growing the system size. For example, in our previous $1D$ TFIM example, we can start with $10$ copies of the 2-site system $S_2$, then applying $\omm{2}$ to each copy will sample a corresponding $\opmap{\omm{2}}{S_2}$. This results in $10$ sets of relevant operators $\opmap{\omm{2}}{S_2}$ based on a probability distribution defined by $\omm{2}$. The loss function $\mathcal{L}_2$ is instead evaluated as the average the sampled index batch $b$ of these 10 systems
\begin{equation}
\begin{aligned}
    \mathcal{L}_2 &=\\
 &\frac{1}{10 \batch} \sum_{S_2} \sum_{\bm{i,t,\sigma}} \norm{\chi_{\bm{i,t,\sigma}}(S_2, T) - \chi_{\bm{i,t,\sigma}}(\opmap{\omm{2}}{S_2}, T)}.
\end{aligned}
\end{equation}
Then, we can apply $G_1$ to the $10$ sets of relevant operators to obtain the ensemble of 3 sites. Other steps stay the same as the general algorithm. 

Treating $\ansatz{n_q}$ as a compression function from an input operator matrix to an output operator matrix aligns well with the idea of generative models in deep learning, where the model generates a set of outputs from a given input and a noise, which models a conditional probability distribution. For readers familiar with computer vision, this problem is similar to an image compression, generation, or manipulation problem, where we generate a new image based on an input image. Under this context, the linear map in NRG and DMRG can be seen as a similar method of principal component analysis (PCA) for image compression~\cite{pearson1901liii}. More modern image generation in deep learning utilize more powerful generative models including Generative Adversarial Networks (GANs)~\cite{goodfellow2020generative}, Variational AutoEncoders (VAEs)~\cite{kingma2013autoencoding}, normalizing flows~\cite{tabak2013family,george2021normalizing} and diffusion models~\cite{sohl2015deep,NEURIPS2020_4c5bcfec}.

In our demonstration, For simplicity, we use the same neural network for each step of the OLRG, thus $\omm{n_q} = \omm{}$. This requires the compression function always reduce the size from $2^{n+l}\times 2^{n+l}$ to $2^n\times 2^n$ to match the input size for next OLRG step. As depicted in \cref{fig:nn}, we employ the simplest toy neural network architecture used in GAN~\cite{goodfellow2020generative} that takes the operator matrix $X$ and a noise vector $\bm{z}$ sampled from Gaussian distribution $\mathcal{N}(0, 1)$ as input and generates an isometric matrix $V$ as output. The isometric matrix $V$ then applies to the operator matrix $X$ to generate the transformed operator. This guarantees the function does not change $I$ and the trace of the operator. Thus, it may have better numerical stability. In the following, we denote this operator map as $\omm{}(X,\bm{z})$. The neural network part of the operator map is a feed-forward neural network (FFNN) using \textit{ReLu}~\cite{fukushima1969visual,10.1007/978-3-642-46466-9_18} activation, each layer $\text{layer}_i(\mathbf{x})$ defined as following
\begin{equation}
	\text{layer}_i(\mathbf{x}) = \text{ReLU}(\mathbf{W}_i\mathbf{x} + \mathbf{b}_i),
\end{equation}
where $\mathbf{W}_i$ and $\mathbf{b}_i$ are the weight matrix and bias vector of the $i$-th layer. The neural network's input is the operator matrix reshaped into a vector concatenated with the noise vector sampled from the Gaussian distribution. The neural network's output is reshaped into a square matrix and then performs QR decomposition to generate an isometric matrix $V$. Then $V$ is applied to the input operator as $\isometric{V}{X}$.

In evaluating the loss function, the exact solver for solving TOBCs is an Ordinary Differential Equation (ODE) solver. Thus, because the same $\omm{n_q}$ is shared as $\omm{}$ between OLRG steps $D_{\step}$, the automatic differentiation needs to go through an ODE solver. Practically, this differentiation is typically achieved using the adjoint method, as detailed in various sources~\cite{pontryagin1962mathematical,Hager_2000,10.1115/DETC2005-85597,chen2018neural,kidger2022neural}. If $\omm{n_q}$ is not shared then only trivial linear algebra rules are needed for automatic differentiation.

We opted for a product state as the initial state, primarily due to the clear and well-defined nature of the growing operator in this context. This decision was influenced by the straightforward representation of a $n+k$-site product state as a composition of smaller system product states. For instance, a $n+k$-site zero state can be written as the following composition of a smaller system and thus defines its growing operator:
\begin{equation}
	G_k(\underbrace{\ket{0\cdots 0}\bra{0\cdots 0}}_\text{$n$ sites}) = \underbrace{\ket{0\cdots 0}\bra{0\cdots 0}}_\text{$n$ sites} \otimes \underbrace{\ket{0\cdots 0}\bra{0\cdots 0}}_\text{$k$ sites}.
\end{equation}

It is unclear how to write a $n+\step$-site state for a non-trivial state as the composition of smaller system states. Intuitively, MPS might be suitable for constructing such a formalism.

\begin{figure}[t]
	\includegraphics[scale=0.72]{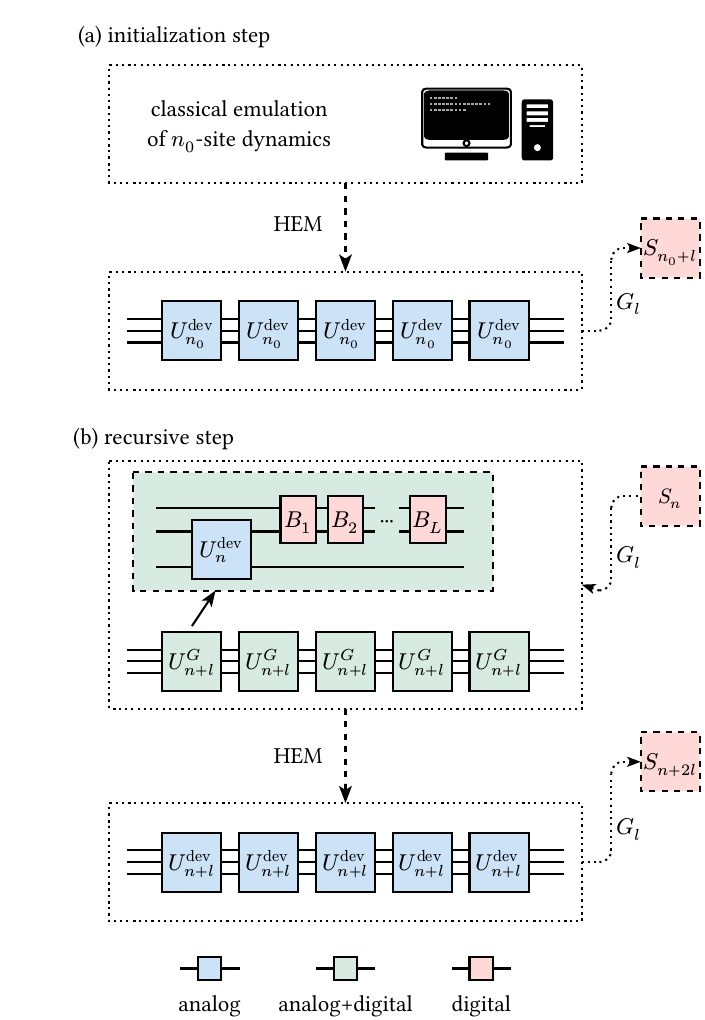}
	\caption{
		Illustration of HEM. (a) In the initialization step, HEM maps the emulation of $n_0$-site problem Hamiltonian dynamics into the $n_0$-site device Hamiltonian dynamics. Then the device dynamics is used to build grown system $S_{n_0 + l}$, forwarding to recursive steps; (b) In recursive steps, HEM maps dynamics $U_{n+l}^{G} = \exp[-it \opmap{G_{\step}}{H_{n}^{\text{dev}}}]$ to the device Hamiltonian dynamics $U_{n+l}^{\text{dev}} = \exp[H_{n+l}^{\text{dev}}]$. $B_i$ are the $w$-qubit digital gates, $L = \norm{(\partial H_n)^{G_l}}$ is the size of saturated boundary. $U_n^{\text{dev}}$ is the dynamics of $n$-site device Hamiltonian.}\label{fig:circuit}
\end{figure}

\subsection{Quantum Algorithm: Hamiltonian Expression Map}\label{sec:quantum-algo}

An alternative approach to simulate the real-time dynamics of quantum many-body systems is to use a quantum computer. The development of quantum simulation~\cite{feynman2018simulating,harrow2017quantum,boixo2018characterizing,neill2018blueprint} demonstrated that a quantum computer can efficiently simulate the real-time evolution of a quantum system. While algorithms based on Hamiltonian simulation~\cite{childs2009universal,gilyen2019quantum,doi:10.1137/18M1231511,blanes2017concise,suzuki1991general, PhysRevX.11.011020, haah2021quantum} have been proposed with rigorous bounds and polynomial complexity, the resource requirement~\cite{wecker2014gate} of these algorithms is beyond the capabilities of near-term intermediate-scale quantum (NISQ) computers~\cite{preskill2018quantum}. For example, the requirement of circuit depth and noise level are still beyond the capabilities of current devices~\cite{Bluvstein2022quantum,bluvstein2023logical,wurtz2023aquila,kim2023evidence,arute2019quantum}. As a result, heuristic algorithms such as variational quantum algorithms (VQA)~\cite{benedetti2019parameterized,ostaszewski2021structure,sim2019expressibility,haug2021capacity,haug2022natural,tilly2022variational,kandala2017hardware,liu2019variational,wang2019accelerated,farhi2014quantum,peruzzo2014variational,wang2015quantum,o2016scalable,shen2017quantum,mcclean2016theory,paesani2017experimental,colless2018computation,santagati2018witnessing,kandala2019error,hempel2018quantum,kokail2019self,li2017efficient,nakanishi2019subspace,PhysRevLett.125.010501} have been proposed for near-term devices. % with consideration of the resource constraints. 
Notably, digital, analog, and logical resources typically coexist in near-term devices. Evidence in digital-analog quantum algorithms (DAQA)~\cite{parra2020digital,PhysRevResearch.2.013012,lu2024digitalanalog} show the potential advantages of using the entire device capabilities. Yet, achieving practical quantum advantage remains an open problem for these heuristic algorithms. These challenges motivate us to search for an alternative framework that can inherit the advantages of the above frameworks and potentially lead to different perspectives on the simulation problem.

In our quantum algorithm, because one can utilize real quantum dynamics, the storage complexity is no longer a concern. Instead, the main objective is to translate the problem of Hamiltonian dynamics into the dynamics of the quantum device. This involves finding the appropriate control parameters of the device Hamiltonian that can closely replicate the dynamics of the problem Hamiltonian. Rather than viewing $\ansatz{n_q}$ as an Operator Matrix Map, $\ansatz{n_q} = \hem{n_q}$ now maps the input expression of a Hamiltonian into device Hamiltonian expression at each system size $n_q = n,n+l,\cdots,N$, leaving other operators untouched. The expressions are parameterized by control parameters in the device pulse sequence. The process begins with the relevant set of operators for $n$-site problem $S_n = \{H_n, \boundary_n^i, \rho_n, O_n\}$, applying $\hem{n}$, we have $\opmap{\hem{n}}{S_n} = \{\device{n}, \boundary_n^i, \rho_n, O_n\}$. $\device{n}$ is the device Hamiltonian with control parameters $\theta_n$ and an input time $t_n$. Thus the effect of $\hem{n}$ is swapping the operator expression from $H_n$ to $\device{n}$. Then we can sample a batch of indices $\bm{i,t,\sigma}$ with batch size $\batch$, evaluate the TOBCs by running a classical solver for the problem Hamiltonian and the device Hamiltonian to compute the first loss function $\mathcal{L}_n$. Next, applying the growing operator $G_{\step}$ on $\opmap{\hem{n}}{S_n}$ result in $S_{n+\step} = \{H_{n+\step}, \boundary_{n+\step}, \rho_{n+\step}, O_{n+\step}\}$, where $\boundary_{n+\step}, \rho_{n+\step}, O_{n+\step}$ stays the same as problem system, and $H_{n+\step}$ is defined by following,
\begin{equation}
	\begin{aligned}
		H_{n+\step} &= \opmap{G_{\step}}{\device{n}}\\
								&= \device{n}\otimes I + \sum_i \boundary_n^i\otimes R_{\step}(\boundary_n^i).\\
	\end{aligned}
\end{equation}

From the second step, we can evaluate the TOBCs for the dynamics described by $H_{n+\step}$ using the quantum device. If $\step \ll n$, then a large component of the dynamics is governed by the device Hamiltonian. We can then use a product formula, such as trotterization to simulate the dynamics of $H_{n+\step}$ using the quantum circuit depicted in \cref{fig:circuit}. Each trotter step results in the following unitary
\begin{equation}
	\begin{aligned}
		&\exp(-i\delta \opmap{G_{\step}}{\device{n}})\\
		&= [\exp(-i\delta \device{n}) \otimes I] \cdot \prod_i \exp(-i\delta \boundary_n^i\otimes R_{\step}(\boundary_n^i)).
	\end{aligned}
\end{equation}

If the problem system is $\locality$-local, then $\boundary_n^i\otimes \opmap{R_{\step}}{\boundary_n^i}$ only applies on $\locality$ qubits. Thus, the circuit only requires $\locality$-qubit high-quality gates at the boundary of the $n$-site system. Next, we can evaluate the TOBCs for the dynamics described by $\opmap{G_{\step}}{\device{n}}$ and $\device{n+\step}$ to obtain the next loss function $\mathcal{L}_{n+\step}$. Other steps stay the same as the general algorithm. In summary, the quantum device here plays the role of an exact solver, and $\hem{n_q}$ generates the parameterized device Hamiltonian expressions. The control parameters in the device Hamiltonian expressions are optimized to minimize the error of the target property. 

Like previous, we explain this algorithm by simulating the real-time dynamics of two-point correlation function $\expect{Z_1 Z_2}_T$ in $1D$ TFIM using a Rydberg atom device, as described in recent experimental demonstrations~\cite{Bluvstein2022quantum,wurtz2023aquila}. The 2-level Rydberg Hamiltonian is defined as follows
\begin{equation}
	\begin{aligned}
		& \rydberg{n}\\
		&= \sum_{\langle i,j \rangle} V_{ij}^{\theta_n} n_i n_j + \Omega^{\theta_n}(t_n) \sum_i X_i - \Delta^{\theta_n}(t_n) \sum_i n_i
	\end{aligned}
\end{equation}
where $V_{ij}^\theta$ denote the strength of interaction, and $\Omega^{\theta}(t)$ and $\Delta^{\theta}(t)$ are two time-dependent functions, commonly called pulse functions. Note that the 2-level Rydberg Hamiltonian is not universal and thus is restricted in the expressiveness of representing arbitrary dynamics. We begin with the set of relevant operators for the 2-site system $S_2 = \{H_2, \boundary_2 = IZ, \rho_2 = \ket{00}\bra{00}, O_2 = ZZ\}$, where $H_2$ is the 2-site TFIM Hamiltonian with $h = 1.0$. We use two FFNNs as the parameterized pulse function $\Omega^{\theta}(t)$ and $\Delta^{\theta}(t)$ and another FFNN representing the strength $V^{\theta}_{i,j}$ so that the effective duration of the device can be controlled. Thus our $\hem{n_q}$ now maps a given Hamiltonian to a $1D$ Rydberg Hamiltonian with the pulse functions $V^{\theta}_{i,j}$, $\Omega^{\theta}(t)$ and $\Delta^{\theta}(t)$. We first run a classical ODE solver to evaluate the TOBCs. This results in the same loss function in $\cref{eq:general-loss-2}$, where $\chi_{\bm{i,t,\sigma}}(\hem{2}(S_2), T)$ is the TOBCs for the 2-site parameterized Rydberg Hamiltonian. Next, we apply the growing operator $G_1$ to the 2-site parameterized Rydberg Hamiltonian to obtain the set of relevant operators for the 3-site system $S_3 = \{H_3, \boundary_3, \rho_3, O_3\}$, where,
\begin{equation}
	\begin{aligned}
		H_3 &= \opmap{G_1}{\rydberg{2}}\\
			&= \rydberg{2}\otimes I + \boundary_2\otimes Z + I\otimes (h\cdot X)\\
		\boundary_3 &= I^{\otimes 2}\otimes Z\\
		\rho_3 &= \ket{000}\bra{000}\\
		O_3 &= ZZ\otimes I.
	\end{aligned}
\end{equation}
Next, we apply $\hem{3}$ on $S_3$ result in
\begin{equation}
	\opmap{\hem{3}}{S_3} = \{\rydberg{3}, \boundary_3, \rho_3, O_3\}.
\end{equation}
The TOBCs for $\opmap{\hem{3}}{S_3}$ can be evaluated on a standard analog Rydberg atom device. We use the circuit in \cref{fig:circuit} to evaluate the TOBCs of $H_3 = \opmap{G_1}{\rydberg{2}}$. Each trotter step results in the following unitary
\begin{equation}
	\begin{aligned}
		\exp(-i\delta \opmap{G_1}{\rydberg{2}}) =
		 & [\exp(-i\delta \rydberg{2}) \otimes I] \cdot\\
		 & [I\otimes\exp(-i\delta Z\otimes Z)]\cdot \\
		 & [I^{\otimes 2}\otimes\exp(-i\delta h\cdot X)].
	\end{aligned}
\end{equation}
Denote $U_G(t) = \exp(-it \opmap{G_1}{\rydberg{2}})$. We can write down the 1st order TOBC $\expect{\chi_{\bm{i,t},\{-\}}}$ as an example of the full circuit
\begin{equation}
	\begin{aligned}
		&\expect{\chi_{\bm{i,t},\{-\}}}(S_3, T)\\
		&= \tr[\rho_3 U_G(t_2)^{\dagger} Z_3 U_G(T - t_2)^{\dagger} Z_1 Z_2 U_G(T)] -\\
		&\tr[\rho_3 U_G(T)^{\dagger} Z_1 Z_2 U_G(T - t_2) Z_3 U_G(t_2)].
	\end{aligned}
\end{equation}

After obtaining the TOBCs for $S_3$, we can calculate the loss function $\mathcal{L}_3$ and repeat the steps until we reach our target size. Like the general algorithm, we optimize the total loss function until convergence to search for the optimal pulse functions. The HEM-based OLRG is not limited to the product state because the operator map $\hem{n_q}$ does not alter the state operator. Thus, we do not need an explicit growing operator for the state operator.

Through HEM, OLRG allows us to leverage large analog and a few digital resources. Moreover, by adjusting the $\step$ in the growing operator, we can trade off the digital-analog resources. For example, if we grow the system by 1 site at each step, the algorithm is closer to a VQA, and if we grow the system by $1\ll\step$ sites at each step, the algorithm is closer to a product formula. Lastly, OLRG also bridges classical algorithms for simulating dynamics in the first step. Instead of competing with classical algorithms, HEM-based OLRG allows us to use the results from classical algorithms in $n$-site system as a starting point and then use the quantum device to grow into $N$-site system where $n < N$. Thus, improvements in the first-step classical simulation will improve HEM-based OLRG, allowing both communities to push the limits of quantum dynamics simulation together.

\subsection{Error and Resource Estimation}\label{sec:error}
Theoretically, the source of error in our framework originates from the estimation and optimization of the e2e-style loss function, as well as the expressiveness of operator maps. This aligns with e2e learning. However, it is important to note that the error bound presented in \cref{thm:rt-scaling-consistency} is not a tight bound. In practice, it intends to predict a large error. Combined with the truncation error arising from estimating the series expansion, the actual error estimation is often inaccurate in our current algorithm. The primary purpose of the theorem is to direct us toward defining a systematically improvable loss function rather than to provide an exact error estimation. A more precise error estimation requires finding a tighter bound in \cref{thm:rt-scaling-consistency}. We discuss intuitions and potential methods to improve this in \cref{appx:improving-loss-function}.

In the OMM-based OLRG, denote the time complexity of the evaluation of OMM as $W_{\omm{}}$ and the time complexity of the small-system solver as $Q$, for $L$ growing steps, the time complexity of evaluating the loss function is $O(L(W_{\omm{}}+2Q))$. The time complexity of evaluating the derivative of the loss function depends on whether $\omm{}$ is shared between different scales. We denote the time complexity of differentiating the operator map evaluation as $W_{\omm{}}^{\prime}$ and the adjoint method as $Q^{\prime}$. Heuristically, $Q^{\prime} = 2Q$~\cite{kidger2022neural}. Thus, if $\omm{}$ is shared, the time complexity of evaluating the derivative of the loss function is $O(L(W_{\omm{}}^{\prime}+2Q^{\prime}))\approx O(L(W_{\omm{}}^{\prime}+4Q))$. However, if $\omm{}$ is not shared, then there is no need to differentiate through the exact solver. The time complexity becomes $O(L W_{\omm{}}^{\prime})$. In terms of storage complexity, aside from the batch and sampling size, we denote the storage complexity of evaluating $\omm{}$ as $S_{\omm{}}$. If $\omm{}$ is shared, since the pure system ODE is reversible, the best algorithm solving the derivative has a constant overhead by using reversibility~\cite{griewank2008evaluating,kidger2022neural}. We denote this constant overhead as $C_{Q}$. The total storage complexity is only $O(S_{\omm{}} + C_Q L)$. However, if $\omm{}$ is not shared, the storage complexity becomes $O(L S_{\omm{}})$. Thus, in the OMM-based OLRG, one can trade storage for time complexity and vice versa by deciding how many $\omm{}$ are shared. For simplicity, we do not discuss the complexity of estimating the $k$-th order TOBCs in the OMM-based OLRG here because it only requires $O(k)$ times matrix multiplication in the small system. When considering the batch and sampling size, they create a constant factor over the time and storage complexity. It is worth noting that the overhead created by batch and sampling size can be easily reduced by parallelization and distributed storage due to their simplicity. This fits well into the modern processor architecture designed for single-program-multiple-data (SPMD)~\cite{DAREMA198811,cuda}.

In HEM-based OLRG, the classical computation components are relatively cheaper. Thus, we focus on discussing the cost of quantum operations. Because our algorithm involves analog circuits, we use the effective pulse duration (i.e., the scaling of the pulse duration to execute the circuit) as the measure instead of using circuit depth. We assume that the optimization only creates a constant prefactor in terms of the pulse duration for simulating $G_l(H_n)$ as $C_\theta\tau(\norm{(\partial H_n)^{G_l}})$. $C_\theta$ is the overhead caused by variational optimized pulse sequence. $\tau(\norm{(\partial H_n)^{G_l}})$ is the overhead caused by product formula, e.g. for 1st-order trotterization $\tau(\norm{(\partial H_n)^{G_l}}) = \norm{(\partial H_n)^{G_l}}$. And there are $M\gg 1$ checkpoints and total time $T$, for evaluating one $k$-th order TOBC using 1st-order trotterization, the average effective pulse duration is $O(C_{\theta} \norm{(\partial H_n)^{G_l}}k T)$ and the worst effective pulse duration is $O(2 C_{\theta} \norm{(\partial H_n)^{G_l}}k T)$ (see the detailed derivation in \cref{appx:complexity}). Due to our Hamiltonian has a large component of the dynamics governed by the device Hamiltonian, there is no dependencies of the total number of sites $N$ in the effective pulse duration. However, this does not mean we break existing gate depth bounds~\cite{haah2021quantum}. This complexity is moved into $C_{\theta}$. Further analysis is required to understand $C_{\theta}$ in our effective pulse duration after reaching the optimal point. As for the digital resources, our HEM circuit requires $O(\norm{\sbset} T)$ digital gates on $w$ qubits at the boundary depicted in~\cref{fig:circuit}.

Next, we discuss the complexity of shots. For each $k$-th order TOBC, there are $O(2^k)$ expectations to evaluate. Thus for batch size $b$ there are $O(b 2^k)$ expectations to evaluate. Assuming each expectation requires $E$ shots for $L$ growing steps, the total number of shots required is $O(bLE 2^k)$. Last, without loss of generality, we discuss the complexity of evaluating the gradients using the parameter shift rule or finite difference. For other ways of evaluating the gradients~\cite{banchi2021measuring,wierichs2021general}, one can derive in the same fashion. The complexity of evaluating the gradient of the loss function using finite difference depends on the number of parameters in device Hamiltonian, such as the $\Omega$ and $\Delta$ in our Rydberg Hamiltonian case. The rest of the parameters in the classical pulse function can be calculated via classical automatic differentiation. Thus, for $P$ parameters in device Hamiltonian, we require $O(bLEP 2^{k+1})$ shots in total.

%For a typical setup where $b = 50, E = 10^3, P = 2, k = 2, L = 10$, for the Rydberg atom device described in~\cite{doi:10.1126/science.abo6587} with shot rate assumed at $10^5$ shots per 15 hours of continuous operation (2 shots/sec), 500 epochs training still requires 68.5 years. Thus, while this algorithm offers the opportunity to utilize the full device and classical algorithms, it still requires real hardware to improve the shot rate to have a reasonable estimation of observables. However, we see this may be trivially improved by running more machines in parallel or being able to trap more atoms in the same array.

\section{Results}\label{sec:results}

To illustrate the convergence of OLRG as a variational principle and the associated classical and quantum algorithms, we applied our algorithm to the TFIM model as previously discussed. We investigated various hyperparameters to understand the algorithm's performance better. The implementation is available at the author's GitHub repository as an early-stage Python package~\cite{roger2024teal}. Additionally, for other hyperparameters and training dynamics, we discuss the training dynamics in \cref{appx:training-history} for different orders of loss functions. For other hyperparameters without much impact on our reported results, we include the extra results in tuning batch and sampling sizes in \cref{appx:hyper-batch-sampling-size} and the training using different step sizes of checkpoints in \cref{appx:hyper-step-size}.

\subsection{OMM}\label{sec:result-classical}

For our implementation of OMM, we initialize the system size at $n = 4$ and aim for a target system size of $N = 10$, setting the field parameter at critical point $h = 1.0$. The initial state is $\rho_0 = \ket{0000}$. The results of observable predictions are taken at the epoch with minimum moving average loss of window size 10. OMM is implemented by a neural network as discussed in \cref{sec:classical-algo}, referred to as neural OMM in the following. The loss function is optimized via the gradient obtained from the adjoint method via \texttt{jax.experimental.ode}. The simulation utilizes only a single GPU. Our results are obtained from various different GPUs, including P100, V100, and A100.

\begin{figure}[t]
	\includegraphics[scale=0.34]{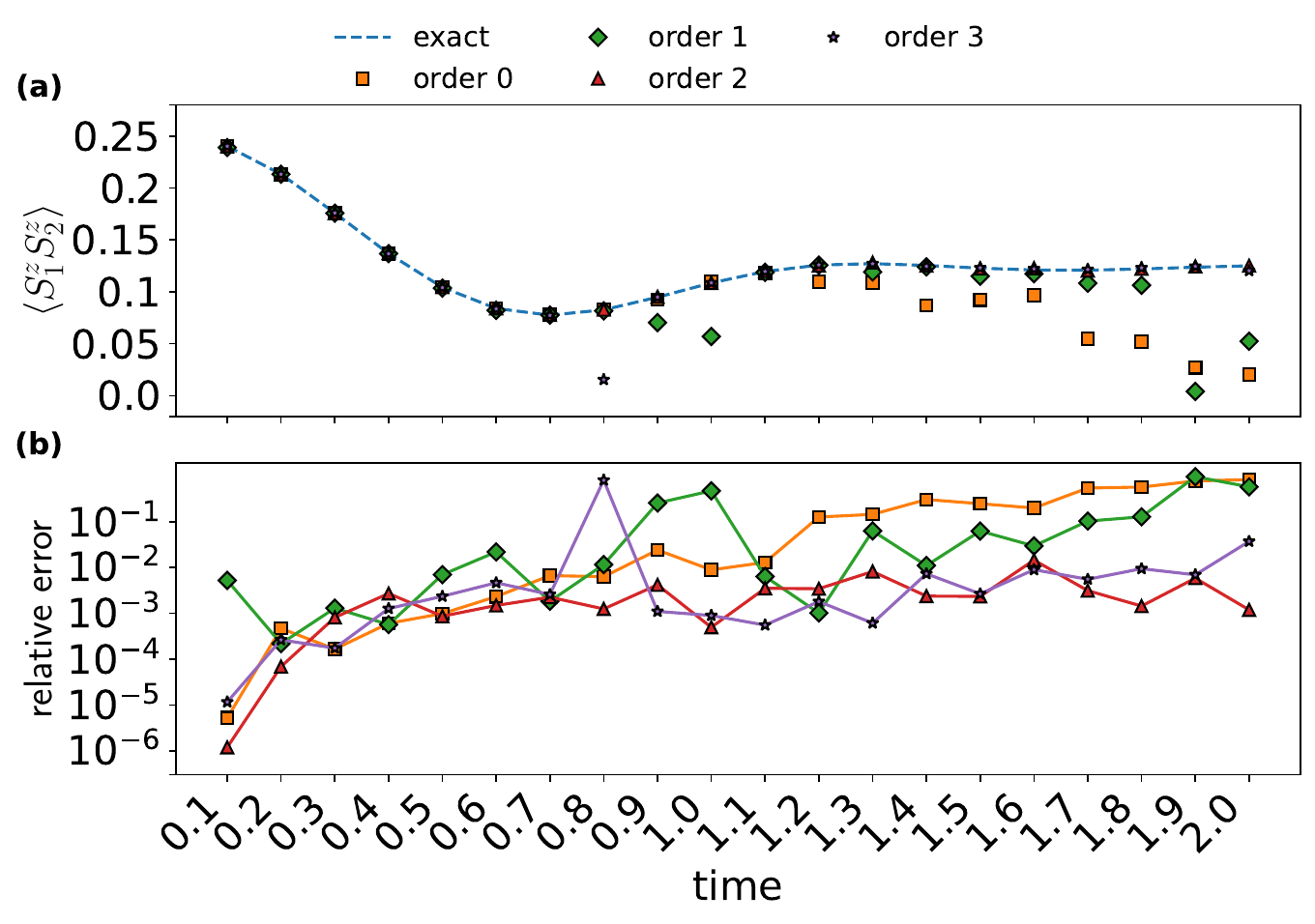}
	\caption{
		Comparison of OMM optimized at different loss function orders.
		(a) two-point correlation function $\langle S_1^z S_2^z\rangle$;
		(b) The relative error of the two-point correlation function $\langle S_1^z S_2^z\rangle$.
	}
	\label{fig:ensemble-order}
\end{figure}

We first evaluate the performance of loss functions at different TOBC orders. Theoretically, increasing the order should enhance the precision of the loss function in estimating discrepancies, thus resulting in better performance. To test this, we measure the relative error of the time-evolved two-point correlation function $\langle S_z^1 S_z^2 \rangle_t$ against the exact result. In our study, the depth of neural OMM is 8. We train the neural OMM with 6000 epochs at each time point, starting from randomly initialized parameters. As depicted in \cref{fig:ensemble-order}, we observed that at short-time intervals, the order of the loss function does not significantly impact the results. However, at longer times, the 0-order and 1-order loss functions failed to produce the correct results in the OMM-based OLRG.

\begin{figure}[t]
	\includegraphics[scale=0.34]{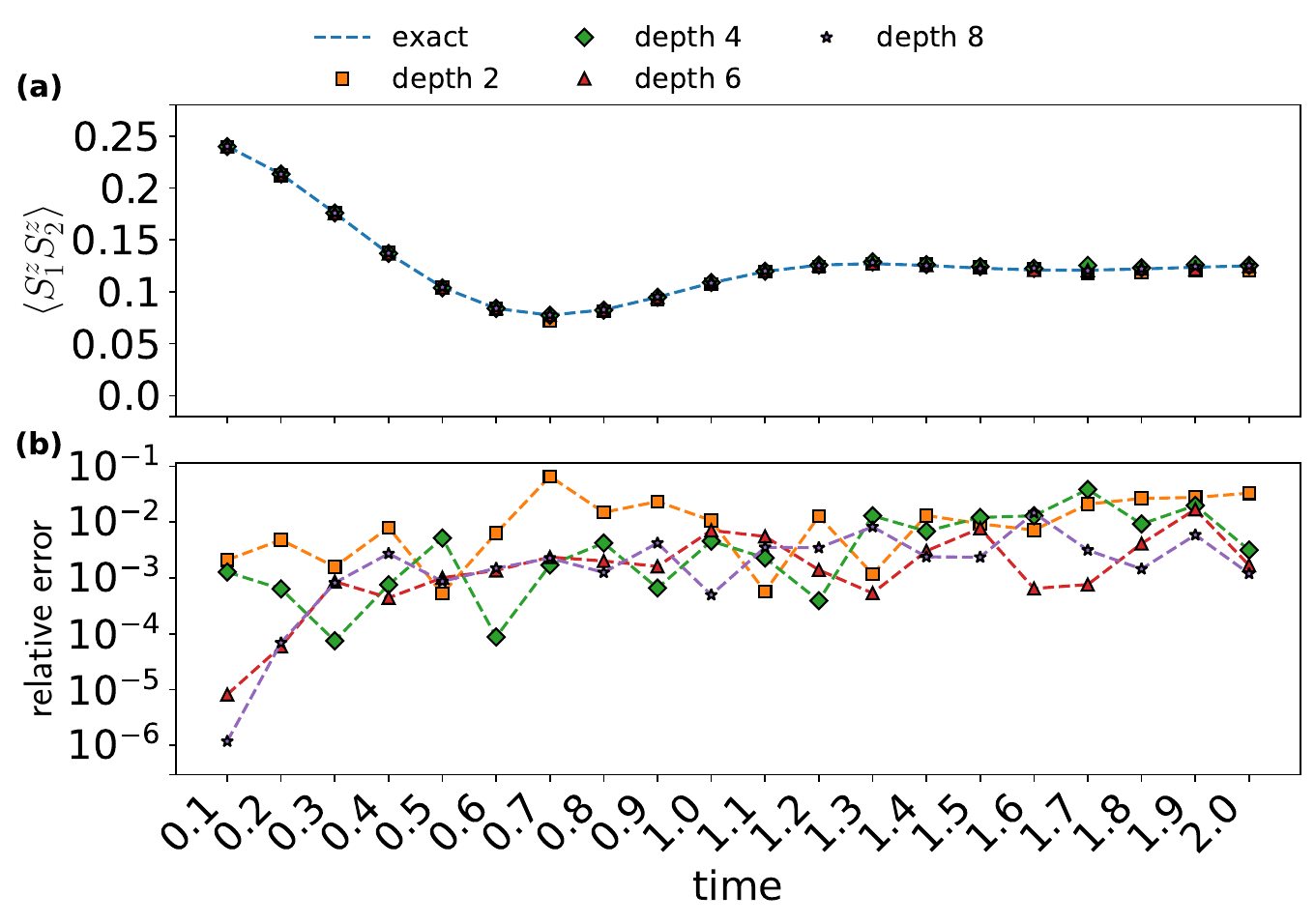}
	\caption{
		Comparison of different depths of the neural network in OMM optimized with 2nd order loss function.
		(a) The two-point correlation function $\langle S_1^z S_2^z\rangle$;
		(b) The relative error of the two-point correlation function $\langle S_1^z S_2^z\rangle$.
	}
	\label{fig:ensemble-depth}
\end{figure}

\begin{figure}[t]
	\includegraphics[scale=0.34]{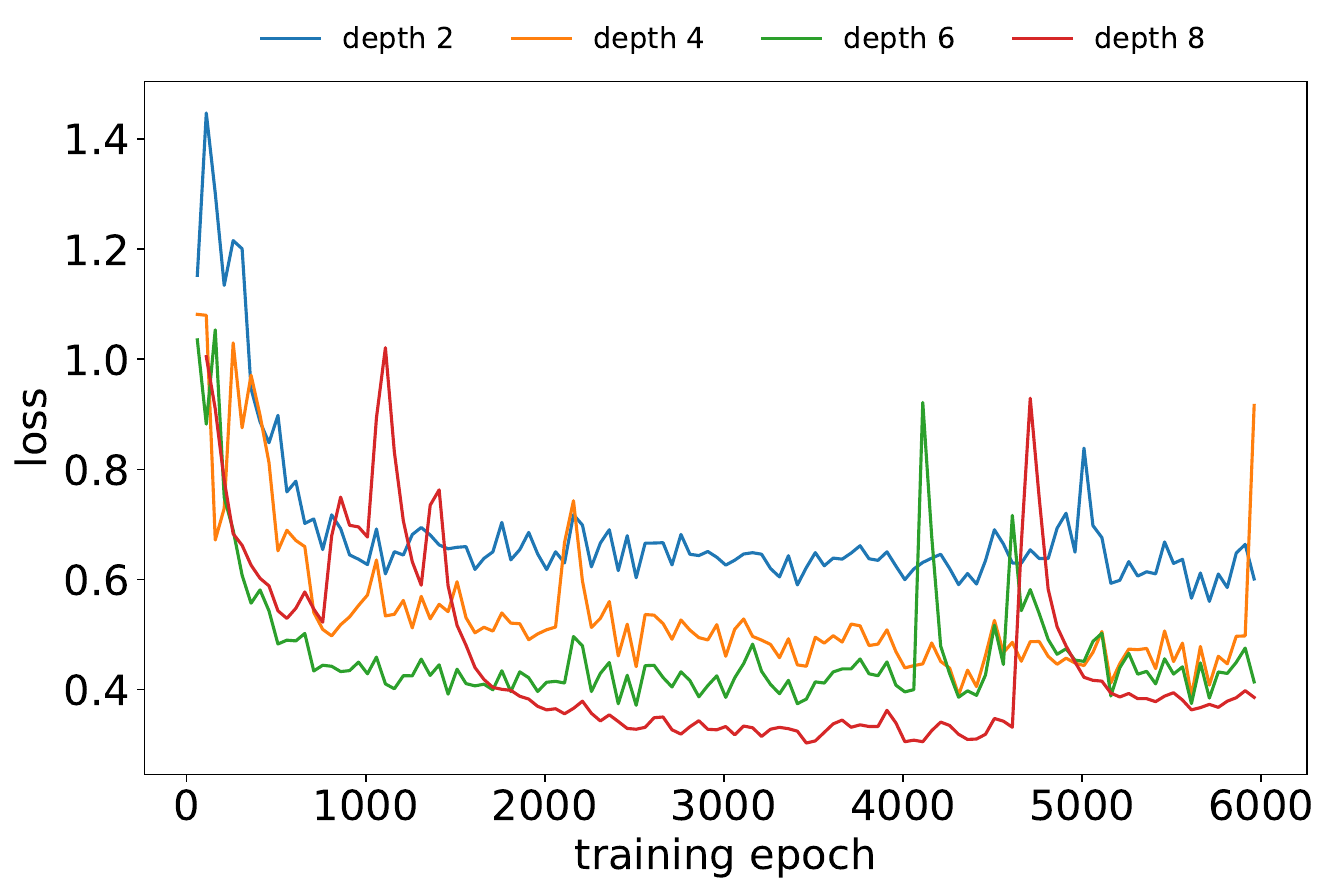}
	\caption{
		The loss function of different depths of neural OMM with 2nd order loss function at $T=2.0$ with a moving average of window size 5.
	}
	\label{fig:ensemble-depth-loss}
\end{figure}

In neural OMM, the depth of the neural network corresponds to the expressiveness of the operator map. As depicted in \cref{fig:ensemble-depth}, we find that the depth of the neural network influences the relative error as well as the speed of convergence as shown in \cref{fig:ensemble-depth-loss}. Deeper networks tend to converge faster and with a lower relative error. This is likely because deeper networks are more expressive and have better local minimums, thus allowing the algorithm to converge to a better solution faster.

\subsection{HEM}\label{sec:result-quantum}

For our implementation of HEM, we initialize the system size at $n = 2$ and aim for a target system size of $N = 6$, setting the field parameter at critical point $h = 1.0$. The initial state is chosen as $\rho_0 = \ket{0\cdots 0}$. The results of observable predictions are taken at the epoch with the minimum moving average loss with window size 10. We use a 2-level Rydberg Hamiltonian as the target device Hamiltonian. The pulse function is represented by a small feedforward neural network that takes the clock $t$ as input and returns the corresponding pulse value at time $t$. The simulation of the HEM algorithm is conducted on a single CPU. The loss function is also optimized via the gradient obtained from the adjoint method via \textit{jax.experimental.ode}. In practice, the gradient could also utilize quantum gradient~\cite{mitarai2018quantum,schuld2019evaluating,wierichs2021general,banchi2021measuring,leng2022differentiable}, finite difference, or other optimization algorithms suitable for the real device. 

\begin{figure}[b]
	\includegraphics[scale=0.34]{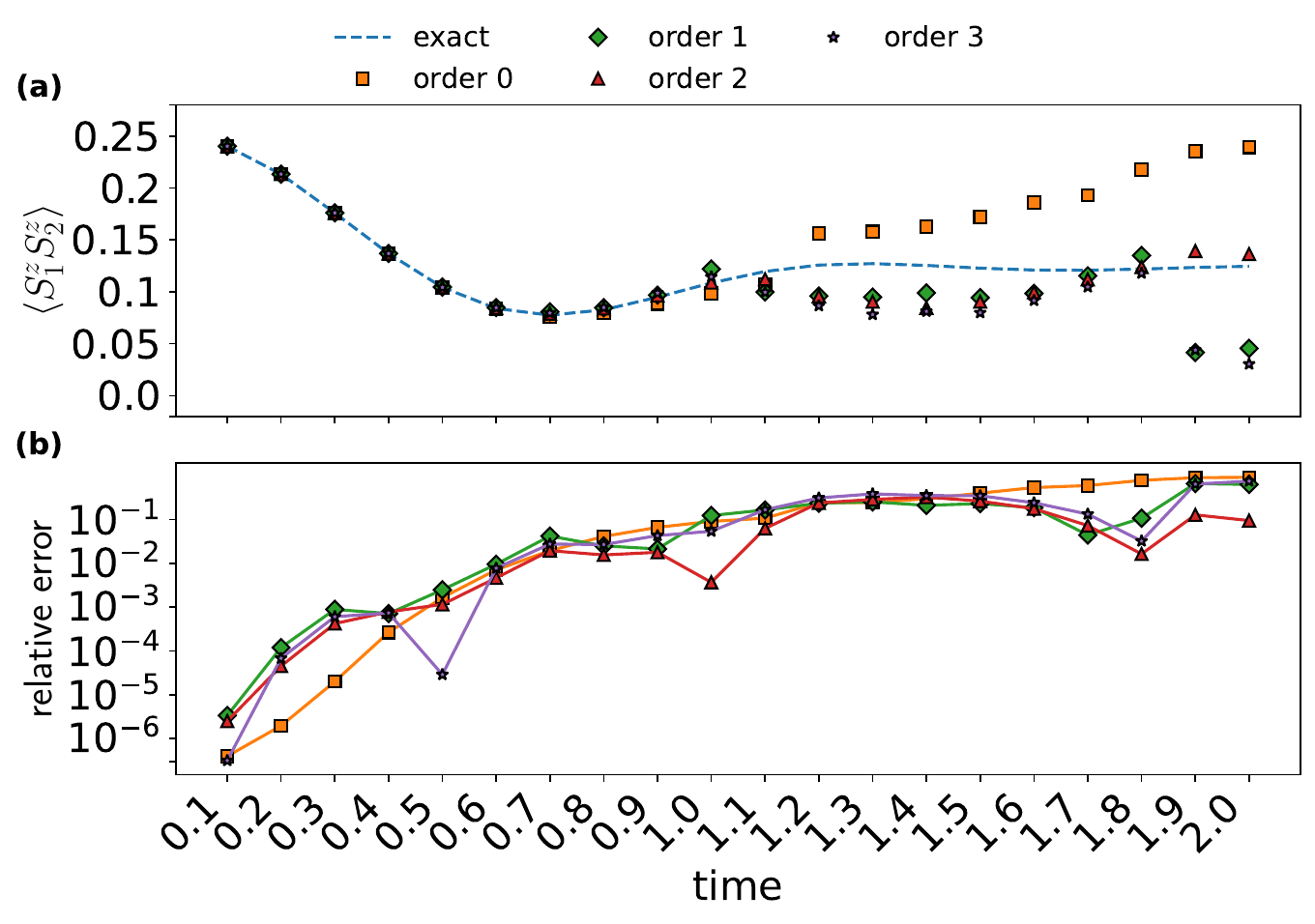}
	\caption{
		Comparison of HEM optimized at different loss function orders.
		(a) The two-point correlation function $\langle S_1^z S_2^z\rangle$;
		(b) The relative error of the two-point correlation function $\langle S_1^z S_2^z\rangle$.
	}
	\label{fig:quantum-order}
\end{figure}

\begin{figure}[t]
	\includegraphics[scale=0.34]{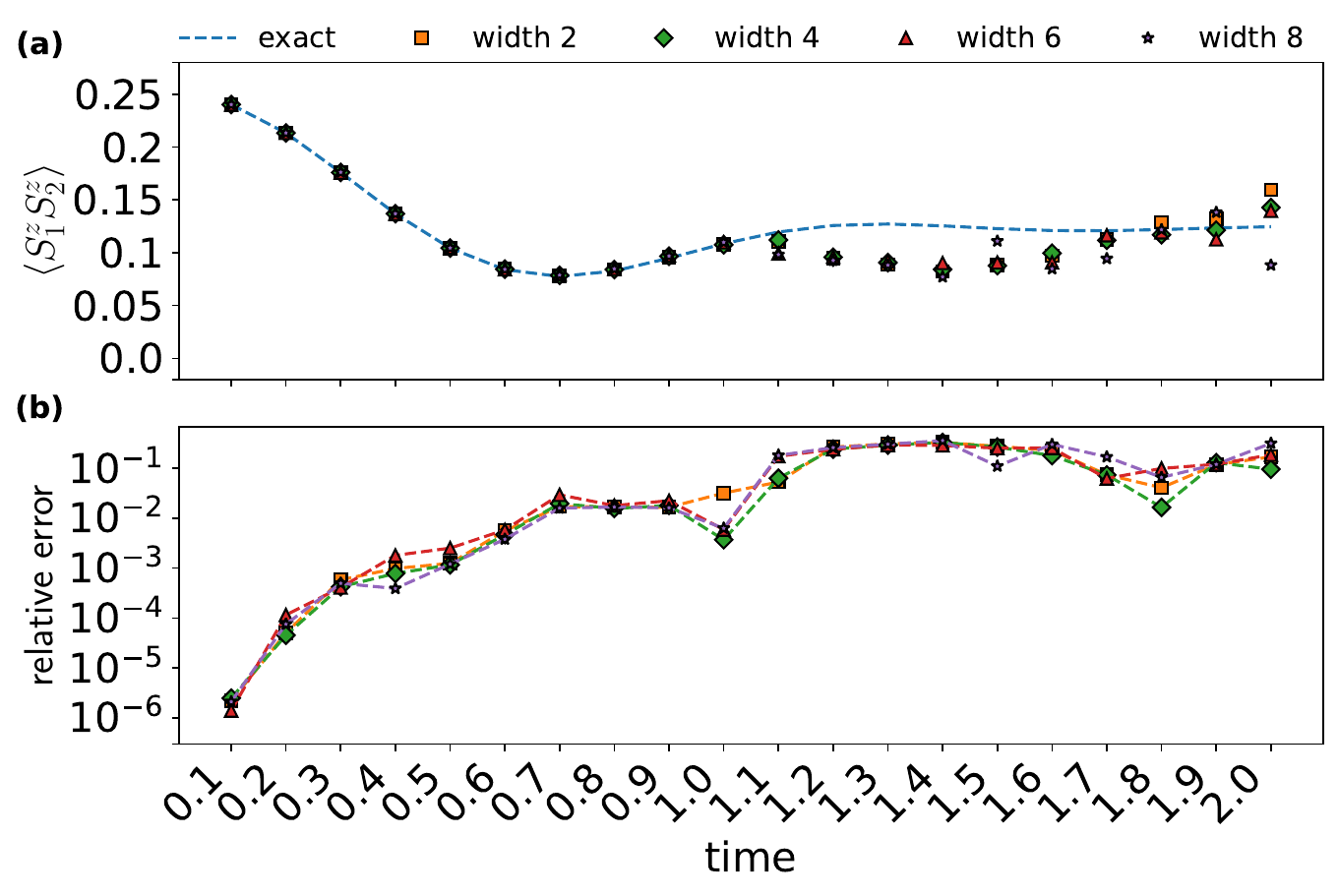}
	\caption{
		Comparison of the HEM optimized at different widths of neural networks with depth 4.
		(a) The two-point correlation function $\langle S_1^z S_2^z\rangle$;
		(b) The relative error of the two-point correlation function $\langle S_1^z S_2^z\rangle$.
	}
	\label{fig:quantum-width}
\end{figure}

\begin{figure}[t]
	\includegraphics[scale=0.34]{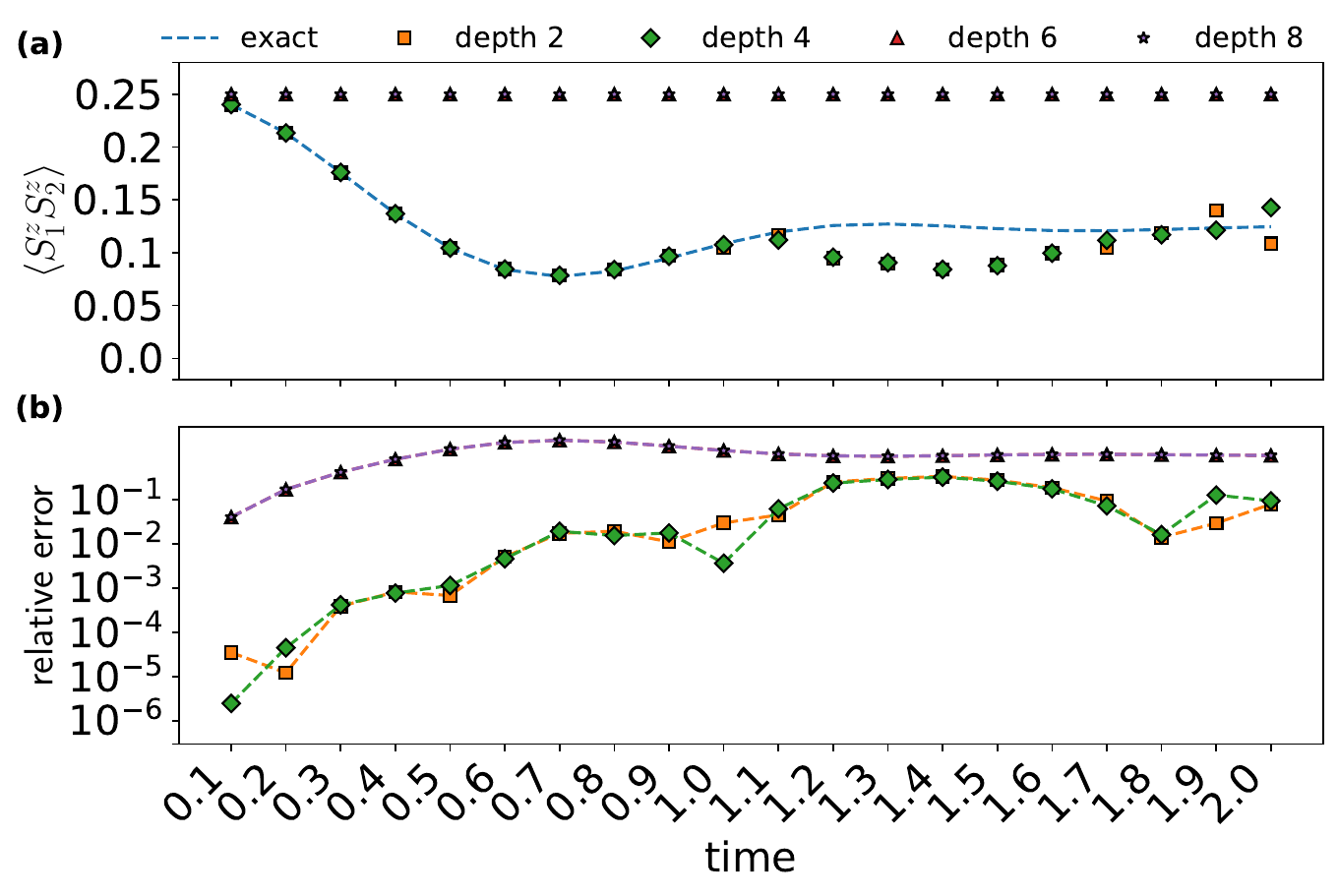}
	\caption{
		Comparison of HEM optimized at different depths of neural networks with width 4.
		(a) The two-point correlation function $\langle S_1^z S_2^z\rangle$;
		(b) The relative error of the two-point correlation function $\langle S_1^z S_2^z\rangle$.
	}
	\label{fig:quantum-depth}
\end{figure}

We also evaluate the performance of HEM at different orders of the loss function. As depicted in \cref{fig:quantum-order}, similar to the classical algorithm, we observe that at short-time intervals, the order of the loss function does not significantly impact the results. At longer times, the 0-order loss functions drifts more from the exact result. However, the 3-order loss also drifts in $t=1.9,2.0$. We suspect this is due to insufficient optimization, because higher orders requires optimizing more discrepancies.

For HEM, the expressiveness of representing a quantum dynamical process is mainly provided by the device Hamiltonian. Thus, the hyperparameters of the neural network affect the optimization rather than the expressiveness. We conduct a comparative analysis of the neural network's width and depth. As illustrated in \cref{fig:quantum-width}, we find that the width of the neural network (set at a depth of 4) does not significantly influence the algorithm's performance. In contrast, the depth of the neural network (set at a width of 4) shows a notable impact. As depicted in \cref{fig:quantum-depth}, a deeper neural network leads to diminished performance, likely due to a vanishing gradient that does not provide a better landscape. Conversely, shallower networks are more successful in identifying an appropriate pulse function. All the results of HEM start drifting after $T=1.1$. We hypothesize that this is due to the 2-level Rydberg Hamiltonian not being universal.

\subsection{Transfer Learning between Time Points}\label{sec:results-transfer-learning}

\begin{figure}[h]
	\includegraphics[scale=0.36]{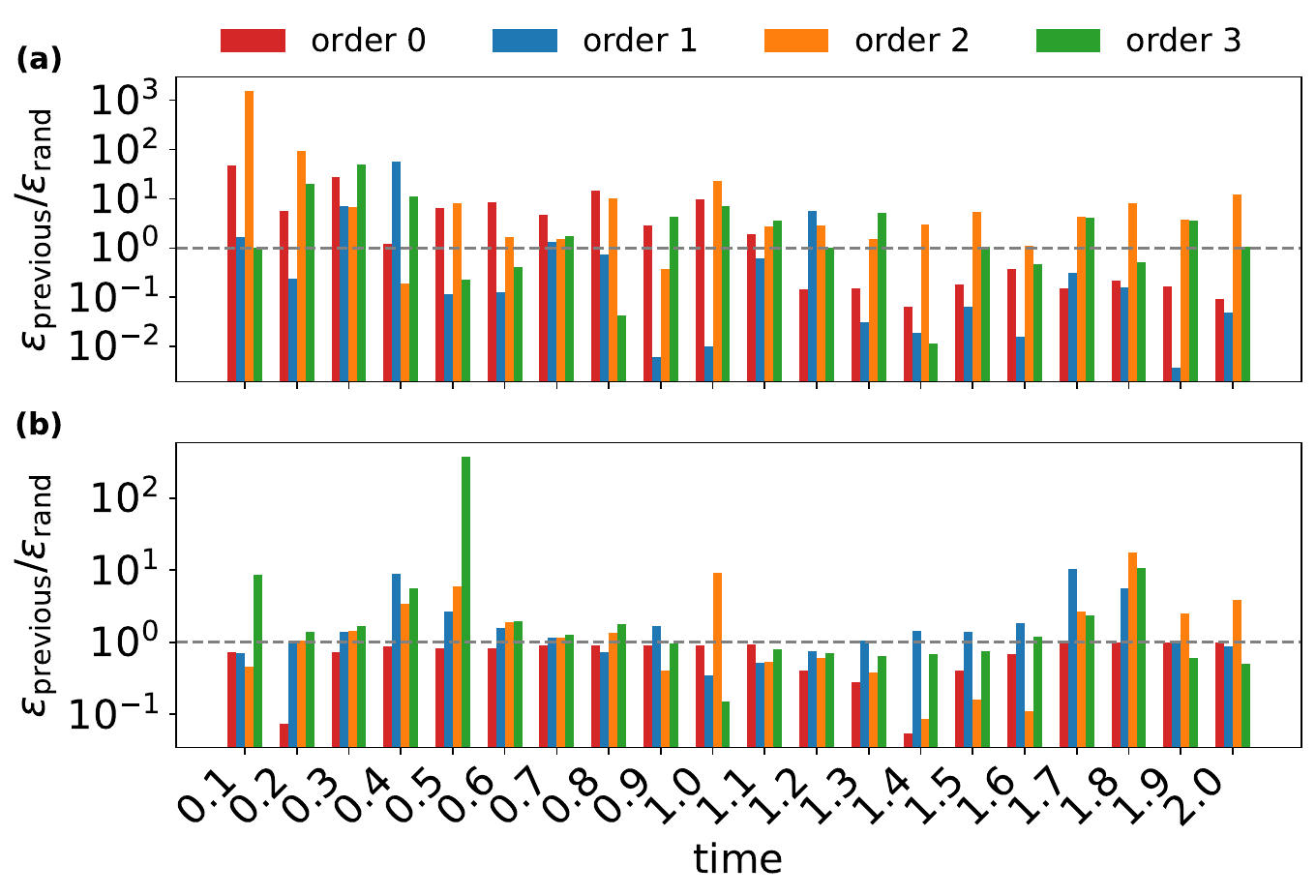}
	\caption{
		Transfer learning to different time points. Compared by a different order of loss function. The y-axis is the ratio between the relative error of initialization from previous time point $\epsilon_{\text{previous}}$ and random initialization $\epsilon_{\text{rand}}$. Above the line $y=10^0$ means random initialization is better; below the line means initialization from the previous time point is better.
		(a) neural OMM;
		(b) HEM targeting Rydberg Hamiltonian;
	}
	\label{fig:sweep-order}
\end{figure}

We investigate the transfer learning between time points with uniformly training each time point for a fixed number of epochs. For neural OMM, we allocated 1000 epochs at each time point, while for HEM, we allocate 500 epochs. As depicted in \cref{fig:sweep-order}, this approach reduces the number of epochs needed compared to initialization from random parameters, yet it still delivers similar performance levels. Additionally, initializing from the parameters of the previous time point results in the lower order loss function achieving a better relative error than when starting from random parameters. This improved performance can be attributed to the smooth nature of this specific time evolution, which allows high-order correlations to propagate through the parameter initialization. In contrast, when OMM represents a pure isometry, it is possible to express an explicit state as a MPS. Therefore, initializing from the parameters of the previous time point can be viewed as utilizing an implicit quantum state as input.

\section{Discussion}\label{sec:discussion}
In this work, we have introduced an algorithmic framework named OLRG, an alternative variational principle that generalizes Wilson's NRG and White's DMRG. The algorithm's e2e-style loss function directly bounds the real-time dynamics of target observables. The OLRG framework allows us to introduce different categories of ansatzes for an operator map between a real and a virtual system. We designed operator maps for real-time dynamics simulation on conventional computers and quantum devices. This includes the Operator Matrix Map (OMM) and the Hamiltonian Expression Map (HEM). OMM opens up new possibilities to provide more expressiveness than the linear operator map in NRG and DMRG. This could lead to opportunities to explore challenging real-time dynamics problems, e.g.~in higher dimension lattices, by exploring different forms of the growing operator $G_k$ (\cref{appx:higher-dimension-lattice}).
As a side product of this work, we also see OLRG as a potentially complementary variational principle to address high-order and long-time correlations for Matrix Product State (MPS) time-dependent variational principle (TDVP) in \cref{appx:mps-tdvp}.
In addition, our HEM-based OLRG provides a digital-analog quantum algorithm that integrates the product formula, VQA, and classical simulators for simulating quantum dynamics. Finally, we discussed tuning different hyperparameters and training schedules to improve the algorithm's performance in calculating two-point correlations for TFIM undergoing real-time dynamics.

Advancement to the OLRG framework can be made by enhancing the operator map and loss function. For the loss function, one could derive a more specific loss function for the target problem and further explore the relationship between superblock formalism from DMRG and series expansion (\cref{appx:improving-loss-function}). Because \cref{thm:sc-error} is general for any properties. Another future direction is finding the loss function directly for other properties such as ground state properties, phase transition points, entanglement entropy, etc. (\cref{appx:loss-for-other}). This will align the framework further with e2e learning for calculating other properties. Such loss functions likely exist due to the success of NRG and DMRG in evaluating various properties especially in solving ground-state problems.

For the operator map, as a further step one can consider the implementation of OMM including tensor network ensembles and deep neural network architectures. The target device Hamiltonian of HEM could be expanded to universal neutral atom arrays, ion traps, and superconducting circuits with different control capabilities. We discuss various ansatz designs under the OLRG framework in \cref{appx:ansatz-improvements}. In this paper, we only investigated the simplest OMM implemented by a feedforward neural network and a HEM targeting the non-universal 2-level Rydberg Hamiltonian. More powerful operator maps remain to be explored in future research. For real quantum devices, skipping the step of compiling the Hamiltonian terms into gates and directly using the pulse sequence may result in a non-trivial pulse sequence that is more efficient than 1 or 2-qubit gates. This is because, for real devices, certain global unitaries are easier to implement with shorter pulse sequences than decomposing into gates~\cite{levine2019parallel}. Thus, one may expect the effective pulse duration to be shorter than performing small-qubit gate compilation. The HEM-based OLRG can thus be also viewed as a quantum-assisted quantum compilation algorithm~\cite{Khatri2019quantumassisted}. A future direction is to benchmark the effective pulse duration in this case. Another interesting direction is exploring the generalization capability of the operator map for a larger system size trained at a small system size. By utilizing previous theoretical work about finite size error~\cite{wang2021bounding}, one may derive the e2e-style loss function for infinite-size systems. Then, one could attempt to train the operator map to predict properties directly for infinite-size systems.

% For optimization, differentiable programming is a powerful technique for training arbitrary operator maps. One should not expect gradient-based optimization always to work well. Besides the advantages we have demonstrated in this paper, disadvantages of differentiable programming includes longer convergence time and larger cost of evaluation comparing to iterative optimization such as utilizing eigensolvers. Future directions in improving gradient-based optimization include co-designing optimization and operator maps by incorporating symmetries and guidance from real data in medium-size systems~\cite {PhysRevB.105.205108,moss2023enhancing}. Because a large component in the time complexity can be parallelized, potential technical improvements may be seen in utilizing distributed gradient descent~\cite{aji2017sparse}. 

\section{Acknowledgments}

The authors acknowledge helpful discussions with Hsin-Yuan Huang, Jing Chen, Pan Zhang, Lei Wang, E.M. Stoudenmire, Matthew Fishman, Fangli Liu, Juan Carrasquilla, Jinguo Liu, Qi Yang, Hai-Jun Liao, Hao Xie, Phillip Weinberg, Jonathan Wurtz, Shengtao Wang, Hong-Ye Hu, Sebastian Wetzel, Ejaaz Merali, and Roeland Wiersema. Special thanks are due to Lei Wang for generously providing GPU resources, a contribution that significantly propelled our computational efforts. Xiuzhe Luo also appreciates Fan Zhang for her love and support during the research. Xiuzhe Luo is also grateful for helpful discussions during Swarma-Kaifeng Workshop 2016,2017. Di Luo acknowledges support from the National Science Foundation under Cooperative Agreement PHY-2019786 (The NSF AI Institute for Artificial Intelligence and Fundamental Interactions, \url{http://iaifi.org/}). This material is based upon work supported by the U.S. Department of Energy, Office of Science, National Quantum Information Science Research Centers, Co-design Center for Quantum Advantage (C2QA) under contract number DE-SC0012704. R.G.M. acknowledges financial support from NSERC.
This research was enabled in part by computational support provided by the Shared Hierarchical Academic Research Computing Network (SHARCNET) and the Digital Research Alliance of Canada. We acknowledge financial support from the Perimeter Institute. Research at Perimeter Institute is supported in part by the Government of Canada through the Department of Innovation, Science and Economic Development Canada and by the Province of Ontario through the Ministry of Economic Development, Job Creation and Trade.

\bibliography{main}

\begin{thebibliography}{173}
\bibitem [{rog()}]{roger2024duck}%
  \BibitemOpen
  \href@noop {} {}\bibinfo {note} {Duck typing was the initial inspiration for this work, thus also called duck RG.}\BibitemShut {Stop}%
\end{thebibliography}

\appendix

\clearpage

\onecolumngrid
\begin{center}
	\noindent\textbf{Supplementary Material}
	\bigskip

	\noindent\textbf{\large{}}
\end{center}

\section{Scaling Consistency}\label{appx:scale-consistency}

The scaling consistency condition is generic to arbitrary properties of the system. To demonstrate this, we first introduce the definition of a system, property function, connecting operator, growing operator, and scaling consistency. Then, we prove the error upper bound of the OLRG process

\begin{definition}[Many-body Hilbert space]
	Denote the many-body Hilbert space with $d$ local states and $n$ sites as $\hilbert{n}$ and the self-adjoint operators on $\hilbert{n}$ as $\hermitian{n}$.
\end{definition}

\begin{definition}[Property function]
	A property function $p_n$ is a function that maps a set of self-adjoint operators to a real value quantity, denoting the domain of $p$ as $\dom{p_n} \subseteq \hermitian{n}\times\cdots\times \hermitian{n}$, we have $p_n: \dom{p_n} \rightarrow \mathbb{R}$.
\end{definition}

Property functions include the expectation value of an observable, the correlation function, the entanglement entropy, energy, etc. As an example, the two-point correlation function on 1st and 2nd sites at time $T$ is defined as
\begin{equation}
	\expect{Z_1 Z_2}_{T} = \tr(\rho_0 \opte{U(T)}{Z_1 Z_2})
\end{equation}
where $\rho_0$ is the initial state, $U(T)$ is the time-evolution operator, and $Z_1, Z_2$ are the Pauli operators acting on the 1st and 2nd sites respectively. Thus, it can be defined as a function on $\hermitian{n}\times\hermitian{n}$ where one input operator is the initial state $\rho_0$ and the other is the Hamiltonian $H$.

\begin{definition}[Connecting operator]
	A connecting operator is the superoperator $R_{\step}: \hermitian{n} \rightarrow \hermitian{\step}$ such that given an operator $L\in \hermitian{n}$ we have $\conn{L}\in \hermitian{\step}$. The expression $L\otimes \connsub{\step}{L}$ characterizes the connection between the $n$-site system and $\step$-site system.
\end{definition}

To elucidate this concept, consider the following $1D$ TFIM Hamiltonian:
\begin{equation}
	H = \sum_i Z_i Z_{i+1} + h\sum X_i
\end{equation}

For $1D$ TFIM, given a $n+\step$-site TFIM, the connection between $n$-site TFIM and $\step$-site TFIM is $Z_n Z_{n+1}$, and the operator on the $n$-site TFIM is $L = Z_n = I^{\otimes n-1}\otimes Z$, thus the connecting operator on this operator is defined as $R_1(L) = Z$. Similarly, for the Heisenberg Model:
\begin{equation}
	H = \sum_i X_i X_j + Y_i Y_j + Z_i Z_j
\end{equation}
There are three types of connections thus $L = X_n, Y_n, Z_n$, and the connecting operators are defined as $R_1(L) = X, Y, Z$ respectively. It is worth noting that although the name "Connecting Operator" was not mentioned in the literature to the best of our knowledge, the concept of the connecting operator has been widely used in the implementation of the DMRG algorithm~\cite{simple-dmrg,ITensor}. The connecting operator describes how one can add new physical sites into an existing system, thus this allows the definition of the growing operator.

\begin{definition}[Growing operator]
	A growing operator is the superoperator $G_{\step}: \hermitian{n} \rightarrow \hermitian{n+\step}$ such that given an operator $X\in \hermitian{n}$ we have $\opmap{G_{\step}}{X}\in \hermitian{n+\step}$. The growing operator has a general form defined using connecting operator $R_{\step}$ as:
	\begin{equation}
		\opmap{G_{\step}}{X} = X\otimes \connsub{\step}{X} + \sum_i B_i\otimes \connsub{\step}{B_i}
	\end{equation}
	where $\{B_i \in S_n \mid G_{\step}(X) \in \hilbert{n+\step}\}$, and we call $B_i$ the boundary operators.
\end{definition}

The growing operator is defined with a connecting operator $R_{\step}$. The summation $\sum_i$ does not limit the number of $B_i$. Thus, such decomposition exists for any operator $X$. For a Hamiltonian with a general form of $n$ such as the TFIM or Heisenberg Hamiltonian, where the Hamiltonian has a definition over arbitrary $n$ sites, the definition of the growing operator is straightforward. 

\begin{corollary}
	For finite operators with definition on a fixed number of sites, because assuming there exists $X_0,X_1,\cdots, X_n$ and $X_0\in\mathcal{H}(\mathbb{C}^d)$, we have the following relationship
	\begin{equation}
		\begin{aligned}
			X_1 &= X_0 \connsub{1}{X_0} + \sum_i L_i^1 \connsub{1}{L_i^1}\\
			X_2 &= X_1 \connsub{1}{X_1} + \sum_i L_i^2 \connsub{1}{L_i^2}\\
			&\cdots\\
			X_n &= X_{n-1} \connsub{1}{X_{n-1}} + \sum_i \connsub{1}{L_i^n}\\
		\end{aligned}
	\end{equation}
	thus, the final operator $X_n$ is a summation of single operator strings. Without limiting the summation $\sum_i$ to be polynomial, we can always decompose a given operator on the summation of $n$ single operator strings denoted as $X_n$. This allows the definition of all previous operators $X_0,\cdots,X_{n-1}$. Thus, following this procedure, we can define the growing operator for any operator $X$ on a fixed number of sites, such as the two-point correlation function on $n$-site system at a specific location $i,j$ as we introduced in \cref{sec:framework}.	
\end{corollary}

From a different perspective, inspired by DMET~\cite{doi:10.1021/ct301044e,doi:10.1021/acs.accounts.6b00356}, one can see such definition as a process of creating fragments of the operator like in DMET. We define the growing operator as the process of adding fragments back. This leads to the definition of the rescalable operator.

\begin{definition}[Rescalable Operator]
	With connecting operator $R_{\step}$ and growing operator $G_{\step}$, we can define the rescalable operator $\mathcal{X}_n$ as the set $\mathcal{X}_n = \{X_n, \partial X_n, R_{\step}\}$, where $X_n$ is the operator at current scale, $\partial X_n$ is a set of operators describing the effect of environment on the system, and thus the growing operator $G_{\step}$ of such operator can be recursively defined as
	\begin{equation}\label{eq:rescalable-operator}
		\opmap{G_{\step}}{X_n} = X_n\otimes I^{\otimes \step} + \sum_{B\in \partial X_n} B\otimes \connsub{\step}{B}\\
	\end{equation}
	where $X_0$ is a constant operator, $B\in \partial X_n$.
\end{definition}

For example, we can define the rescalable Hamiltonian $\mathcal{H}_n$ as the set $\mathcal{H}_n = \{H_n, \partial H_n, R_{\step}\}$, where $H_n$ is the Hamiltonian at current scale, $\partial H_n$ is a set of operators describing the effect of environment on the system referred as the boundary set in the following context, and thus the growing operator $G_{\step}$ of such Hamiltonian operator can be recursively defined as
\begin{equation}\label{eq:rescalable-hamiltonian}
	\opmap{G_{\step}}{H_n} = H_n\otimes I^{\otimes \step} + \sum_{B\in \partial H_n} B\otimes \connsub{\step}{B}\\
\end{equation}
where $H_0$ is a constant operator, $B\in \partial H_n$.

\begin{definition}[Rescalable System]
	Given a property $p_N$, where $N$ is the number of sites, we can define the system $S_N$ as a set of operators such that $S_N\in \mathrm{dom}(p_N)$. Then for $n\leq N$, we can define the rescalable system $S_n$ as the set $S_n = \{S_n, \partial S_n, R_{\step}\}$, where $S_n$ is the operator at current scale, $\partial S_n$ is a set of boundary operators, and thus the growing operator $G_{\step}$ of such system can be recursively defined as
	\begin{equation}
		S_{n+\step} = \opmap{G_{\step}}{S_n} = \{\opmap{G_{\step}}{X}\mid X \in S_n\}
	\end{equation}
\end{definition}

For example, for the two-point correlation function $\expect{Z_1 Z_2}_T$ at time $T$ for 4-site $1D$ TFIM with $\ket{0\cdots 0}$ as initial state, we have
\begin{equation}
	\begin{aligned}
		S_4 &= \{\ket{0000}\bra{0000}, H_4, Z\otimes Z\otimes I\otimes I \} & \partial S_4 = \{Z_4\}\\
		S_3 &= \{\ket{000}\bra{000}, H_3, Z\otimes Z\otimes I\} & \partial S_3 = \{Z_3\}\\
		S_2 &= \{\ket{00}\bra{00}, H_2, Z\otimes Z\} & \partial S_2 = \{Z_2\}\\
		S_1 &= \{\ket{0}\bra{0}, H_1, Z\} & \partial S_1 = \{Z_1\}\\
	\end{aligned}
\end{equation}

where $H_i,i=1,2,3,4$ is the TFIM Hamiltonian of $i$ sites. The boundary set $\partial S_i, i=1,2,3,4$ only contains the $\partial H_i$ because $\ket{0\cdots 0}\bra{0\cdots 0}$ and $Z,Z\otimes Z,Z\otimes Z\otimes I,Z\otimes Z\otimes I\otimes I$ have no boundary operators. Now, with the definition of the rescalable system, we can study the behavior of an operator map $\ansatz{n}: \mathcal{A}_n \rightarrow \mathcal{A}_n$ where $\theta$ is the parameter of the operator map.

\begin{definition}[OLRG step]
	Given an operator map $\ansatz{n}: \mathcal{A}_n \rightarrow \mathcal{A}_n$, we define the one OLRG step $D_k$ as applying $\ansatz{n}$ on all operators in $S_n$ and then growing the system to $S_{n+\step}$, thus we have $D_{\step} = G_{\step}\circ \ansatz{n}$. Here we assume $D_l$ is adaptive on the system size $n$.
\end{definition}

\begin{definition}[$\epsilon$-scaling consistency]\label{def:appx-epsilon-scaling-consistency}
	An operator map $\ansatz{n}: \mathcal{A}_n \rightarrow \mathcal{A}_n$ is said to satisfy $\epsilon$-scaling consistency for system $S_n$ and property $p_N$ where $N\geq n$, if $\exists \epsilon > 0, \forall q = 1,2,\cdots,(N - n)/\step$ we always have
	\begin{equation}
		\norm{p_N[\opmap{G_{\step}^q}{S_n}] - p_N[\opmap{(G_{\step}^{q-1}\circ D_{\step})}{S_n}]} \leq \epsilon
	\end{equation}
\end{definition}

The $\epsilon$-scaling consistency condition allows us to bound the error of the OLRG process. While the OLRG process does not necessarily use the same $\ansatz{n_q}$ at each step $D_{\step}$, without loss of generality, we present the following theorem by assuming $\ansatz{n_q} = \ansatz{}$ is the same between $S_n$ and $S_{n+\step}$ for convenience.

\begin{theorem}[System scaling error]\label{thm:formal-system-scaling-error}
	For target system size $N$ and starting system size $n$, where $N\geq n$, if the operator map $\ansatz{}$ satisfy the $\epsilon$-scaling consistency condition for $S_n,S_{n+1},\cdots,S_{N-1}$, and $q = (N - n)/\step$ then
	\begin{equation}
		\norm{p_N[\opmap{G_{\step}^q}{S_n}] - p_N[\opmap{D_{\step}^q}{S_n}]} \leq q\epsilon
	\end{equation}
\end{theorem}

\begin{proof}
	denote $\eta_{i} = p_N[\opmap{(G_{\step}^{q-i} \circ D_{\step}^i)}{S_n}]$, where $G_{\step}^{q-i} = \underbrace{G_{\step}\circ \cdots \circ G_{\step}}_{q-i \text{ times}}$ and $D_{\step}^i = \underbrace{D_{\step}\circ\cdots\circ D_{\step}}_{i \text{ times}}$
	\begin{align}
		&= \norm{\eta_{0} - \eta_{q}} = \norm{\eta_{0} - \eta_{1} + \eta_{1} - \eta_{q}}\\
		&=\norm{\sum_{i=0}^{q-1} \eta_{i} - \eta_{i+1}} \leq \sum_{i=0}^{q-1} \norm{\eta_{i} - \eta_{i+1}} =q\epsilon & \text{(triangular inequality)}
	\end{align}
\end{proof}

The above theorem breaks the system error of OLRG into errors between each step at target size $N$. This allows us to further bound the error of the OLRG process by looking at more specific Hamiltonians and properties.

\section{Growing Operator of Rescalable Local Hamiltonians}\label{appx:grow-op}

In general, the $\epsilon$-scaling consistency cannot be evaluated on a small system directly because \cref{thm:formal-system-scaling-error} requires evaluating the property function at size $N$. However, intuitively, the discrepancy caused by applying $f$ on system $S_n$ can be traced back to the change of some operators in the system of size $n$. If the interaction of the Hamiltonian is local, the propagation of the discrepancy should not be far. This motivates us to study the rescalable local Hamiltonians defined as follows.

\begin{corollary}[Rescalable Local Hamiltonian]
	For a rescalable Hamiltonian $\mathcal{H}_n$, if $\forall B\in \partial H_n$, $B$ act on $x$ sites and $R(B)$ acts on $\locality-x$ sites for $x = 0,1,\cdots,\locality$, then this Hamiltonian is a $\locality$-local Hamiltonian at every scale.
\end{corollary}

Notably, for local Hamiltonian, $I^{\otimes n} \in \partial H$ because $\conn{B}$ can act on $m$ at most. Physically, this represents the terms that only affect the environment but not the system. For example, in the TFIM, the $h\cdot X$ term only appears in the environment.

\begin{corollary}[Local Hamiltonian]
    For $\locality$-local Hamiltonian of $N$ sites, one can always define the corresponding rescalable $\locality$-local Hamiltonian up to $N$ sites.
\end{corollary}

This is because one can always cut the $N$-site $\locality$-local Hamiltonian into fragments, then we can create the definition of $\mathcal{H}_n$ recursively by defining $\mathcal{H}_1$. Define the $H_1$ as one fragment, $\partial H_1$ as the interaction terms between $\mathcal{H}_1$ and another fragment. Thus, we define $H_2$ as the composition of two fragments and repeat until we have $\mathcal{H}_N$.

\begin{lemma}[Boundary set of geometrically local Hamiltonian]
	For a geometrically local Hamiltonian $\mathcal{H}_{n+\step}$, the boundary set $\partial H_{n+\step}$ has the following form
	\begin{equation}
		\partial H_{n+\step} = \{I^{\otimes \step}\otimes B_i \mid B_i \in \partial H_n\}
	\end{equation}
\end{lemma}

\begin{proof}
	Without loss of generality, we can always assume $B = \bigotimes_{i=1}^x X_i$ where $X_i\in \hilbert{}$, because if $B$ is not a tensor product, we can always decompose it onto Pauli basis with coefficients $B = \sum_{\mathbf{b}} c_{\mathbf{b}}\cdot P_{b_1}\otimes P_{b_2}\otimes \cdots \otimes P_{b_x}$, where $P_{i}$ is a Pauli operator. Thus resulting redefinition of $B$ as $c_{\mathbf{b}}\cdot P_{b_1}\otimes \cdots \otimes P_{b_x}$. The effect of the environment will not change by rescaling, and new terms cannot be applied outside of the $m$ sites at the boundary by the definition of geometrically local, thus $\partial H_{n+\step} = \{I^{\otimes \step}\otimes B_i \mid B_i \in \partial H_n\}$
\end{proof}

As shown in \cref{fig:grow-demo}, for geometrically $\locality$-local Hamiltonian, applying the growing operator $q > \locality$ times on the Hamiltonian will saturate the boundary set. This motivates us to define the following concept.

\begin{corollary}[Saturated Boundary Set for geometrically local Hamiltonian]
	For a geometrically $\locality$-local Hamiltonian $\mathcal{H}$, the boundary set $(\partial H_n)$ will saturate for $G_{\step}$ as $\step$ increases. Denote as $(\partial H_n)^{G_{\step}}$. For $\step > \locality$, $\norm{(\partial H)^G_{\step}}$ scales with the boundary size for geometrically local Hamiltonians as
	\begin{equation}
		\mathcal{O}(nL^{n-1})
	\end{equation}
	where $L = \max{\abs{dim_i}},i=1,\cdots,n$, e.g in $1D$ it scales as $\mathcal{O}(1)$,
	and in 2D scales as $\mathcal{O}(2L)$.
\end{corollary}

For 1-D geometrically $\locality$-local Hamiltonian, with $G_{\step}$ always adding sites on one side of the original system, we have
\begin{equation}
	\begin{aligned}
		\norm{(\partial H_n)^{G_{\step}}} = \begin{cases}
			\step\norm{\partial H_n}	& \step < \locality - 1    \\
			(\locality - 1)\norm{\partial H_n} 	& \step \geq \locality - 1
		\end{cases}
	\end{aligned}
\end{equation}
where $m$ is the number of species of connecting operators in $G_{\step}$.

\begin{corollary}\label{cor:saturated-grow-op}
	the growing operator $G_{\step}^q$ defined on a geometrically $\locality$-local Hamiltonian can be rewritten as the following form
	\begin{equation}\begin{aligned}
		\opmap{G_{\step}^{q}}{H_n} = H\otimes I^{\otimes q\step} + \sum_{B_i\in (\partial H)^{G_{\step}}} B_i\otimes \conn{B_i} + I^{\otimes n} \otimes K
		\end{aligned}\end{equation}
\end{corollary}

In \cref{sec:sc-rt-loss}, we mention this is a property of geometrically local Hamiltonian, which can be generalized to rescalable local Hamiltonian with constant non-geometrically local terms. This can be shown by constructing a system with periodic boundary conditions, where the interaction term at the boundary is not geometrically local. Still, there are only a constant number of them. Thus, we have the following example

\begin{example}[Saturated Boundary Set for Periodic Boundary]
    Consider the $1D$ periodic boundary TFIM Hamiltonian of $n$ sites
    \begin{equation}
        H_n = \sum_{i = 1}^{n-1} Z_i Z_{i+1} + Z_n Z_1 + h\cdot \sum_i X_i
    \end{equation}
    We can define $G_1$ as follows
    \begin{equation}
        \begin{aligned}
						\opmap{G_1}{H_n} = &H_{n}\otimes I + \underbrace{I^{\otimes n-1} \otimes Z \otimes Z}_{\text{connection with new site}} + \underbrace{I^{\otimes n}\otimes h\cdot X}_{\text{field term on the new site}}\\
            &- \underbrace{Z\otimes I^{\otimes n-2}\otimes Z \otimes I}_{\text{old interaction at $n-1$-site boundary}}
            + \underbrace{Z\otimes I^{\otimes n-1} \otimes Z}_{\text{new interaction at $n$-site boundary}}\\
        \end{aligned}
    \end{equation}
    applying $G_1$ twice, we have
    \begin{equation}
        \begin{aligned}
						\opmap{G_1^2}{H_n} = &H_{n}\otimes I^{\otimes 2} + \underbrace{I^{\otimes n-1} \otimes Z \otimes Z\otimes I + I^{\otimes n} \otimes Z \otimes Z}_{\text{connection with new site}} + \underbrace{I^{\otimes n}\otimes h\cdot X + I^{\otimes n+1}\otimes h\cdot X}_{\text{field term on the new site}}\\
            &- \underbrace{Z\otimes I^{\otimes n-2}\otimes Z \otimes I^{\otimes 2}}_{\text{old interaction at $n-1$-site boundary}}
            + \underbrace{Z\otimes I^{\otimes n} \otimes Z}_{\text{new interaction at $n$-site boundary}}\\
        \end{aligned}
    \end{equation}
    which still results in a saturated boundary set, equivalent to the saturated boundary set for open boundary $1D$ TFIM Hamiltonian (the geometrically local Hamiltonian) plus the operator at $n$-site periodic boundary $\{-Z\otimes I^{\otimes n-2}\otimes Z, Z\otimes I^{\otimes n-1}\}$. Note that, unlike geometrically local Hamiltonian, in this case, the result of connecting the operator on  $R(Z\otimes I^{\otimes n-1})$ is changing as the system grows $I\otimes Z, I^{\otimes 2}\otimes Z, \cdots$. But this will not affect the set of $B_i$ on the system's boundary.
\end{example}

Thus, for a more general case, we have the following

\begin{corollary}[Saturated Boundary Set of Rescalable Local Hamiltonian]
    For the rescalable $\locality$-local Hamiltonian $\mathcal{H}_n$, if there is only a constant number of non-geometrically local terms in $\partial H_n$, then applying $G_{\step}^q$ for arbitrary $q$ times, will result in a saturated set $(\partial H_n)^{G_{\step}}$.
\end{corollary}

\section{Scaling Consistency Condition for Real Time Evolution}\label{appx:sc-rt-proof}

To prove the scaling consistency condition for real-time evolution, we need to find a
series expansion for our time-evolved observables. We first introduce the following
notation of commutators and anti-commutators.

\begin{notation}[Adjoint]
	We denote the commutator for operator $A,B$ as $\lie_{A,-1}(B) = [A, B] = AB - BA$,
	and the anti-commutator as $\lie_{A,+1}(B) = \{A, B\} = AB + BA$, and for $\sigma = \pm 1$, we denote
	$\lie_{A,\sigma}(B) = AB + \sigma BA$.
\end{notation}

Then we have the following lemma due to linearity of the commutator and anti-commutator.

\begin{lemma}[Adjoint expansion]\label{lemma:adjoint-expand}
	We can expand the adjoint of the sum of operators $\sum_{i=1}^n A_i$
	with an operator $B$ as following:
	\begin{equation}
		\lie_{\sum_{i=1}^n A_i,\sigma}(B) = \sum_{i=1}^n \lie_{A,\sigma}(B)
	\end{equation}
\end{lemma}

\begin{proof}
	This is due to the linearity of the commutator and anti-commutator.
\end{proof}

Furthermore, we can denote the composition of the adjoints as following

\begin{notation}[Composition of Adjoints]
	We have the following notation for the adjoint of the composition of operators

	\begin{align}
		\lie_{A,\sigma}(\lie_{B,\sigma}(C)) &= \lie_{A,\sigma}\lie_{B,\sigma}(C)\\
		\lie_{A,\sigma}^k(B) &= \lie_{A,\sigma}(\lie_{A,\sigma}^{k-1}(B))
	\end{align}
\end{notation}

And we have the following lemma

\begin{lemma}[Adjoint power]\label{lemma:adjoint-power}
	We can expand the power of the adjoint of the sum of operators $\sum_{i=1}^n A_i$
	with an operator $B$ as following:
	\begin{equation}
		\lie_{\sum_{i=1}^n A_i,\sigma}^k(B)
		= \sum_{k_1,\cdots,k_n} \prod_{i=1}^n \lie_{A_{k_i},\sigma}
		= (\sum_{k=1}^n\lie_{A_{k},\sigma})^k
	\end{equation}
\end{lemma}

\begin{proof}
	\begin{align}
		& \lie_{\sum_{i=1}^n A_i,\sigma}^k(B)                                             \\
		& = \lie_{\sum_{i=1}^n A_i,\sigma}^{k-1}(\lie_{\sum_{k_1=1}^n A_{k_1},\sigma}(B)) \\
		& = \sum_{k_1=1}^n \lie_{\sum_{i=1}^n A_i,\sigma}^{k-1}(\lie_{A_{k_1},\sigma}(B)) \\
		& = \sum_{k_1=1}^n\cdots\sum_{k_n=1}^n (\prod_{i=1}^k \lie_{A_{k_i},\sigma})(B)
	\end{align}
\end{proof}
We can verify the correctness by checking for $k=2$, $n=2, \sigma=-1$, denote $\lie_{A,-1} = \lie_{A}$, we have

\begin{align}
	& (\lie_{A_1} + \lie_{A_2})^2                                                            \\
	& =\lie_{A_1}^2 + \lie_{A_2}^2 + \lie_{A_1}\lie_{A_2} + \lie_{A_2}\lie_{A_1}             \\
	& \lie_{A_1+A_2}^2(B)                                                                    \\
	& =\lie_{A_1+A_2}(\lie_{A_1+A_2}(B))                                                     \\
	& =\lie_{A_1+A_2}(\lie_{A_1}(B) + \lie_{A_2}(B))                                         \\
	& =\lie_{A_1+A_2}(\lie_{A_1}(B)) + \lie_{A_1+A_2}(\lie_{A_2}(B))                         \\
	& =\lie_{A_1}^2(B) + \lie_{A_2}\lie_{A_1}(B) + \lie_{A_1}\lie_{A_2}(B) + \lie_{A_2}^2(B)
\end{align}

\begin{lemma}[Adjoint of Tensor Product]\label{lemma:adjoint-kron}
	The power of adjoint of tensor product of operators
	$\lie_{A\otimes B,\sigma}^k$ can be expanded as following:
	\begin{equation}
		\begin{aligned}
			\lie_{A\otimes B,\sigma}^k & = \frac{1}{2^k} \sum_{\sigma_1,\sigma_2,\cdots,\sigma_k \in \{+,-\}}
			(\prod_{i=1}^k \lie_{A,\sigma_i}) \otimes (\prod_{i=1}^k \lie_{B,\sigma \sigma_i})                                              \\
			                           & = \frac{1}{2^k} (\sum_{\sigma_i \in \{+,-\}} \lie_{A,\sigma_i} \otimes \lie_{B,\sigma \sigma_i})^k
		\end{aligned}
	\end{equation}
	where $\lie_A\otimes\lie_B$ is defined as $\lie_{A}(X)\otimes \lie_{B}(Y) = (\lie_A\otimes\lie_B) (X\otimes Y)$
\end{lemma}

\begin{proof}
	It can be checked that
	\begin{equation}
		\lie_{A\otimes C,\sigma}(B\otimes D) = \frac{1}{2}\sum_{\sigma = +,-}\lie_{A,\sigma}(B)\otimes \lie_{C,\sigma \sigma}(D)
	\end{equation}
	by iterating this equation,
	\begin{equation}
		\begin{aligned}
			 & \lie_{A\otimes C,\sigma}(\lie_{A,\sigma_1}(B)\otimes \lie_{C,\sigma \sigma_1}(D))                                                    \\
			 & =\frac{1}{2}\sum_{\sigma_2=+,-} \lie_{A,\sigma_2}(\lie_{A,\sigma_1}(B))\otimes \lie_{C,\sigma \sigma_2}(\lie_{C,\sigma \sigma_1}(D)) \\
		\end{aligned}
	\end{equation}
	we can get
	\begin{equation}
		\begin{aligned}
			 & \lie_{A\otimes C,\sigma}^k(B\otimes D)                                                                                                            \\
			 & = \frac{1}{2^k} \sum_{\sigma_1,\sigma_2,\cdots,\sigma_k} (\prod_{i=1}^k \lie_{A,\sigma_i})(B) \otimes (\prod_{i=1}^k \lie_{C,\sigma \sigma_i})(D) \\
			 & = \frac{1}{2^k} (\sum_{\sigma_i} \lie_{A,\sigma_i} \otimes \lie_{C,\sigma \sigma_i})^k(B\otimes D)
		\end{aligned}
	\end{equation}
\end{proof}

\begin{lemma}[Lie-Trotter product formula~\cite{hall2013lie}]\label{lemma:lie-trotter-formula}
	For arbitrary operators $A,B \in \hilbert{n}$, we have
	\begin{equation}
		\exp[A+B] = \lim_{n\rightarrow\infty} (\exp[A/n]\exp[B/n])^n
	\end{equation}
	where $\exp[A] = \sum_{k=0}^{\infty} \frac{A^k}{k!}$.
\end{lemma}

\begin{lemma}[Baker-Campbell-Hausdorff formula~\cite{rossmann2006lie}]\label{lemma:bch-formula}
	For arbitrary operators $X,Y \in \hilbert{n}$, we have
	\begin{equation}
		\exp[X] Y \exp[-X] = \sum_{k=0}^{\infty} \frac{1}{k!} \lie_{X,-}^k(Y)
	\end{equation}
\end{lemma}

\begin{lemma}[Von Neumann's trace inequality~\cite{neumann1962inequalities}]\label{lemma:trace-inequality}
	if $A,B$ are complex $n\times n$ matrices with singular values
	\begin{equation}
		\alpha_1 \geq \alpha_2 \geq \cdots \geq \alpha_n \geq 0,\quad
		\beta_1 \geq \beta_2 \geq \cdots \geq \beta_n \geq 0
	\end{equation}
	then
	\begin{equation}
		\abs{\tr(AB)} \leq \sum_{i=1}^n \alpha_i \beta_i
	\end{equation}
\end{lemma}

Equipped with the above lemmas, we can now prove an important series expansion for the time-evolved observable that splits the observable into system and environment parts. Although the following lemma can be seen as a variant of the Dyson series on operators. To the best of our knowledge, we did not find a similar lemma in the literature. Thus, we will introduce the proof of this lemma in the following.

\begin{lemma}[Growing Dyson series]\label{lemma:expansion}
	Given observable defined as $O_{S}\otimes O_{E}$ where $O_S \in \mathcal{A}_n$ is the observable of the system and $O_E \in \mathcal{A}_{N-n}$ is the observable of the environment. Providing the rescalable local Hamiltonian $\mathcal{H}_n = \{H_n, \partial H_n, R_{\step}\}$, denote the corresponding growing operator as $G_{\step}$
	\begin{equation}
		\opmap{G_{\step}}{H_n} = H_n\otimes I + \sum_{B_i\in \partial H_n} B_i\otimes \connsub{\step}{B_i} + I^{\otimes n} \otimes K\\
	\end{equation}	
	and total evolution time as $T$, we can expand the time-evolved observable $O_S(T)\otimes O_E(T)$ as following:
	\begin{equation}
		\begin{aligned}
			\exp{iT \opmap{G_{\step}}{H_n}} O_S\otimes O_E \exp{-iT \opmap{G_{\step}}{H_n}}
			&=\lim_{M\rightarrow\infty} \sum_k
			\frac{\delta^k}{2^k k!} (
			\sum_{m=0}^{M-1} \mathcal{T}_{B(m\delta)}
			)^k (O_S(T)\otimes O_E(T))\\
		\end{aligned}
	\end{equation}
	where $\delta = t / M$, $\mathcal{T}_{B(t)} = \sum_{i,\sigma} \lie_{B_i(t),\sigma}\otimes \lie_{\connsub{\step}{B_i}(t),-\sigma}$, and $\forall t_1 \leq t_2 \in \mathbb{R}$ we define $\mathcal{T}_{B(t_2)} \mathcal{T}_{B(t_1)} = \mathcal{T}_{B(t_1)}\mathcal{T}_{B(t_2)}$, thus the product of $\mathcal{T}$ is time-ordered. And
	\begin{equation}
		\begin{aligned}
			O_S(T) &= \exp{it H} O_S \exp{-it H},\quad O_E(T) = \exp{it K} O_E \exp{-it K}\\
			B_i(t) &= \exp{it H_n} B_i \exp{-it H_n},\quad R[B_i](t) = \exp{it K} R[B_i] \exp{-it K}\\
		\end{aligned}
	\end{equation}
\end{lemma}

\begin{proof}
	The proof uses previous lemmas to expand the operator onto system and environment parts, then simplify the series by reorganizing the summation. First by using \cref{lemma:lie-trotter-formula}, we divide our evolution into small time steps $t/M$
	\begin{equation}
		e^{it \opmap{G_{\step}}{H_n}}O_S\otimes O_E e^{-it \opmap{G_{\step}}{H_n}} = \lim_{M\rightarrow\infty} (e^{it/M \opmap{G_{\step}}{H_n}})^M O_S\otimes O_E (e^{-it/M \opmap{G_{\step}}{H_n}})^M
	\end{equation}
	We can see this product as $M$ steps of time evolution with time step $\delta = t/M$.
	\begin{equation}
		\lim_{M\rightarrow\infty} e^{i\delta \opmap{G_{\step}}{H_n}} (e^{i\delta \opmap{G_{\step}}{H_n}}\cdots (e^{i\delta \opmap{G_{\step}}{H_n}} O_S\otimes O_E e^{-i\delta \opmap{G_{\step}}{H_n}})\cdots e^{-i\delta \opmap{G_{\step}}{H_n}}) e^{-i\delta \opmap{G_{\step}}{H_n}}
	\end{equation}
	Because $\delta \rightarrow 0$, we can move the terms only depending on the system or environment onto the observables, leaving only the boundary terms in the time evolution.
	\begin{equation}
		e^{i\delta \opmap{G_{\step}}{H_n}} O_S\otimes O_E e^{-i\delta \opmap{G_{\step}}{H_n}} = e^{i\delta\sum_i B_i\otimes \connsub{\step}{B_i}} O_S(\delta)\otimes O_E(\delta) e^{-i\delta\sum_i B_i\otimes \connsub{\step}{B_i}}
	\end{equation}
	to further expand the boundary terms, using \cref{lemma:bch-formula} we have
	\begin{equation}
		= \sum_k \frac{(i\delta)^k}{k!} \lie_{\sum_i B_i\otimes \connsub{\step}{B_i},-}^k(O_S(\delta)\otimes O_E(\delta))
	\end{equation}
	and \cref{lemma:adjoint-kron} we have
	\begin{equation}
		= \sum_k \frac{(i\delta)^k}{2^k k!} (\sum_{i,\sigma} \lie_{B_i,\sigma}\otimes \lie_{\connsub{\step}{B_i},-\sigma})^k(O_S(\delta)\otimes O_E(\delta)) = \sum_k \frac{(i\delta)^k}{2^k k!} \mathcal{T}^k_{B(0)}(O_S(\delta)\otimes O_E(\delta))
	\end{equation}
	and because the product of $\mathcal{T}$ is time-ordered, we always do the
	multiplication in the order of time steps, making the product commutative. Now, if we apply
	$e^{i\delta \opmap{G_{\step}}{H_n}} X e^{-i\delta \opmap{G_{\step}}{H_n}}$ again, because $e^{-it H_n}$ cancels
	$e^{it H_n}$, they can be merged into the time evolution of each separate system. Resulting in the following
	\begin{equation}
		\sum_{k_1,k_2} \frac{(i\delta)^{k_1+k_2}}{2^{k_1+k_2} k_1!k_2!} \mathcal{T}^{k_2}_{B(0)} \mathcal{T}^{k_1}_{B(\delta)} (O_L(2\delta)\otimes O_R(2\delta))
	\end{equation}
	Because the product of $\mathcal{T}$ is commutative, reorganizing the summation index as $k=k_1+k_2$ we have
	\begin{equation}\label{eq:timestep-decomp}
		= \sum_k \frac{(i\delta)^k}{2^k k!} (\mathcal{T}_{B(0)} + \mathcal{T}_{B(\delta)})^k (O_L(2\delta)\otimes O_R(2\delta))
	\end{equation}
	By re-using \cref{eq:timestep-decomp} iteratively, we reach the general form
	\begin{equation}
		\exp{it \opmap{G_{\step}}{H_n}} O_S\otimes O_E \exp{-it \opmap{G_{\step}}{H_n}}
		=\lim_{M\rightarrow\infty} \sum_k
		\frac{\delta^k}{2^k k!} (
		\sum_{m=0}^{M-1} \mathcal{T}_{B(m\delta)}
		)^k (O_S(t)\otimes O_E(t))
	\end{equation}
\end{proof}

\cref{lemma:expansion} is exactly the series expansion we are looking for. Before proceeding into the proof, we introduce the notion of multi-index for convenience.

\begin{notation}[Multi-index]
	The multi-index sum and product are defined as the following
	\begin{equation}
		\begin{aligned}
			\sum_{\mathbf{i}}                         & = \sum_{i_1,i_2,\cdots,i_k}                              \\
			\prod_{\mathbf{i}} A_i                    & = A_{i_1}A_{i_2}\cdots A_{i_k}                           \\
			\sum_{\mathbf{i}}\prod_{\mathbf{i}} A_{i} & = \sum_{i_1,i_2,\cdots,i_k} A_{i_1}A_{i_2}\cdots A_{i_k}
		\end{aligned}
	\end{equation}
\end{notation}

Now, we can check if summing up the components of the environment converges to a value.

\begin{definition}[Time-Ordered Boundary Correlator]\label{def:appx-tobc}
	Given a rescalable Hamiltonian $\mathcal{H}_n = \{H_n, \partial H_n, R_{\step}\}$ and an observable $O$ and an initial state $\rho$ on the current scale, denote the corresponding growing operator as $G_{\step}$, the total evolution time as $T$, we can define the $k$-th order Time-Ordered Boundary Correlator (TOBC) as following:
	\begin{equation}
		\expect{\chi_{\bm{i,t,\sigma}}(\mathcal{S}_n, T)} = \expect{\chi_{\bm{i,t,\sigma}}(\rho, \mathcal{H}_n, O, T)} = tr(\rho [\mathcal{T}\prod_{\bm{i,t,\sigma}}\lie_{B_i(t),\sigma}] O(T))
	\end{equation}
	where $\mathcal{S}_n$ is the corresponding rescalable system, $B_i\in(\partial H_n)^{G_{\step}}$, $t\in[0,T]$, $\sigma\in\{+1,-1\}$, and $\bm{i,t,\sigma}$ is a multi-index of size $k$. The product is time-ordered, i.e. $\lie_{B_i(t_1),\sigma_1}\lie_{B_j(t_2),\sigma_2} = \lie_{B_j(t_2),\sigma_2}\lie_{B_i(t_1),\sigma_1}$ for $t_1 > t_2$.
\end{definition}

From a physics perspective, TOBC describes how the environment affects the system. Higher order TOBC corresponds to longer-time, longer-distance correlations. It is derived from the following theorem.

\begin{theorem}[Real-time $\epsilon$-scaling consistency]\label{thm:appx-rt-scaling-consistency}
	Given a $\locality$-local rescalable Hamiltonian $\mathcal{H}_n = \{H_n, \partial H_n, R_k\}$ that has a saturated boundary set $(\partial H_n)^{G_{\step}}$. If $\exists \epsilon > 0$ for $\rho = \rho_S\otimes\rho_E, O = O_S\otimes O_E$ and $\forall \bm{i,t,\sigma}$ such that
	\begin{equation}
		\norm{\expect{\chi_{\bm{i,t,\sigma}}(\mathcal{S}_n, T)} - \expect{\chi_{\bm{i,t,\sigma}}(\opmap{\ansatz{}}{\mathcal{S}_n}, T)}} \leq \epsilon
	\end{equation}
	then for $N=n+kq$, the error of expectation $\opmap{p_N}{\opmap{G_k^q}{S_n}} = \expect{\rho e^{itH_{N}} O e^{-itH_{N}}}$ is bounded by
	\begin{equation}
		\norm{\opmap{p_N}{\opmap{G_k^q}{S_n}} - \opmap{p_N}{\opmap{(G_k^{q-1}\circ D_k)}{S_n}}} \leq \epsilon C \exp{T \norm{(\partial H_n)^{G_{\step}}} C / 2}
	\end{equation}
	where $C$ is the maximum of $\max \{\norm{\conn{B_i}}_{\infty}\mid B_i \in \partial H_n\}$ and $\norm{O_R}_{\infty}$.
\end{theorem}

\begin{proof}
	Using \cref{lemma:expansion} and \cref{cor:saturated-grow-op} on $G_{\step}^q$ we have
	\begin{equation}
		\exp(it \opmap{G_{\step}^q}{H_n}) O_S\otimes O_E \exp(-it \opmap{G_{\step}^q}{H_n}) = \sum_k \frac{(i\delta)^k}{2^k k!} (\sum_{m=0}^{M-1}\mathcal{T}_{B(m\delta)})^k (O_S(T)\otimes O_E(T))
	\end{equation}
	where $B_i \in (\partial H_n)^{G_{\step}}$ instead of $\partial H_n$, checking the $k$-th order, where the $\mathcal{T}$ on the right denotes the products are time-ordered, we have
	\begin{equation}
		(\sum_{m=0}^{M-1}\mathcal{T}_{B(m\delta)})^k =\mathcal{T}(\sum_{i,m,\sigma} \lie_{B_i(m\delta),\sigma}\otimes \lie_{\conn{B_i}(m\delta),-\sigma})^k
	\end{equation}
	we can expand the power of sum into the sum of tensor products on the system and environment
	\begin{equation}
		\sum_{\mathbf{i,m},\bm{\sigma}} \mathcal{T}\prod_{\mathbf{i,m},\bm{\sigma}}\lie_{B_i(m\delta),\sigma}\otimes \lie_{\conn{B_i}(m\delta),-\sigma}
	\end{equation}
	here, the multi-index notion is defined as the following
	\begin{equation}
		\sum_{\mathbf{i,m},\bm{\sigma}}\mathcal{T}\prod_{\mathbf{i,m},\bm{\sigma}} A_{i,m,\sigma} = \sum_{i_1,i_2,\cdots,i_k}\sum_{m_1,m_2,\cdots,m_k}\sum_{\sigma_1,\sigma_2,\cdots,\sigma_k} \mathcal{T}A_{i_1,m_1,\sigma_1}A_{i_2,m_2,\sigma_2}\cdots A_{i_k,m_k,\sigma_k}
	\end{equation}
	applying the $k$-th order back to the observable $O_S(T)\otimes O_R(T)$, we obtained the observables on the system and environment
	\begin{equation}
		\sum_{\bm{i,m,\sigma}} [\mathcal{T}\prod_{\bm{i,m,\sigma}}\lie_{B_i(m\delta),\sigma}(O_S(T))]\otimes [\mathcal{T}\prod_{\bm{i,m,\sigma}}\lie_{\conn{B_i}(m\delta),-\sigma}(O_E(T))]
	\end{equation}
	taking the expectation, the left-hand side of the tensor product is the $k$-th order TOBC, which we assume to be bounded by $\epsilon$, and the right-hand side is the observable on the environment. Thus, we can write the expectation as
	\begin{equation}
		\begin{aligned}
			 & \opmap{p_N}{\opmap{G_k^q}{S_n}} = tr([\rho_E\otimes\rho_S] \exp(it \opmap{G_k^q}{H_n}) [O_S\otimes O_E] \exp(-it \opmap{G_k^q}{H_n}))                                                                                                        \\
			 & = \sum_k \frac{(i\delta)^k}{2^k k!} \sum_{\bm{i,m,\sigma}} tr[\rho_S\mathcal{T}\prod_{\bm{i,m,\sigma}}\lie_{B_i(m\delta),\sigma}(O_S(T))]\cdot tr[\rho_E\mathcal{T}\prod_{\bm{i,m,\sigma}}\lie_{\conn{B_i}(m\delta),-\sigma}(O_E(T))] \\
			 & = \sum_k \frac{(i\delta)^k}{2^k k!} \sum_{\bm{i,m,\sigma}} \expect{\chi_{\bm{i,m,\sigma}}(\mathcal{S}_n, T)}\cdot tr[\rho_E\mathcal{T}\prod_{\bm{i,m,\sigma}}\lie_{\conn{B_i}(m\delta),-\sigma}(O_E(T))]                \\
		\end{aligned}
	\end{equation}
	Next, because $\ansatz{}$ only applies to the operators in the system, leaving the environment untouched, we can write the error of expectation as
	\begin{equation}
		\begin{aligned}
			&\norm{\opmap{p_N}{\opmap{G_k^q}{S_n}} - \opmap{p_N}{\opmap{(G_k^{q-1}\circ D_k)}{S_n}}} =\\
			 & \norm{\sum_k \frac{(i\delta)^k}{2^k k!} \sum_{\bm{i,m,\sigma}} (\expect{\chi_{\bm{i,t,\sigma}}(\mathcal{S}_n, T)} - \expect{\chi_{\bm{i,t,\sigma}}(\opmap{\ansatz{}}{\mathcal{S}_n}, T)})\cdot tr[\rho_E\mathcal{T}\prod_{\bm{i,m,\sigma}}\lie_{\conn{B_i}(m\delta),-\sigma}(O_E(T))]} \\
		\end{aligned}
	\end{equation}
	Using triangular inequality, we have
	\begin{equation}
		\begin{aligned}
			&\norm{\opmap{p_N}{\opmap{G_k^q}{S_n}} - \opmap{p_N}{\opmap{(G_k^{q-1}\circ D_k)}{S_n}}}\leq\\
			& \sum_k \frac{\delta^k}{2^k k!} \norm{\sum_{\bm{i,m,\sigma}} (\expect{\chi_{\bm{i,t,\sigma}}(\mathcal{S}_n, T)} - \expect{\chi_{\bm{i,t,\sigma}}(\opmap{\ansatz{}}{\mathcal{S}_n}, T)})\cdot tr[\rho_E\mathcal{T}\prod_{\bm{i,m,\sigma}}\lie_{\conn{B_i}(m\delta),-\sigma}(O_E(T))]}
		\end{aligned}
	\end{equation}
	Because we assume the TOBC is bounded by $\epsilon > 0$ and $\norm{a\cdot b} =\norm{a}\cdot \norm{b}$, we have
	\begin{equation}
		\norm{\opmap{p_N}{\opmap{G_k^q}{S_n}} - \opmap{p_N}{\opmap{(G_k^{q-1}\circ D_k)}{S_n}}}\leq \epsilon \sum_k \frac{\delta^k}{2^k k!} \norm{\sum_{\bm{i,m,\sigma}}tr[\rho_E\mathcal{T}\prod_{\bm{i,m,\sigma}}\lie_{\conn{B_i}(m\delta),-\sigma}(O_E(T))]}
	\end{equation}
	Notice that sum over all possible commutator and anti-commutator in $\sum_{\bm{i,m,\sigma}}$ recovers the product of operators in application order. Applying the triangular inequality again, we have
	\begin{equation}
		\norm{\sum_{\bm{i,m,\sigma}}tr[\rho_E\mathcal{T}\prod_{\bm{i,m,\sigma}}\lie_{\conn{B_i}(m\delta),-\sigma}(O_E(T))]} \leq \sum_{\mathbf{i,m}}\norm{tr[\rho_E\mathcal{T}(\prod_{\mathbf{i,m}}\conn{B_i}(m\delta)) O_E(T)]}
	\end{equation}
	Using \cref{lemma:trace-inequality} on the trace, we have
	\begin{equation}
		\norm{tr[\rho_E\mathcal{T}(\prod_{\mathbf{i,m}}\conn{B_i}(m\delta)) O_E(T)]} \leq \sum_a \alpha_a \beta_a
	\end{equation}
	where $\alpha_a$ is the eigenvalues of $\rho_E$ and $\sum_a \alpha_a = 1$ by definition of density matrix, $\beta_a$ is the eigenvalues of $\mathcal{T}\prod_{\bm{i,m,\sigma}}\lie_{\conn{B_i}(m\delta),-\sigma}(O_E(T))$, taking the maximum of $\beta_a$ we have
	\begin{equation}
		\norm{tr[\rho_E\mathcal{T}(\prod_{\mathbf{i,m}}\conn{B_i}(m\delta)) O_E(T)]} \leq
		\beta_{\text{max}} \sum_a \alpha_a = \beta_{\text{max}}
	\end{equation}
	$\beta_{\text{max}}$ is equivalent to the operator 2-norm of the operator, thus
	\begin{equation}
		\norm{\opmap{p_N}{\opmap{G_k^q}{S_n}} - \opmap{p_N}{\opmap{(G_k^{q-1}\circ D_k)}{S_n}}} \leq \epsilon \sum_k \frac{\delta^k}{2^k k!} \sum_{\mathbf{i,m}} \norm{\mathcal{T}(\prod_{\mathbf{i,m}}\conn{B_i}(m\delta)) O_E(T)}_{\text{op}}
	\end{equation}
	Next, we can replace the variable $\delta \rightarrow \frac{\delta}{2}$ and rescale the environment Hamiltonian $K\rightarrow 2K$, thus $\conn{B_i}(m\delta)\rightarrow \conn{B_i}(m\delta/2)$ and $O_E(T)\rightarrow O_E(T/2)$, this allows us to remove the factor of $2^k$ in the summation so that we can find the summation later, now we have
	\begin{equation}
		\norm{\opmap{p_N}{\opmap{G_k^q}{S_n}} - \opmap{p_N}{\opmap{(G_k^{q-1}\circ D_k)}{S_n}}} \leq \epsilon \sum_k \frac{(\delta/2)^k}{k!} \sum_{\mathbf{i,m}} \norm{\mathcal{T}(\prod_{\mathbf{i,m}}\conn{B_i}(m\delta)) O_E(T)}_{\text{op}}
	\end{equation}
	note that the Hamiltonian for $\conn{B_i}(m\delta/2)$ and $O_E(T/2)$ here is $2K$ instead of $K$. Next, we can use the sub-multiplicative of norm $\norm{A B}\leq \norm{A}\norm{B}$ This allows us further to break the product into the norm of each operator
	\begin{equation}
		\norm{\opmap{p_N}{\opmap{G_k^q}{S_n}} - \opmap{p_N}{\opmap{(G_k^{q-1}\circ D_k)}{S_n}}} \leq \epsilon \sum_k \frac{(\delta/2)^k}{k!} \sum_{\mathbf{i,m}} \mathcal{T}(\prod_{\mathbf{i,m}}\norm{\conn{B_i}(m\delta/2)}_{\text{op}}) \norm{O_E(T/2)}_{\text{op}}
	\end{equation}
	Because time evolution is unitary, the norm of the operator is preserved, thus $\norm{\conn{B_i}(m\delta/2)}_{\text{op}} = \norm{\conn{B_i}}_{\text{op}}$, and $\norm{O_E(T/2)}_{\text{op}} = \norm{O_E(T)}_{\text{op}}$, thus if we have $C = \max\{\norm{\conn{B_i}}_{\text{op}}\mid B_i \in \partial H_n\}$ and $\norm{O_R}_{\text{op}}$, we have
	\begin{equation}
		\norm{\opmap{p_N}{\opmap{G_k^q}{S_n}} - \opmap{p_N}{\opmap{(G_k^{q-1}\circ D_k)}{S_n}}} \leq \epsilon \sum_k \frac{(\delta/2)^k}{k!} \sum_{\mathbf{i,m}} C^{k+1}
	\end{equation}
	Now we can sum over the multi-index $\mathbf{i,m}$, the sum of $\mathbf{m}$ is $M^k$ and the sum of $\mathbf{i}$ is $|\partial H_n|^k$, thus we have
	\begin{equation}
		\norm{\opmap{p_N}{\opmap{G_k^q}{S_n}} - \opmap{p_N}{\opmap{(G_k^{q-1}\circ D_k)}{S_n}}} \leq \epsilon C \sum_k \frac{(\delta M/2)^k}{k!} (|\partial H_n|C)^k = \epsilon C \exp{T \norm{(\partial H_n)^{G}} C / 2}
	\end{equation}
\end{proof}

\section{Effective Pulse Duration for the Quantum Algorithm}\label{appx:complexity}

Denote the checkpoints for $k$-th order TOBC as $t_1,\cdots,t_k$, and assuming the constant overhead in product formula for implementing $O(\norm{(\partial H_n)^{G_{\step}}})$ number of $w$-qubit gates result in total. The total effective pulse duration is $O(C_{\theta} \tau(\norm{(\partial H_n)^{G_l}}) t)$ for the evolution $\exp{-it G_{\step}(H_n^{\text{dev}})}$. $C_{\theta}$ is the overhead due to variational optimized pulse in effective pulse duration. The prefactor $\tau(\norm{(\partial H_n)^{G_l}})$ depends on the product formula, e.g for 1st-order trotterization $\tau(\norm{(\partial H_n)^{G_l}}) = \norm{(\partial H_n)^{G_l}}$. Thus the total effective pulse duration for a single $k$-th order TOBC with 1st-order trotterization is $C_{\theta} \norm{(\partial H_n)^{G_l}} (2 t_1 + 2 t_2 + \cdots + t_k + T - t_k + T) = C_{\theta} \norm{(\partial H_n)^{G_l}} (2T + 2\sum_{i=1}^{k-1} t_i)$. The worst case effective pulse duration is $O(2 C_{\theta} \norm{(\partial H_n)^{G_l}} kT)$. Because we are sampling $b$ TOBCs for each loss function, we now analyze the average effective pulse duration for a single loss function. For $M$ checkpoints, the average effective pulse duration is $C_{\theta}\norm{(\partial H_n)^{G_l}} \frac{2T + 2\sum_{i=1}^{k-1} t_i}{M^k}$ thus summing over all the $k$-th order TOBCs, we have
\begin{equation}
	\begin{aligned}
		\sum_{t_1,\cdots,t_k} \frac{2T + 2\sum_{i=1}^{k-1} t_i}{M^k}
		&= \sum_{t_2,\cdots,t_k} \frac{2MT + 2 (T(1+M)/2 + M\sum_{i=2}^{k-1} t_i)}{M^k}\\
		&= \sum_{t_3,\cdots,t_k} \frac{2M^2T + 2 (TM(1+M)/2 + TM(1+M)/2 + M^2\sum_{i=3}^{k-1} t_i)}{M^k}\\
		&= \frac{2M^k T + (k-1) TM^{k-1}(1+M)}{M^k} = (2 + (k-1) \frac{1+M}{M})T
	\end{aligned}
\end{equation}
Taking $M\gg 1$, we have the average effective pulse duration as $O(C_{\theta}\norm{(\partial H_n)^{G_l}} (k+1)T) = O(C_{\theta}\norm{(\partial H_n)^{G_l}} kT)$ when using 1st order trotterization. This removes $n$ from the effective pulse duration when comparing to pure trotterization. However, this does not mean we break the optimal bounds such as~\cite{haah2021quantum}. Part of the complexity is moved into $C_{\theta}$ which becomes heuristic.

\section{Improving Loss Function}\label{appx:improving-loss-function}

\begin{figure}[H]
	\includegraphics[scale=0.4]{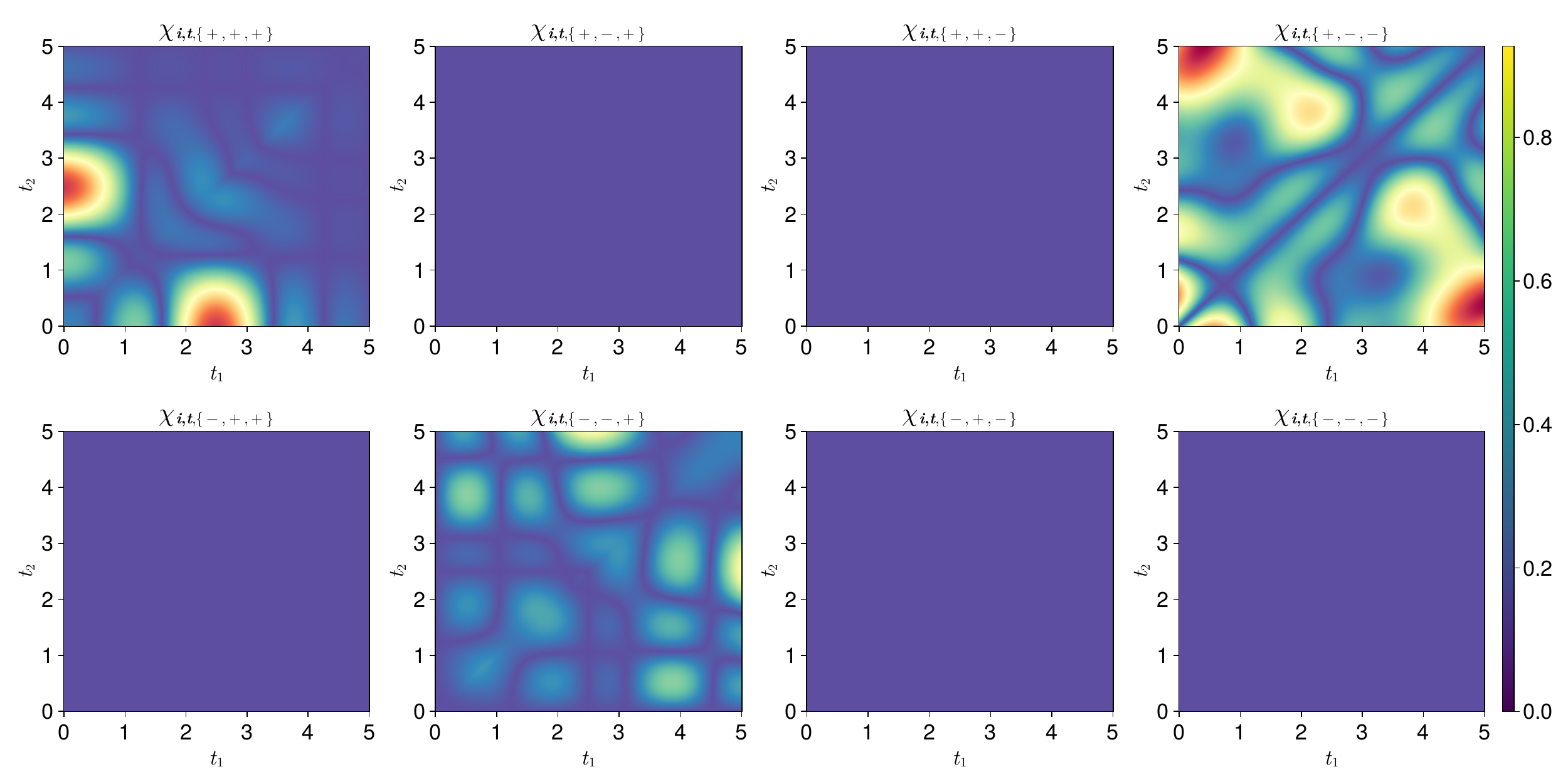}
	\caption{
		3rd-order TOBC for 5-site $1D$ TFIM at $T=5.0$ for the two-point correlation function $\langle Z_1 Z_2\rangle_{T=5.0}$ with $\ket{00000}$ as initial state and $h=1.0$. We fix $t_3 = 2.5$ and plot the TOBC for $t_1,t_2\in[0,5.0]$.
	}\label{fig:3rd-tobc}
\end{figure}

The theoretical bounds we present for real-time evolution in \cref{thm:rt-scaling-consistency} is a general estimation for arbitrary geometrically local Hamiltonian. Thus, it is rather a loose bound considering more specific system properties. We believe that a tighter bound can be established for specific system properties. This may result in a better loss function and a more efficient algorithm. Furthermore, the global loss function and modeling by e2e provide advantages in that every learning step optimizes the target problem but also has limitations~\cite{glasmachers2017limits}. The usage of e2e heavily relies on optimization and thus may result in a slow convergence and ill-conditioned optimization. Our framework also allows theoretical improvements through a better theoretical understanding of the problem, such as the analytical or heuristic understanding of the TOBCs. This will incorporate the theoretical knowledge into the loss function and thus may potentially improve the algorithm's efficiency. For example, as shown in \cref{fig:tobc-2} and \cref{fig:3rd-tobc}, some TOBCs can be almost perfectly zero, and the nonzero TOBCs are also very sparse in $1D$ TFIM dynamics, where many points are relatively small thus result in a small contribution to the loss function. This suggests that by looking into specific Hamiltonian and TOBCs, we may be able to design a better loss function that can be more efficient in practice.

On the other hand, there is a different possible style of constructing the loss function by reusing existing scales. Using the same superblock construction as DMRG, one can see such loss function as the following concept. If we can prepare good representations of the systems at size $n_1, n_2$, then concatenating them into a system at size $n_1 + n_2$ should be a good representation of the system at size $n_1 + n_2$. This is a very natural assumption, and it is also the core idea of superblocks. From a series expansion perspective, assuming we have the series expansion of a property $p$ as

\begin{equation}
	p(S_{n_1}\otimes S_{n_2}) = \sum_{i=0}^{\infty} \alpha_i \langle A_i\rangle\langle B_i\rangle
\end{equation}
Here, we can use the infinite DMRG style loss function as an example. If we are promised to have a good representation of $S_{n_1}$,
copying it then the concatenating system $S_{n_1}\otimes S_{n_1}$ should be a good representation of $2 n_1$ system. This is exactly the description of infinite DMRG~\cite{schollwock2005density} in the traditional fashion. If we assume the property we are calculating is the same observable. The series expansion becomes a sum of square expressions:

\begin{equation}
	p(S_{n_1}\otimes S_{n_1}) = \sum_{i=0}^{\infty} \alpha_i \langle A_i\rangle^2
\end{equation}

If the loss function is then set as the error of observables on this superblock, we can see we are optimizing the error of some high-order terms in the Dyson series.

\begin{equation}
	\begin{aligned}
		L & = \norm{p(S_{n_1}\otimes S_{n_1}) - p(S_{n_1}^{\prime}\otimes S_{n_1}^{\prime})}                                        \\
		  & = \norm{\sum_{i=0}^{\infty} \alpha_i \langle A_i\rangle^2 - \sum_{i=0}^{\infty} \alpha_i \langle A_i^{\prime}\rangle^2}
	\end{aligned}
\end{equation}

However, it remains uncertain whether this loss function is upper-bounded in a manner that would rigorously ensure the scaling consistency condition. Utilizing superblocks in practice will improve the efficiency of evaluating the loss function, as now one does not need to solve time-evolved operators but can directly evaluate the discrepancies between expectation values in superblocks instead. This is similar to using the superblock in DMRG to evaluate the system's energy.

\section{Improving Operator Maps}\label{appx:ansatz-improvements}

Like all other variational algorithms, the expressiveness of the operator map is crucial to the algorithm's performance. In our current implementation, we only use some vanilla operator maps without much careful design.

For OMM, the power of optimizing operator maps, such as deep neural networks or tensor network ensembles, is yet to be explored. Furthermore, one can use different operator maps for larger system sizes for different scales and only share at closer scales. This naturally creates a hierarchical structure of the operator map, similar to how depth of neural networks are used in deep learning~\cite{LeCun2015deep}. As we now utilize an operator map for operator mapping rather than a state, the expressiveness of such an operator mapping remains to be determined. More specifically, although our neural OMM has the potential of expressiveness representing MPS with exponential large bonds. Because they are no longer ansatzes for states, it is unclear what the limit of such operator maps is.

For HEM, we only use a simple and small neural network to parameterize the pulse, which does not consider more realistic pulse shapes. Thus, the result pulse shape does not necessarily execute on real hardware due to violation of hardware constraints. On the other hand, our HEM targets a non-universal Hamiltonian, thus resulting in worse performance over a longer time despite increasing the order. It is important to explore more realistic pulse shapes, universal Hamiltonians, and more detailed device capabilities to develop a better understanding of the algorithm.

\section{Finding the Loss Function for Other Properties}\label{appx:loss-for-other}
We have demonstrated the existence of a loss function that effectively bounds the real-time dynamics of local observables. However, this methodology might not be entirely end-to-end (e2e) when dealing with target properties that are complex functions not directly derived from local observables, such as phase transition points or entanglement entropy. Additionally, our \cref{thm:rt-scaling-consistency} does not extend to imaginary-time evolution or ground-state simulations. Because the series expansion we derived does not hold in these cases. Despite these limitations, our framework is not intrinsically confined to real-time dynamics alone, as suggested by \cref{thm:sc-error}. The proven effectiveness of NRG and DMRG inspires the possibility that suitable loss functions for imaginary time and ground state challenges may also exist. To further follow the e2e principle in solving real-world quantum many-body simulation problems, we seek if there is a loss function that guarantees scaling consistency for other properties such as entanglement entropy, phase transition points, etc.

\section{Higher Dimension Lattice and Other Geometry}\label{appx:higher-dimension-lattice}

While our numerical results are confined to a $1D$ lattice in \cref{sec:results}, it's important to note that, like NRG and DMRG, the variational principle, OLRG framework is not inherently limited to this geometry. Indeed, the OLRG framework can be applied to any geometric configuration. However, in geometries other than $1D$, the implementation of the growing operator presents a range of alternative strategies that have yet to be fully explored. Moreover, by incorporating the growing scheme into the loss function, our operator map no longer necessitates an exponential increase in storage, provided that $\ansatz{n_q}$ is not a dense isometric map. Consequently, techniques developed for navigating $1D$ ansatzes, such as Matrix Product State (MPS)~\cite{stoudenmire2012studying} and autoregressive neural networks~\cite{PhysRevLett.128.090501}, could be adapted and prove beneficial in 2D and other configurations within this framework.

\section{Relation with MPS TDVP}\label{appx:mps-tdvp}
When $\ansatz{n_q}$ is a linear map, the OLRG framework is equivalent to a tensor network. For example, if $\ansatz{n_q}$ is not shared by each OLRG step, and denoting $\ansatz{n_q}$ for $q$-th OLRG step, the set $\{\ansatz{n_q}\}$ represents the tensors in an MPS as shown in the left column of \cref{fig:generic-workflow}. On the other hand, the MPS TDVP algorithm projects an MPS $\ket{\psi(t)}$ at time $t$ to the MPS $\ket{\psi(t+\delta)}$ at time $t + \delta$ by solving the Schrödinger equation in the subspace of MPS. Assuming the bond dimension does not change from $\ket{\psi(t)}$ to $\ket{\psi(t+\delta)}$, the MPS TDVP should find the optimal MPS representation for $\ket{\psi(t+\delta)}$. By optimal $\ket{\psi(t+\delta)}$ we mean this state has the minimum error for arbitrary observables at $t+\delta$. Thus, this is equivalent to optimizing support of observables at $t+\delta$ using OLRG starting from the initial point $\ket{\psi(t)}$, which is the transfer learning algorithm we introduced in \cref{sec:results-transfer-learning}.

From this perspective, plugging the small duration $\delta$ into \cref{thm:appx-rt-scaling-consistency}, we can see that the loss function of the MPS TDVP algorithm directs the optimization towards the $t+\delta$ time observables instead of the final time $T$ observables. Thus, only the error of $\chi_{\bm{i},\bm{t}=\{\delta\}, \sigma}(S_n, t+\delta)$ are optimized in the MPS TDVP algorithm for an arbitrary observable $O(t)$, thus missing longer time correlations in the optimization target. This is consistent with the recent analysis of the MPS TDVP algorithm in the ancillary Krylov subspace TDVP~\cite{Yang2020TimedependentVP}.

With this observation, we can see that the OLRG framework can be a complementary approach to the MPS TDVP algorithm. A potential improvement to the MPS TDVP algorithm is to add the loss function of the OLRG framework as a regularization term for long-time TOBCs, and thus help the MPS TDVP algorithm to include long-time correlations in the optimization target and increase the efficiency of the MPS TDVP algorithm for longer time by increasing the step size $\delta$. However, the gradient-based optimization in OLRG also has its limitations, as it may not be able to adjust bond dimension variationally, thus for pure MPS, it can only be used as a regularization term instead of the main optimization target.

\section{Implementation}\label{appx:impl}

Both classical and quantum algorithms were optimized using the ADAM optimizer~\cite{loshchilov2017decoupled} and implemented via the recent automatic differentiation frameworks and GPU computing \texttt{jax}~\cite{jax2018github} and \texttt{flax}~\cite{flax2020github} frameworks. Due to the absence of sufficient sparse matrix support when the author implements the software in \texttt{jax}, a brute-force solver was employed to compute the observables. This limitation restricted the quantum algorithm's simulation to no more than 6 sites due to memory limitation. The classical algorithms use single NVIDIA GPUs, while the quantum algorithms are executed on 1 CPU cores. From an implementation perspective, the OLRG framework opens the door to adapting good small-system solvers to larger systems. Thus, like DMRG, all the technologies developed for small-system solvers can be transferred into large system calculations. We believe that by integrating with better small-system solvers, the practical performance of the OLRG framework can be further improved. We thank the support of the open-source community in the development of the following software, they contributed directly in producing our results: \texttt{jax}~\cite{jax2018github}, \texttt{flax}~\cite{flax2020github}, \texttt{optax}~\cite{deepmind2020jax}, \texttt{tqdm}, \texttt{wandb}, \texttt{matplotlib}~\cite{Hunter:2007}, \texttt{Yao}~\cite{Luo2020yaojlextensible}, \texttt{Makie}~\cite{DanischKrumbiegel2021}.

\newpage

\section{Additional Results}\label{appx:results}
\subsection{Training History}\label{appx:training-history}

\begin{figure}[H]
	\includegraphics[scale=0.75]{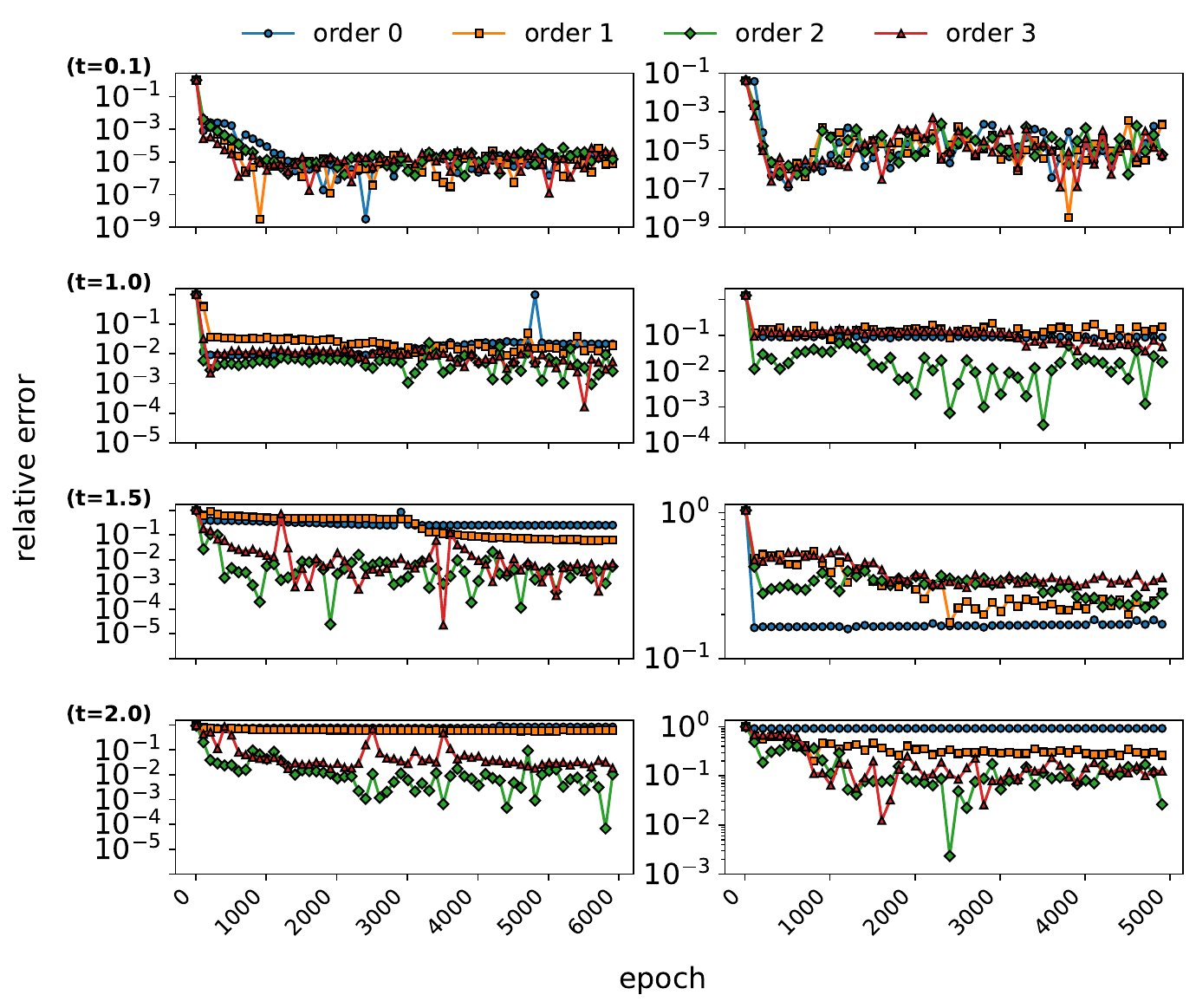}
	\caption{
		Training history of relative error. Left is the training history of the classical algorithm, and right is the training history of the quantum algorithm.
	}
	\label{fig:training-history-order-error}
\end{figure}

\begin{figure}[H]
	\includegraphics[scale=0.75]{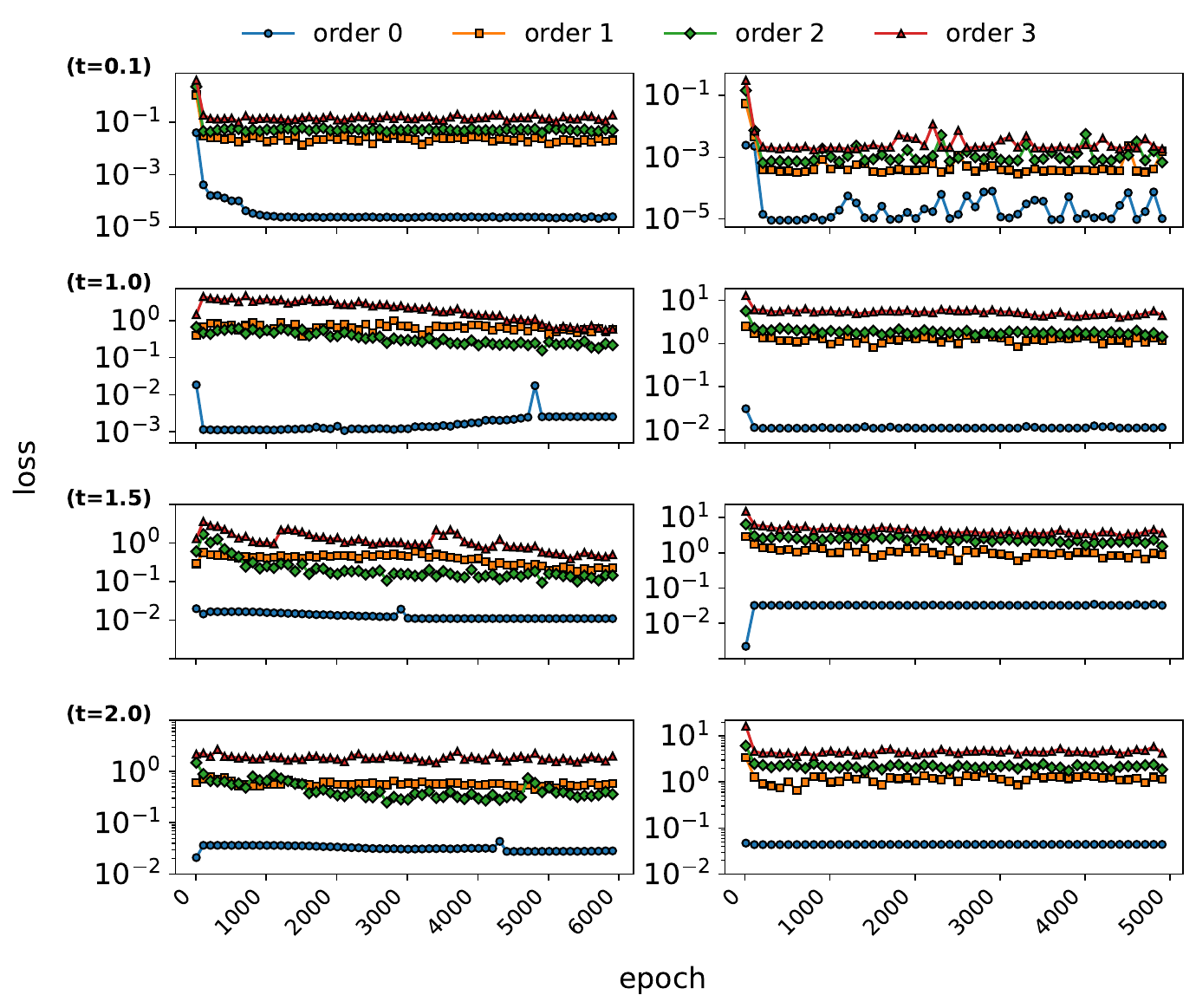}
	\caption{
		Training history of the loss function. Left is the training history of the classical algorithm, and right is the training history of the quantum algorithm.
	}
	\label{fig:training-history-order-loss}
\end{figure}

The training history at different orders of OMM and HEM are shown in \cref{fig:training-history-order-error} and \cref{fig:training-history-order-loss}. The left column is OMM, and the right is HEM. Each row is ordered by time. In a short time, there isn't a significant difference between each order. However, with time increase, the higher order loss function approaches higher precision faster and can reach significantly higher precision over a long time than the lower order loss function.

\begin{figure}[H]
	\includegraphics[scale=0.65]{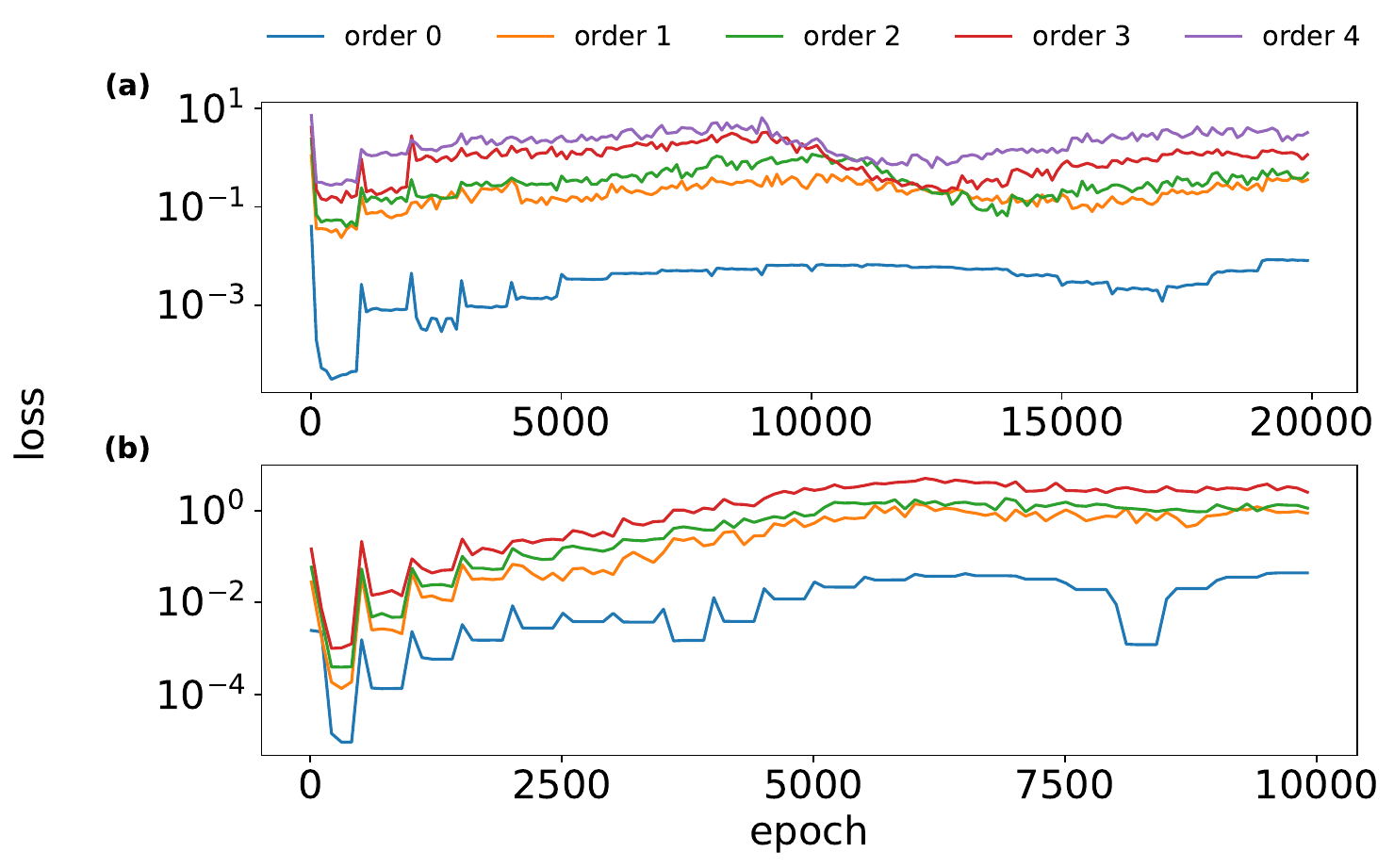}
	\caption{
		Training history of the training by reusing previous time point's parameters.
		(a) The history of loss function for OMM. (b) The history of loss function for HEM.
	}
	\label{fig:training-history-sweep-order}
\end{figure}

We also show the history of the loss function when we reuse the previous time point's parameters in \cref{fig:training-history-sweep-order}. The left column is the OMM, and the right is the HEM. Each row is ordered by time. The loss function is larger at higher order, we suspect the higher order TOBC is harder to optimize because the search space is larger than the lower order thus resulting in a worse absolute value of the loss function but a better relative error.

\newpage

\subsection{Batch and Sampling Size}\label{appx:hyper-batch-sampling-size}

\begin{figure}[H]
	\centering
	\begin{minipage}{0.45\textwidth}
		\includegraphics[scale=0.35]{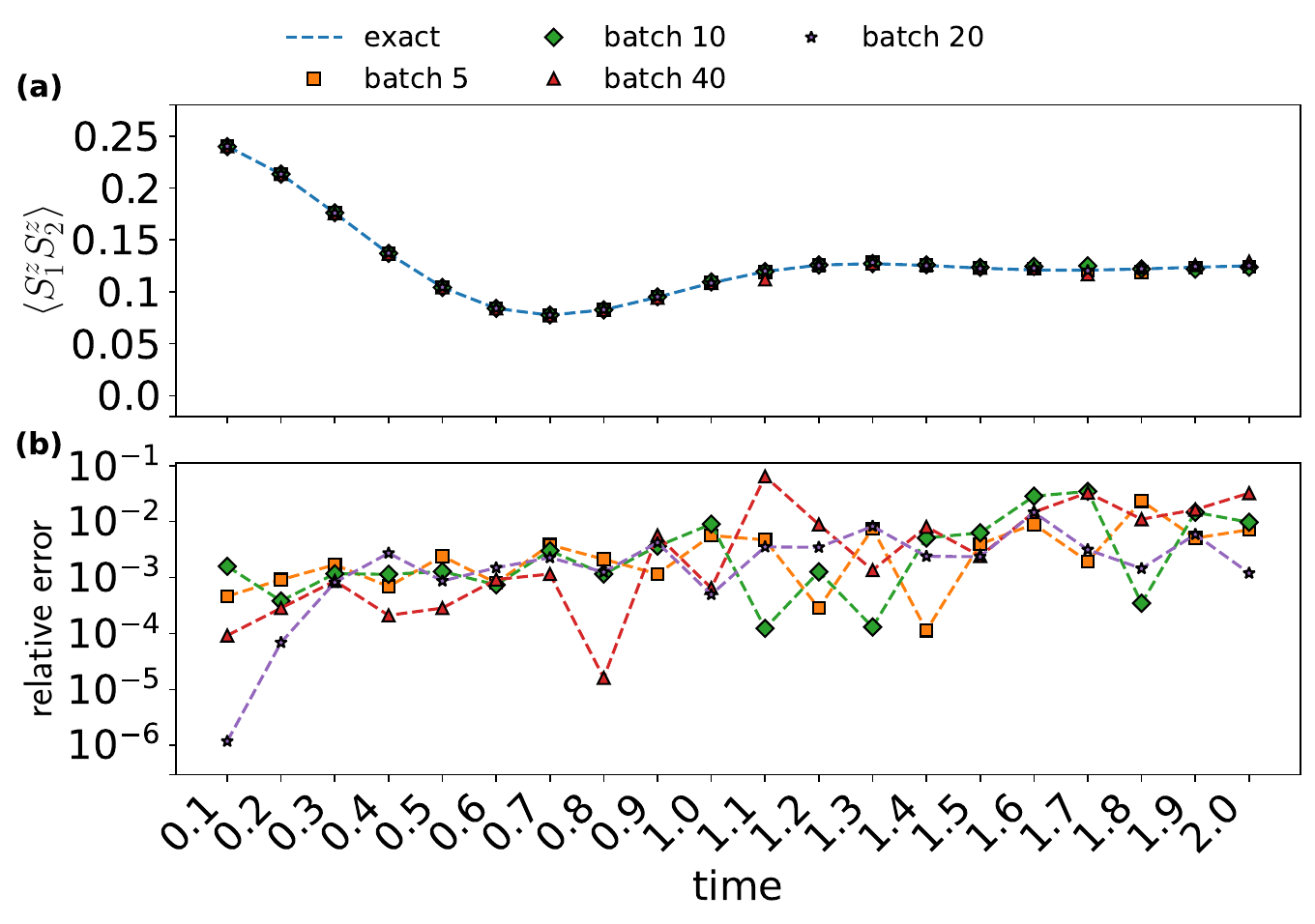}
		\caption{
			Comparison of different batch sizes at order 2, with depth 8 for OMM. (a) The value of $\langle S_1^z S_2^z\rangle$; (b) the relative error.
		}\label{fig:ensemble-batch}
	\end{minipage}\hfill
	\begin{minipage}{0.45\textwidth}
		\includegraphics[scale=0.35]{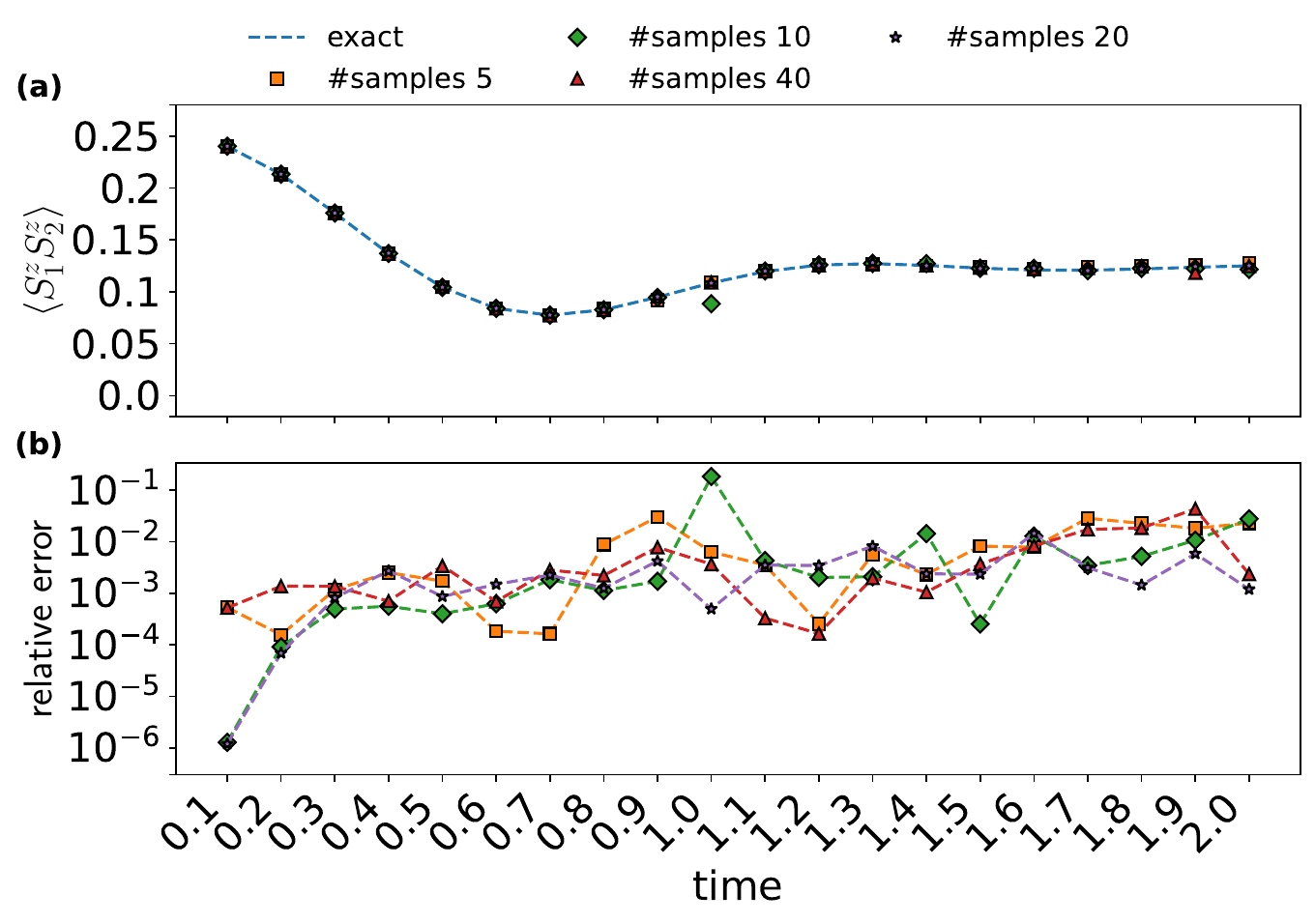}
		\caption{
			Comparison of different sampling sizes at order 2, with depth 8 for OMM. (a) The value of $\langle S_1^z S_2^z\rangle$; (b) the relative error.
		}\label{fig:ensemble-samples}
	\end{minipage}
\end{figure}

We also compared the batch size and sampling size. The batch size controls the sampling error of the neural OMM. Increasing batch size will increase the precision of gradient estimation. We did not observe a significant difference in tuning batch size. The results are shown in \cref{fig:ensemble-batch}. The sampling size controls how many observables are sampled to estimate the loss function at each evaluation. Increasing the sampling size should decrease the variance in the gradient. We did not observe a significant difference in tuning sampling size. The results are shown in \cref{fig:ensemble-samples}. We hypothesize that the variance of the loss function is useful for SGD exploring better minimum. On the other hand, the toy problem we ran our simulation with might be too simple to demonstrate the difference between these two hyperparameters.

\begin{figure}[H]
	\centering
	\includegraphics[scale=0.35]{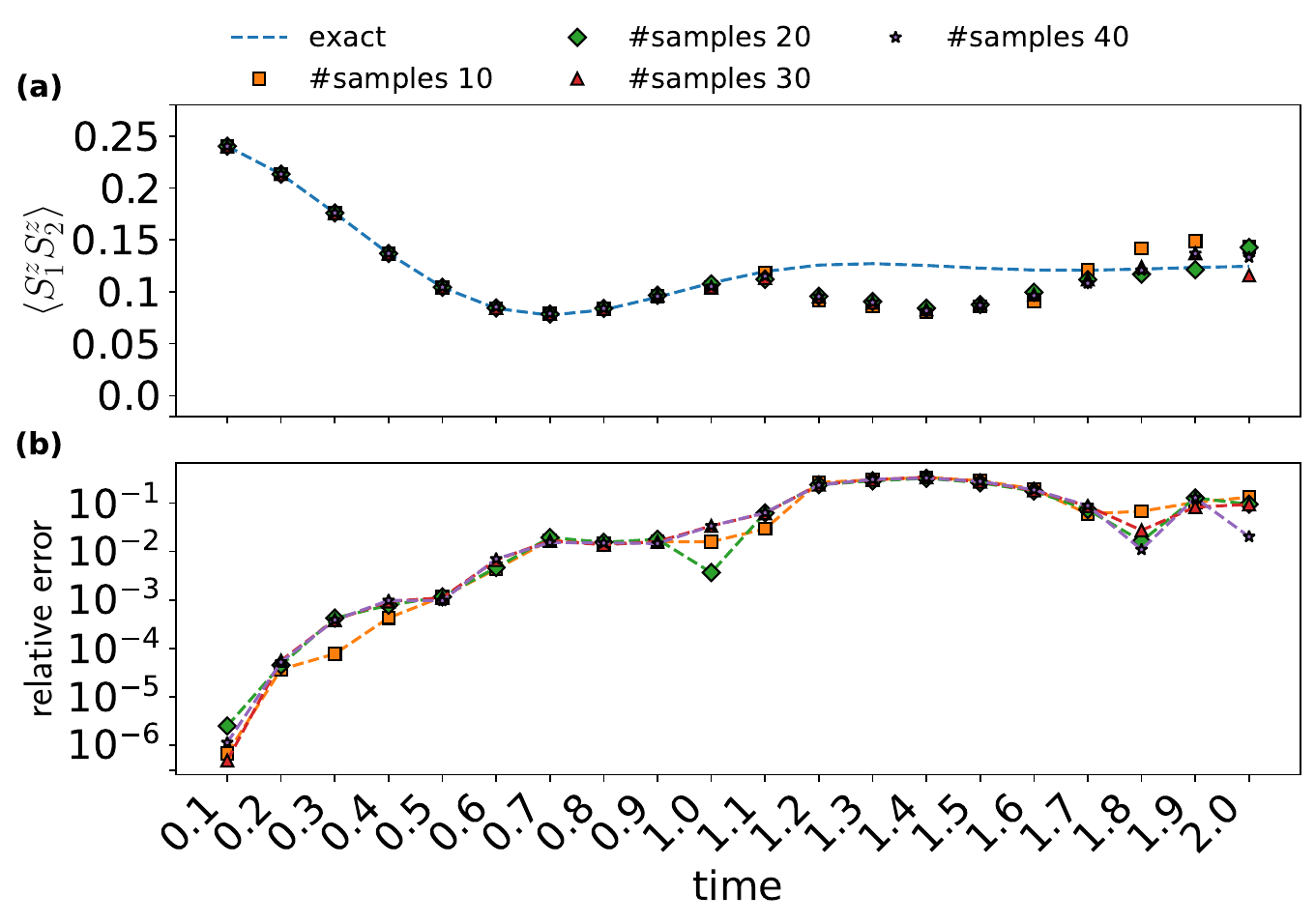}
	\caption{
			Comparison of different sampling sizes at order 2, with depth 8 for HEM. (a) The value of $\langle S_1^z S_2^z\rangle$; (b) the relative error.
	}\label{fig:quantum-samples}
\end{figure}

\newpage

\subsection{Step Size}\label{appx:hyper-step-size}

\begin{figure}[H]
	\includegraphics[scale=0.65]{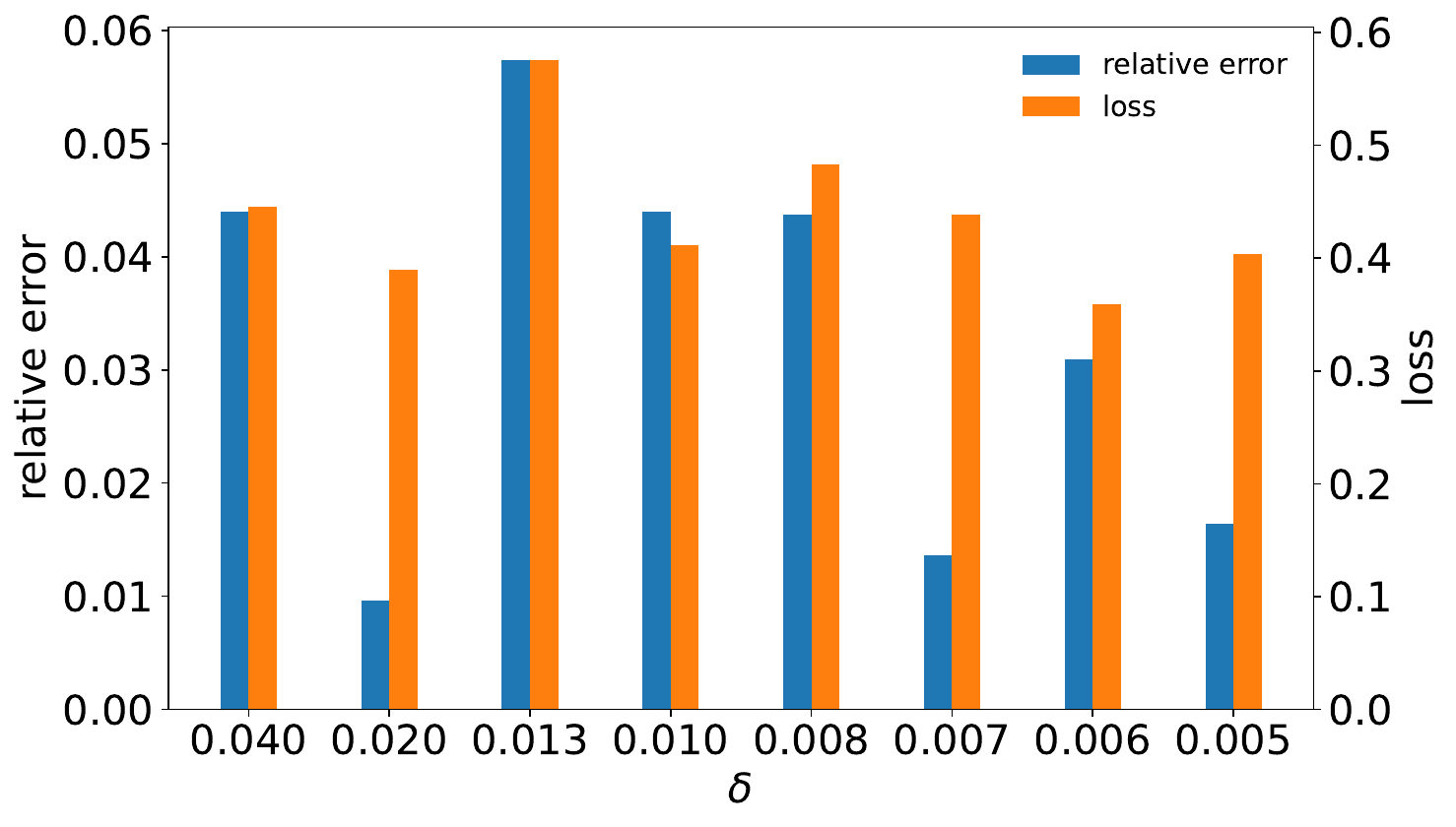}
	\caption{
		Comparison of different step sizes $\delta$ at order 2, with depth 8.
	}
	\label{fig:ensemble-nsteps}
\end{figure}

As for the step size $\delta$ in the sampling, we tune the number of checkpoints $M$ in an ODE solver, which controls the step size as $\delta = T/M$. While smaller step size generally increases the precision of the loss function, we did not observe a significant difference in tuning step size. The results are shown in \cref{fig:ensemble-nsteps}. As discussed in \cref{sec:framework}, this is likely because the TOBC in $1D$ TFIM has many zeros and is quite smooth. Thus, the loss function is not sensitive to the step size.

\end{document}